\documentclass[twocolumn,citeautoscript,pra,superscriptaddress,amsmath,amssymb,longbibliography]{revtex4-1}
 
\usepackage{graphicx}
\usepackage{color}
\usepackage{upgreek}
\usepackage{float}
\usepackage{lipsum}
\usepackage{mathrsfs}
\usepackage{bm}
\usepackage{booktabs}
\usepackage[colorlinks,citecolor=blue,linkcolor=blue]{hyperref}
\usepackage{algorithm} 
\usepackage{algpseudocode}

\AtBeginDocument{
\heavyrulewidth=.08em
\lightrulewidth=.05em
\cmidrulewidth=.03em
\belowrulesep=.65ex
\belowbottomsep=0pt
\aboverulesep=.4ex
\abovetopsep=0pt
\cmidrulesep=\doublerulesep
\cmidrulekern=.5em
\defaultaddspace=.5em
}
 
\newcommand{\ket}[1]{\ensuremath{\left| #1 \right\rangle}}
\newcommand{\bra}[1]{\ensuremath{\left\langle #1 \right|}}
\DeclareMathOperator*{\argmax}{\arg\!\max}
\DeclareMathOperator*{\argmin}{\arg\!\min}
\newcommand{\Tr}[0]{\ensuremath{\text{Tr}}}

\newcommand{\kket}[1]{\ensuremath{\left| #1 \right\rangle\!\rangle}}

\begin{document}
 
\title{Non-Markovian Quantum Process Tomography}
 
\author{G. A. L. White}
\email{white.g@unimelb.edu.au}
\affiliation{School of Physics, University of Melbourne, Parkville, VIC 3010, Australia}

\author{F. A. Pollock}
\affiliation{School of Physics and Astronomy, Monash University, Clayton, VIC 3800, Australia}

\author{L. C. L. Hollenberg}
\affiliation{School of Physics, University of Melbourne, Parkville, VIC 3010, Australia}

\author{K. Modi}
\email{kavan.modi@monash.edu}
\affiliation{School of Physics and Astronomy, Monash University, Clayton, VIC 3800, Australia}

\author{C. D. Hill}
\email{cdhill@unimelb.edu.au}
\affiliation{School of Physics, University of Melbourne, Parkville, VIC 3010, Australia}
\affiliation{School of Mathematics and Statistics, University of Melbourne, Parkville, VIC, 3010, Australia}

\begin{abstract}
Characterisation protocols have so far played a central role in the development of noisy intermediate-scale quantum (NISQ) computers capable of impressive quantum feats. This trajectory is expected to continue in building the next generation of devices: ones that can surpass classical computers for particular tasks -- but progress in characterisation must keep up with the complexities of intricate device noise. 
A missing piece in the zoo of characterisation procedures is tomography which can completely describe non-Markovian dynamics over a given time frame. Here, we formally introduce a generalisation of quantum process tomography, which we call process tensor tomography. We detail the experimental requirements, construct the necessary post-processing algorithms for maximum-likelihood estimation, outline the best-practice aspects for accurate results, and make the procedure efficient for low-memory processes. The characterisation is a pathway to diagnostics and informed control of correlated noise.
As an example application of the hardware-agnostic technique, we show how its predictive control can be used to substantially improve multi-time circuit fidelities on superconducting quantum devices. Our methods could form the core for carefully developed software that may help hardware consistently pass the fault-tolerant noise threshold.

\end{abstract}
 
\maketitle
 
\section{Introduction}

Central to the theme of progress in quantum computing has been the development and application of quantum characterisation, verification, and validation (QCVV) procedures~\cite{Eisert2020,Endo2018,Ferracin2019,white-POST,Harper2020,Jurcevic2021,RBK2017}. These techniques model and identify the presence of errors in a quantum information processor (QIP). These errors may have different origins, such as coherent control noise, decoherence, crosstalk, or state preparation and measurement (SPAM) errors. 
The operational description of open quantum dynamics has been immensely useful in describing the noise present in QIPs \cite{Milz2017,Milz2021}. 
Typified by mappings of density operators, a discrete snapshot of a given noisy process can be characterised through a series of experiments on the QIP. The resulting object is the gold standard for describing two-time errors on a quantum device: a completely positive, trace-preserving (CPTP) map. 

CPTP maps, however, are not sufficient to describe all dynamics present on real quantum devices. 
A generic quantum stochastic process represents many times, and carries correlations across multiple time scales as a rule rather than the exception~\cite{Li2018}. The emergence of adverse effects from temporal correlations in quantum processes is known as non-Markovian noise, and arises from mutual interaction between a system and its complex environment.
Standard CP maps -- such as those characterised by quantum process tomography (QPT) -- cannot describe reduced system-environment ($SE$) dynamics arising from a correlated state.
Famously, this leads to CP-divisibility of a process as a granular measure of non-Markovianity \cite{breuer2016, deVega2017, rivas-NM-review} and prohibits their use in the detailed study of multi-time quantum correlations.

Discourse on device quality is mostly shaped through conventional benchmarks either at the low level -- by emphasis on the fidelity of individual gates; or at the high level -- through holistic measures such as quantum volume~\cite{Cross-QV}. However, there is a discontinuity in these abstractions: the former does not consider the effects of intrinsic \emph{temporal} \emph{context}, by which we mean the tendency of past gate choices to be correlated with future gate outcomes.
Meanwhile, the latter coarsely summarises the average performance of a QIP in high-width, high-depth random circuits.
Indeed, recent progress in quantum computing has led to the engineering of very low error gates~\cite{Arute2019,Pogorelov2021,PhysRevX.11.021058}, but evidence has shown that NISQ devices do not behave like the sum of their parts. Typically, they perform worse than predicted by constituent gate errors alone, which ignore non-Markovian quantum processes present in reality \cite{mirror-benchmarking,Jurcevic2021, white-POST}. A simple example of how this complex noise can impact the outcome of an experiment is that the choice of a gate in the past may influence the action of a gate applied at the present. The two gates do not simply multiply out. Correlated noise can be particularly deleterious not only in its complexity, but in its ability to reduce or even eliminate completely the effectiveness of quantum error correcting codes~\cite{Clader2021, correlated-qec}. Importantly, there is ample evidence that Markov models are insufficient to fully capture the dynamics exhibited by current generation quantum devices~\cite{White-NM-2020,nielsen-gst,Sarovar2020detectingcrosstalk,RBK2017,mirror-benchmarking}, thus motivating the present work.


Characterising general quantum stochastic processes is the first step in explicating correlated noise on quantum processors.
Contextual noise detection has been rigorously studied under a null hypothesis standpoint in prior work. That is, a successful approach has been to probe for a change in measurement statistics -- or some other invariant property of the data -- both with and without the presence of a context variable~\cite{PhysRevX.9.021045, veitia2020macroscopic, veitia2018testing, helsen2019spectral}. This includes non-Markovian noise under its umbrella, where the context is limited to circuit-level decisions.
However, a clear procedure and a systematic framework to characterising contextual noise has been notably absent; there are several technical and fundamental challenges that make this task highly non-trivial. 

In this work, we formally flesh out a part of this missing puzzle piece in quantum tomography, which we call process tensor tomography (PTT). Specifically, PTT addresses the subclass of non-Markovian noise. Dynamical maps -- estimated by QPT -- break down when employed for multi-time processes with memory. Instead, these dynamics are well-described by the process tensor, a recently developed mathematical framework to represent quantum stochastic processes~\cite{Pollock2018a}. We first focus on formalising the tomographic reconstruction of generic multi-time quantum processes and discuss the challenges and pitfalls that lie in the way. We then turn our attention to sparingly characterise processes while overcoming many of these obstacles by integrating PTT with maximum likelihood estimation (MLE). This approach is then applied efficiently for the case of processes with low-memory, described by their Markov order~\cite{taranto1}. We emphasise that this work adopts a constructive approach to solve the problem of statistically-robust non-Markovian quantum process tomography in full generality. Consequently, we consider relatively small numbers of steps on single-qubit systems. Scalability of the method is intrinsically linked to the complexity of the process under scrutiny. For environments with relatively few relevant degrees of freedom, or processes with quickly decaying memory we foresee no fundamental obstacles to simplifying and scaling up the techniques introduced here~\cite{cramer2010efficient,cygorek2022simulation,dang2021process,PhysRevLett.111.020401}

The MLE based PTT (or MLE-PTT) has several advantages over linear inversion PTT (LI-PTT); it is significantly more efficient than standard inversion methods, as well as overcoming the difficulties of positivity and causality conditions. Yet, it grows exponentially in the number of timesteps. To overcome this we accommodate for sparseness in the complexity of the dynamics.
Low memory processes are ubiquitous in nature and our characterisation tools leverage this to offer both an experimentally and computationally more efficient description of the process. In particular, this permits us to quantify the degree to which a limited memory model fails to predict laboratory observations. This significantly enhances previous results which quantify how memoryless models breakdown~\cite{RBH-wildcard,nielsen-gst}. Moreover, our methods establish a trade-off between the level of characterisation complexity and a desired approximation. Our measures have both simple interpretations, and a clear recipe to expand the model should they fail to adequately explain the data.

To show the efficacy of this approach, we report the real device characterisation of multi-time quantum stochastic processes in a way that is both consistent and fully inclusive of non-Markovian dynamics. There are three key motivators for these implementations: to benchmark quantum devices and diagnose non-Markovian noise, to study the structure of quantum stochastic processes, and for precise enhanced control of non-Markovian systems. 
As such, we first show MLE-PTT to be highly reliable at characterising multi-time dynamics. 
The reliability and the efficiency of these methods allows us to implement noise-aware control to significantly reduce the noise present across a variety of contexts. We demonstrate not only the utility of our technique here, but also its necessity by showing that improvement is contingent on the inclusion of higher order (i.e. multi-time) temporal correlations in the model.

\begin{table*}[ht]
\centering
\resizebox{\textwidth}{!}{%
\begin{tabular}{@{}l|l|l|l@{}}
\toprule
                        & Quantum State                  & Quantum Process                                                 & Process Tensor                                                                                                                     \\ \midrule
Characterisation object & Density Operator $\rho$                  & Quantum Channel $\mathcal{E}$                                             & Process Tensor $\mathcal{T}_{k:0}$                                                                                                              \\
Mapping                 & $\mathcal{H}_S\rightarrow \mathcal{H}_S$ & $\mathscr{B}(\mathcal{H}_{S})\rightarrow \mathscr{B}(\mathcal{H}_{S})$    & $\bigotimes_{i=1}^k \mathscr{B}(\mathscr{B}(\mathcal{H}_{S})) \rightarrow \mathscr{B}(\mathcal{H}_S)$                                                   \\
Observed probabilities  & $p_i = \text{Tr}\left[\Pi_i\rho\right]$    & $p_{ij} = \text{Tr}\left[(\Pi_i\otimes \rho_j^\text{T})\hat{\mathcal{E}}\right]$         & $p_{i,\vec{\mu}} = \text{Tr}\left[(\Pi_i \otimes  \mathcal{B}_{k-1}^{\mu_{k-1}\text{T}}\otimes \cdots \otimes \mathcal{B}_0^{\mu_0\text{T}})\Upsilon_{k:0}\right]$ \\
Positivity Constraint   & $\rho \succcurlyeq 0$                    & $\hat{\mathcal{E}} \succcurlyeq 0$                                              & $\Upsilon_{k:0} \succcurlyeq 0$                                                                                                              \\
Affine Constraint       & $\text{Tr}\left[\rho\right] = 1$         & $\text{Tr}_{\text{out}}[\hat{\mathcal{E}}] = \mathbb{I}_{\text{in}}$ & $\text{Tr}_{\text{out}}\left[\Upsilon_{k:0}\right] = \mathbb{I}_{\text{in}}\otimes \Upsilon_{k-1:0}\: \forall\: k$                           \\ \bottomrule
\end{tabular}%
}
\caption{Detail on different levels of the quantum tomography hierarchy, pertinent to experimental reconstruction. QST reconstructs a density operator, $\rho$, a positive matrix with unit trace representing the quantum state. QPT reconstructs a quantum channel, $\mathcal{E}$ through its action on different states. This map must be both CP and TP, conditions which, in Choi form, manifest themselves as positivity and affine constraints on the matrix. Finally, PTT reconstructs a process tensor $\mathcal{T}_{k:0}$ through its action on different control operations. This object must have a positive matrix Choi form, and respect causality.
The information of each column is strictly contained in the column to the right.}
\label{tab:tomography-summary}
\end{table*}

Our paper is structured as follows. In Section~\ref{sec:hierarchy}, we give a brief background on quantum state tomography (QST) and QPT before continuing into the theory of quantum stochastic processes and tomographic reconstruction of a process tensor. This analyses our non-Markovian characterisation in context. In Section~\ref{sec:MLPTT} we derive the main components of a MLE-PTT protocol, including a positive causal projection necessary for reconstructed process tensors to be physical. This procedure is then benchmarked and validated on IBM Quantum devices. Specifically, the number of circuits required for accurate LI-PTT in a real setting scales as $\mathcal{O}(N_{\rm oc}^k)$, where $N_{\rm oc}$ is the number of (overcomplete) inputs per timestep and $k$ is the number of timesteps. The MLE-PTT procedure reduces this to $\mathcal{O}(N_{\rm mle}^k)$, where $N_{\rm oc}=24$ and $N_{\rm mle}=10$. Using the tools of Section~\ref{sec:MLPTT}, in Section~\ref{sec:MO} we motivate and advance the theory of quantum Markov order, in which we determine how to adaptively truncate weaker long-time temporal correlations in the model. This tempers the exponential scaling to a linear scaling, i.e., $\mathcal{O}(k \cdot N_{\rm mle}^\ell)$, where $\ell$ is the fixed Markov order. This is not only practical, it provides an accessible diagnostic to the complexity of device noise.
In Section~\ref{sec:applications} we demonstrate noise-aware control over a non-Markovian process, which is based on noise characterisation methods of previous sections. Namely, we characterise noise for Markov orders of $\ell = \{1,2,3\}$ and use this information to significantly increase fidelity of several NISQ devices by using non-Markovian correlations as a resource. We show that the performance of the machine improves when the Markov order is chosen to be higher. This leads to a trade-off relation between characterisation complexity and accuracy of the characterisation. 
\subsection{Summary of quantum characterisation}
Before we proceed, we will briefly analyse where PTT sits among existing characterisation techniques. Frameworks for dynamics in the literature can be broadly classed under dynamical map or master equation formalisms. The former captures only two-time correlations which, in the presence of non-Markovian noise, will fail to describe multi-time processes when composed together~\cite{Milz2017}. This applies clearly to QPT, but also to any QCVV procedures born out of the umbrella of quantum channels, such as GST, randomised benchmarking (RB), and Hamiltonian tomography~\cite{nielsen-gst, PhysRevA.77.012307,PhysRevLett.113.080401,Wang_2015,Eisert2020}. The latter generally is a function of at most three-time correlations and be reduced to a family of dynamical maps~\cite{pollock-tomographic-equations}.

In the case where these approaches are insufficient to characterise features of non-Markovian noise, ad-hoc extensions techniques exist to detect a departure from the Markov assumptions. These detect some parts of the non-Markovian character but are not generalisable or predictive. Their application is typically as witnesses or for shallow diagnostics, but cannot rigorously measure the memory or be employed systematically to control the system. Common examples include memory kernels for master equations~\cite{PhysRevA.93.052111}, or statistical tests to establish causal connections between environmental factors or gate choices, and system-level dynamics~\cite{PhysRevX.9.021045, veitia2020macroscopic, veitia2018testing, helsen2019spectral, Sarovar2020detectingcrosstalk}. Statistical tests have also been employed to quantify the confidence with which breakdown of Markovianity can be described~\cite{nielsen-gst}.

\par

In a precise way, PTT is a direct generalisation of the QPT framework: instead of estimating a single dynamical map, it estimates a sequence of possibly correlated dynamical maps. These temporal correlations may be arbitrarily strong. Thus it is, to the best of our knowledge, the only procedure demonstrated to fit the criteria of systematically capturing this difficult and important class of noisy dynamics. Moreover, it is fully general, without relying on underlying microscopic models. It carries some of the same associated baggage (exponential scaling in full generality, assumptions about prior calibrations) -- but can also be imbued with the modifications and simplifications that make the QPT framework a rich tool for quantum characterisation. A guiding set of simplifying assumptions can reduce the experimental burden of characterisation at the expense of either a possible sacrifice in accuracy or information gain, for example in compressed sensing or tensor network models~\cite{flammia2012quantum,cramer2010efficient}.
We also foresee that many of the ideas that either arise from or utilise QPT could be applied to PTT, further extending this branch of QCVV and error mitigation~\cite{Endo2018}.


We conclude this section by emphasising the extent to which the methods and theory in this work offer a departure from those introduced in Ref.~\cite{White-NM-2020}. This previous work showed that non-Markovian characterisation is possible on real quantum devices by implementing the process tensor mapping through linear inversion. In particular, it was a demonstration of how to effectively wield the spatiotemporal version of Born's rule (Equation~\eqref{eq:PToutput} and Figure~\ref{fig:trajectory}c) on quantum processors -- thus permitting characterisation and control regardless of the strength of the $SE$ interaction. However, the shortcomings were that the process tensor itself (and thus any rigorous non-Markovian measures) were not estimated. Further, the averaging over sampling statistics was made very expensive through an overcomplete basis -- and no guarantees were made of a completely positive or causal mapping. Finally, the structure of the process was left opaque, such that all possible operational trajectories were considered. This leaves it impossible to leverage any sparseness in the real process.

By fleshing out PTT in a sophisticated way, the present work provides a clear framework by which a process tensor may be estimated using a minimal complete basis in a manner which is both physical and statistically robust. We also show how decaying correlations in the memory may be truncated to offer an efficient characterisation of the non-Markovian process, and finally how all of the above may be employed to provide superior control of a quantum device.

\section{A hierarchy of quantum tomography}
\label{sec:hierarchy}
We start by outlining the fundamentals of QST, QPT, and PTT. These procedures build on each other; for a $d$-dimensional system, QST requires a set of $\mathcal{O}(d^2)$ experiments, QPT is most easily thought of as $d^2$ QSTs and requires $\mathcal{O}(d^4)$ experiments, and PTT $\mathcal{O}(d^{4k})$ experiments, where $k=1,2,\ldots$ is the number of times steps~\cite{RBK2010,PhysRevA.87.062119, White-NM-2020}. The familiarity of the first two lays the groundwork for the latter. The treatment and practical concerns of each technique are similar with respect to real data. One key difference lies in the fact that due to the higher-dimensional superoperator basis, especially for PTT, small errors can become magnified and require closer attention. In addition to an overview of tomography, we present the conceptual developments of PTT in this section and scrutinise its proclivities -- such as with respect to hardware control restrictions and SPAM error.\par
Fundamentally, quantum tomography is an exercise in reconstructing linear maps from experimental data. This can be accomplished by measuring the input-output relations on a complete basis for the input space. A disconnect between theory and experiment occurs when, in practice, the input vectors are faulty in some way (such as noisy preparation) or the measured output frequencies differ from that of the real population (due to a noisy probe or finite sampling error)~\cite{RBK2010}. As well as producing an object that may disagree with experiment for inputs away from the characterisation, the resulting estimate might not even be physical. A variety of different methods may be employed to overcome some of these problems: the collection of more data, the elevation of inputs and outputs to the model~\cite{gst-2013,PhysRevA.87.062119}, employing an overcomplete basis in the characterisation~\cite{intro-GST}, and the treatment of the measured data to fit a physical model~\cite{Hradil2004}. These techniques are applicable, regardless of the model type, and we will discuss their utility in PTT. \par 
Before examining the PTT description of quantum stochastic processes, we will emphasise parallels to more conventional tomography in QST and QPT, such that the content can appear more familiar to readers.
Along our exposition, we emphasise a `hierarchy' in the sense that the information of each level is strictly contained within the characterisation of the next level. That is to say, QPT can describe the reconstructed state of QST, and PTT can describe the dynamical map of QPT. We present a summary of each map in Table~\ref{tab:tomography-summary}, as well as their physical requirements, and continue to flesh out here. 

\subsection{Quantum State and Quantum Process Tomography}
The foundation of most QCVV procedures is the estimation of quantum states and quantum channels on an experimental device. By quantum state, we mean the density matrix representation of a system at a given time.
A quantum channel -- or quantum stochastic matrix, or dynamical map -- then expresses the evolution of some state between two times, and has a freedom in representation.
A convenient choice for QPT employs the Choi representation. Here, using the Choi-Jamiolkowski isomorphism (CJI), CP maps may be given by a positive matrix representation as a quantum state, exploiting the correspondence between $\mathscr{B}(\mathcal{H}_{\text{in}})\rightarrow \mathscr{B}(\mathcal{H}_{\text{out}})$ and $\mathscr{B}(\mathcal{H}_{\text{out}})\otimes \mathscr{B}(\mathcal{H}_{\text{in}})$. Here, $\mathscr{B}(\mathcal{H})$ denotes the space of bounded linear operators on a Hilbert space $\mathcal{H}$. Explicitly, for some channel $\mathcal{E}$, its Choi state $\hat{\mathcal{E}}$ is constructed through the action of $\mathcal{E}$ on one half of an unnormalised maximally entangled state $\ket{\Phi^+} = \sum_{i=1}^d \ket{ii}$, with identity map $\mathcal{I}$ on the other half:
\begin{equation}
\label{eq:channel-choi}
    \hat{\mathcal{E}} := (\mathcal{E}\otimes \mathcal{I})\left[|\Phi^+\rangle\langle\Phi^+|\right] = \sum_{i,j=1}^d \mathcal{E}\left[|i\rangle\langle j|\right]\otimes |i\rangle\langle j|.
\end{equation}

All reconstruction of experimental properties must begin with a probe to read out quantum information. 
This extraction comes from a known POVM $\mathcal{J}:=\{\Pi_i\}_{i=1}^L$ with associated elements called `effects'. To reconstruct any state, $\mathcal{J}$ must span the space of density matrices $\mathscr{B}(\mathcal{H}_S)$, a characteristic known as informational completeness (IC).
For some density matrix $\rho$, a POVM yields observable probabilities for each effect in accordance with Born's rule: 
\begin{equation}
\label{born-rule}
    p_i = \Tr[\Pi_i \rho].
\end{equation}
For a quantum channel, its action is given in terms of its Choi state by:
\begin{equation}\label{ch-born-rule}
    \mathcal{E}\left[\rho_{\text{in}}\right] = \text{Tr}_{\text{in}}\left[(\mathbb{I}\otimes \rho_{\text{in}}^\text{T})\hat{\mathcal{E}}\right] =\rho_{\text{out}}.
\end{equation}

These input-output relations are sufficient to reconstruct both a state  $\rho$ and a channel $\mathcal{E}$. That is, measuring $p_i$ for each element of $\mathcal{J}$ is sufficient to construct $\rho$, and measuring $\{\rho'_i\} := \{\mathcal{E}[\rho_i]\}_{i=1}^n$ for a full basis of inputs is sufficient to construct $\hat{\mathcal{E}}$. This is accomplished with the construction of a \emph{dual} set $\mathcal{D} := \{\Delta_i\}_{i=1}^L$ to linearly independent $\mathcal{J}$, satisfying $\Tr[\Pi_i\Delta_j] = \delta_{ij}$. Similarly, let $\{\omega_j\}$ be the dual set to $\{\rho_i\}$. 
Note that in practice, linear independence may be relaxed with an overcomplete basis in the case where $L,n > d^2$, where a matrix pseudoinverse is used to find the dual set, rather than an inverse -- see Appendix~\ref{appendix:pt-maths} for further details.
Then, we may (respectively) express $\rho$ and $\hat{\mathcal{E}}$ as
\begin{equation}
\label{qst-rho}
    \rho = \sum_{i=1}^L p_i \Delta_i \quad\mbox{and}\quad
    \hat{\mathcal{E}} = \sum_{i=1}^n \rho'_i \otimes \omega_i^\text{T},
\end{equation}
which, by design, are consistent with Equations~\eqref{born-rule} and~\eqref{ch-born-rule}, respectively. 
By combining the two equations above, $\hat{\mathcal{E}}$ may also be decomposed in terms of POVM effects as
\begin{equation}
    \hat{\mathcal{E}} = \sum_{j=1}^L\sum_{i=1}^n p_j'\Delta_j\otimes \omega_i^\text{T}.
\end{equation}
Note that the $\Delta_j$ and $\omega_i$ matrices are not usually positive, but the resulting $\rho$ is both positive with unit-trace, and the resulting $\hat{\mathcal{E}}$ is both positive and trace-preserving, in that its marginal input is maximally mixed. These conditions are written explicitly in the final row of Table~\ref{tab:tomography-summary}.

\subsection{Process tensor tomography}

\begin{figure}
    \centering
    \includegraphics[width=\linewidth]{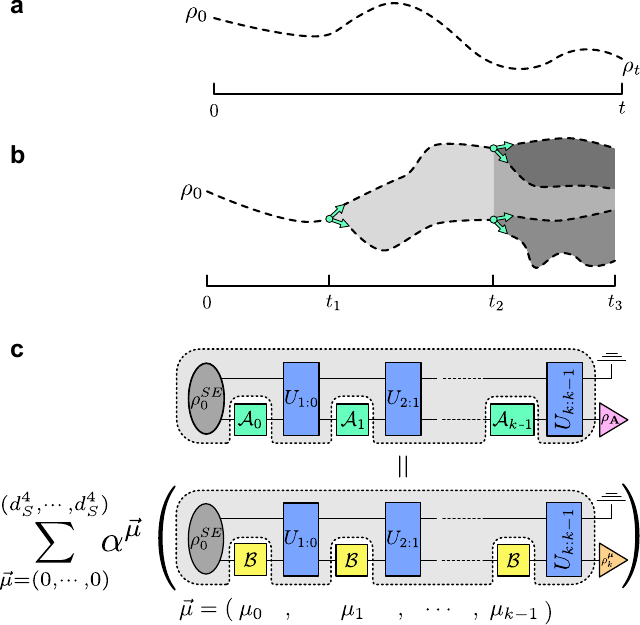}
    \caption{\textbf{a} The conventional model of open quantum systems tracks the state of the system as a function of time, but cannot build up multi-time joint statistics. \textbf{b} The quantum stochastic process picture considers the response of the system to different sequences of gates; by considering all trajectories of the system, correlations between different times may be exactly characterised and quantified.
    \textbf{c} A pictorial description of how linear expansion in a basis by the process tensor preserves intermediate dynamics and expresses an arbitrary sequence. 
    }
    \label{fig:trajectory}
\end{figure}

A typical approach to studying dynamical processes consists of monitoring the state of the system as a function of time, as in Figure~\ref{fig:trajectory}a~\cite{breuer2016}. Although effects such as coherent state oscillation can flag non-Markovianity, any interrogation of the system necessarily disrupts its future evolution. For this reason, joint statistics cannot be measured across time and, consequently, multi-time correlations cannot be characterised. As a result, quantum non-Markovian effects, which can have a variety of different physical sources, have been historically difficult to theoretically describe, much less experimentally capture. Often, non-Markovian effects are quantified in terms of `leftover' error, inferred by the extent to which Markov models break down \cite{RBH-wildcard, nielsen-gst, rivas-NM-review}.\par 
Recently, the process tensor framework~\cite{Pollock2018a} (and the process matrix framework~\cite{1367-2630-18-6-063032}) were proposed as a generalisation of classical stochastic processes to the quantum domain. Importantly, this generalisation allows for the study of multi-step temporal correlations -- or non-Markovianity -- in quantum systems. An important feature of the PTT formalism is that it maps all possible temporal correlations onto spatial correlations over sequences of CPTP channels, leaving non-Markovian measures as operationally well-defined as for any quantum or classical spatial correlation. Applying conventional many-body techniques allows for necessary and sufficient measures of device non-Markovianity, as well as the more fine-grained study of operation-specific context dependence. This fills a gap in the library of QCVV resources~\cite{Eisert2020}.\par
Any continuous-time quantum stochastic process can be discretised in a number of time-steps: $\mathbf{T}_k = \{t_0, t_1, \cdots, t_k\}$ (for example, in the context of a quantum circuit). A finite-time process tensor is then a marginal of the continuous time process tensor~\cite{Milz2020}, which represents all possible correlations in $\mathbf{T}_k$. To capture statistics for each $t_i\in \mathbf{T}_k$, the experimenter applies an IC basis of control operations, which each change the trajectory of the state and maps to an output. In completing this procedure for all times, all trajectories consistent with $\mathbf{T}_k$ may be inferred. This is depicted in Figure~\ref{fig:trajectory}b. This is sufficient both to construct all joint statistics, and to predict the output subject to any generic sequence.  Panels a and b of Figure~\ref{fig:trajectory} contrast the traditional approach to open quantum dynamics and PTT framework.\par
To be precise, we consider the situation where a $k$-step process is driven by a sequence $\mathbf{A}_{k-1:0}$ of control operations, each represented mathematically by CP maps: $\mathbf{A}_{k-1:0} := \{\mathcal{A}_0, \mathcal{A}_1, \cdots, \mathcal{A}_{k-1}\}$, after which we obtain a final state $\rho_k(\mathbf{A}_{k-1:0})$ conditioned on this choice of interventions. 
These controlled dynamics have the form:
\begin{equation}\label{eq:multiproc}
        \rho_k\left(\textbf{A}_{k-1:0}\right) = \text{tr}_E [U_{k:k-1} \, \mathcal{A}_{k-1} \cdots \, U_{1:0} \, \mathcal{A}_{0} (\rho^{SE}_0)],
\end{equation}
where $U_{k:k-1}(\cdot) = u_{k:k-1} (\cdot) u_{k:k-1}^\dag$. Eq.~\eqref{eq:multiproc} can be used to define a mapping from past controls $\mathbf{A}_{k-1:0}$ to future states $\rho_k\left(\textbf{A}_{k-1:0}\right)$, which is the process tensor $\mathcal{T}_{k:0}$:
\begin{equation}
\label{eq:PT}
    \mathcal{T}_{k:0}\left[\mathbf{A}_{k-1:0}\right] = \rho_k(\mathbf{A}_{k-1:0}).
\end{equation}
The logic of the process tensor is depicted in Figure~\ref{fig:trajectory}c, mirroring the trajectory sketch in Figure~\ref{fig:trajectory}b.
\par

\begin{figure*}[ht!]
    \centering
    \includegraphics[width=\linewidth]{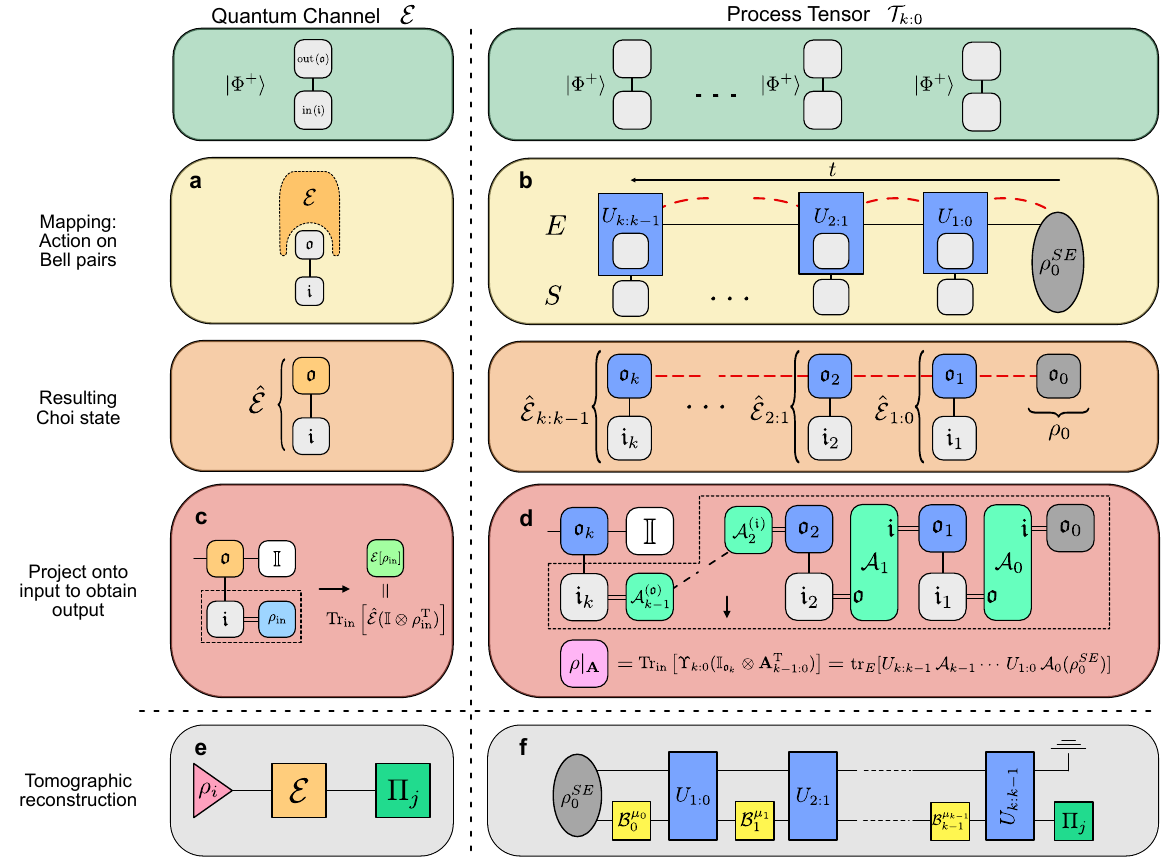}
    \caption{Operation, manipulation, and characterisation analogues between quantum process tomography and process tensor tomography. \textbf{a} The Choi-Jamiolkowski isomorphism represents a quantum process $\mathcal{E}$ by the density matrix $\hat{\mathcal{E}}$ using $\mathcal{E}$'s action on one half of a maximally entangled state. \textbf{b} In the generalised CJI, $SE$ unitaries act on one half of a maximally entangled state per time-step. Correlations between times (denoted in red dashed lines) are then mapped onto spatial correlations between each output leg of a Bell pair. The result is a collection of (possibly correlated) CPTP maps, as well as the average initial state. \textbf{c} The outcome of $\mathcal{E}$ on some $\rho_{\text{in}}$ is obtained by projecting the Choi state onto $\mathbb{I}\otimes \rho_{\text{in}}^\text{T}$ and tracing over the input space. \textbf{d} The outcome of a process conditioned on a sequence of operations $\mathbf{A}_{k-1:0}$ is obtained by projecting the Choi state of $\Upsilon_{k:0}$ onto the Choi state $\bigotimes_{i=0}^{k-1}\mathcal{A}_i$ and tracing over the input. Each $\mathcal{A}_i$ maps the output state of the $i$th CPTP map to the input leg of the $(i+1)$th CPTP map. \textbf{e} To reconstruct $\hat{\mathcal{E}}$ experimentally, prepare a complete set of states $\{\rho_i\}$, apply $\mathcal{E}$, and reconstruct the output state with an IC-POVM $\{\Pi_j\}$. \textbf{f} To reconstruct $\Upsilon_{k:0}$, measure each final state of the system subject to a complete basis of CP maps $\{\mathcal{B}_{j}^{\mu_j}\}$ at each time. Note that the blue unitaries here are symbolic of any $SE$ interactions, and not gates that need performing.}
    \label{fig:PTT-explanation}
\end{figure*}

In this sense, the process tensor is designed to account for intermediate control operations, and quantifies quantum correlations between past operations and future states. Just as in the case of CPTP maps, it can be shown that the process tensor too has a many-body Choi representation~\cite{Pollock2018a}. Both states $\rho$ and channels $\hat{\mathcal{E}}$ have affine conditions ensuring unit probability and trace preservation, respectively. Similarly, the Choi state of the process tensor $\Upsilon_{k:0}$ also has affine conditions: these guarantee causality. That is, any future control operations cannot affect the past statistics. These facets of the process tensor are all possible to experimentally reconstruct using many of the techniques from QPT and QST. We directly employ and build upon these ideas in this work. We start from the direct linear inversion construction of the process tensor Choi state, and then proceed with maximum likelihood estimation and truncated Markov models.
Along the way, we make explicit the parallels between QPT and PTT as a generalisation in Figure~\ref{fig:PTT-explanation} for pedagogical purposes. In particular, Figure~\ref{fig:PTT-explanation}b contrasts the generalised CJI of the process tensor with the standard channel CJI, emphasising the extension made.




\emph{Linear Inversion Construction} --- 
We begin by discussing the construction of the process tensor direct from experimental data, which recently was reported in Ref.~\cite{White-NM-2020}. The estimate here comes from (pseudo)inverting the feature matrix on observed data. The resulting object need not be physical, and though it may be consistent with its measurement basis, it may not even serve as a good indicator for the behaviour of other sequences. 
\par 
The sequence of interventions $\mathbf{A}_{k-1:0}$ is a CP map represented through the CJI as a $2k$-partite quantum state -- i.e. through action on a maximally entangled state at each of the $k$ time steps. When the operations at each time step are chosen independently, the sequence is given by $\mathbf{A}_{k-1:0}=\bigotimes_{j=0}^{k-1}\mathcal{A}_j$. Each time-local operation may be expanded into a basis $\{\mathcal{B}^{\mu_j}_j\}$ such that any CP map can be expressed as $\mathcal{A}_j = \sum_{\mu_j=1}^{d^4} \alpha_j^{\mu_j} \mathcal{B}^{\mu_j}_j$. The subscript $j$ allows for the possibility of a different basis at each time, meanwhile the superscript $\mu_j$ denotes the elements of that particular set. The complete spatio-temporal basis is
\begin{equation}
\{\mathbf{B}_{k-1:0}^{\vec{\mu}}\} = \left\{\bigotimes_{j=0}^{k-1}\mathcal{B}_j^{\mu_j}\right\}_{\vec{\mu}=(1,1,\cdots,1)} ^{(d^4,d^4,\cdots,d^4)}
\end{equation}
with vector of indices $\vec{\mu}$. To construct the process tensor, therefore, it suffices to measure the output $\rho_k^{\vec{\mu}} := \rho_k(\mathbf{B}_{k-1:0}^{\vec{\mu}})$ for each $\vec{\mu}$, see Fig. \ref{fig:PTT-explanation}f. To do so, we make use of the dual set $\{\Delta_j^{\mu_j}\}$ such that $\Tr[\mathcal{B}_j^{\mu_j}\Delta_j^{\nu_j}] = \delta_{\mu_j\nu_j}$. Then, the Choi state $\Upsilon_{k:0}$ of the process tensor $\mathcal{T}_{k:0}$ is given by
\begin{equation}
    \label{choi-pt}
    \Upsilon_{k:0} = \sum_{\vec{\mu}} \rho_k^{\vec{\mu}}\otimes \mathbf{\Delta}_{k-1:0}^{\vec{\mu}\ \text{T}},
\end{equation}
where $\{\mathbf{\Delta}_{k-1:0}^{\vec{\mu}}\} = \{\bigotimes_{j=0}^{k-1}\Delta_j^{\mu_j}\}$ satisfies $\Tr[\mathbf{B}_{k-1:0}^{\vec{\mu}}\mathbf{\Delta}_{k-1:0}^{\vec{\nu}}] = \delta_{\vec{\mu}\vec{\nu}}$. We remark here that the Choi form of a process tensor is an $2k+1$-partite state with alternating input and output indices. 

We use the notation $\mathfrak{o}_j$ to denote an output leg of the process at time $t_j$, and $\mathfrak{i}_j$ for the input leg of the process at time $t_{j-1}$. The collection of indices is therefore $\{\mathfrak{o}_k,\mathfrak{i}_k,\cdots,\mathfrak{o}_2,\mathfrak{i}_2,\mathfrak{o}_1,\mathfrak{i}_1,\mathfrak{o}_0\}$. These correspond to the marginals of the process, $\{\hat{\mathcal{E}}_{k:k-1},\cdots,\hat{\mathcal{E}}_{2:1},\hat{\mathcal{E}}_{1:0},\rho_0\}$ as shown in Figure~\ref{fig:PTT-explanation}b. Note that this ordering is opposite to the arrow of time in quantum circuits, following instead the convention of matrix multiplication.

Once characterised, the action of the process tensor on a sequence of operations is found by projecting the process tensor onto the Choi state of this sequence (up to a transpose). That is,
\begin{equation}\label{eq:PToutput}
    \rho_k(\mathbf{A}_{k-1:0}) \!=\! \text{Tr}_{\overline{\mathfrak{o}}_k} \! \left[\Upsilon_{k:0}\left(\mathbb{I}_{\mathfrak{o}_k}\otimes \mathcal{A}_{k-1}\otimes \cdots \mathcal{A}_0\right)^\text{T}\right],
\end{equation}
where $\overline{\mathfrak{o}}_k$ is every index except $\mathfrak{o}_k$. This equation, reminiscent of the Born rule~\cite{chiribella_memory_2008, Shrapnel_2018}, can determine the output of any sequence of operations (that are consistent with $\mathbf{T}_k$), and is inclusive of all intermediate $SE$ dynamics as well as any initial correlations. Predicting outcomes subject to control sequences naturally makes the process tensor a very useful object for quantum control.\par 

The generalised CJI of the process tensor maps the multi-time process onto a many-body state. Consequently, most of the analytical tools used for the description of spatial correlations can also be employed to describe the temporal correlations of the process. This provides a key motivator for carrying out PTT: the reconstructed process tensor Choi state provides all information about any spatiotemporally correlated behaviour in a quantum device. This makes it a very useful diagnostic tool for near-term quantum devices: one can qualitatively and quantitatively describe the complexity of the system's interaction with its environment. 

Conventional measures of non-Markovianity are well-motivated, but typically only describe a subset of non-Markovian processes. That is to say, they are sufficient but not necessary measures~\cite{PhysRevLett.101.150402, PhysRevLett.103.210401,PhysRevLett.105.050403, PhysRevA.83.052128, vacchini_non-markovian_2013, rivas-NM-review, breuer2016, deVega2017, Li2018}. The process tensor gives rise to both a necessary and sufficient measure of non-Markovianity through all CP-contractive quasi-distance measures between $\Upsilon_{k:0}$ and its closest Markov process tensor according to that distance.
A Markovian process is one without any correlations in its Choi state -- i.e. a product state of some CPTP maps $\hat{\mathcal{E}}_{j+1:j}$ and average initial state $\rho_0$. In general the closest product state is not found with an analytic form, however there exist convenient choices. One example is the relative entropy, $\mathcal{S}[\rho\|\sigma] = \Tr[\rho (\log\rho - \log\sigma)]$. For this, the closest Markov process tensor is obtained by discarding the correlations. That is,
\begin{equation}
\label{eq:qmi}
    \Upsilon_{k:0}^{\text{Markov}} = \text{Tr}_{\overline{k}}\left[\Upsilon_{k:0}\right]\otimes \text{Tr}_{\overline{k-1}}\left[\Upsilon_{k:0}\right]\otimes \cdots \text{Tr}_{\overline{0}}\left[\Upsilon_{k:0}\right],
\end{equation}
where $\overline{j}$ is the trace over every index except $\mathfrak{o}_j$ and $\mathfrak{i}_{j}$. 
These marginals constitute exactly the above average CPTP maps. Once reconstructed with PTT, this informs the user how close their device performs when compared to a temporally uncorrelated (not necessarily noiseless) Markov model. \par

\subsection{Performing PTT on NISQ devices}

To reconstruct the process tensor, a minimal complete basis for the process tensor requires $d_S^4$ operations spanning the superoperator space $\mathscr{B} (\mathscr{B} (\mathcal{H}_S))$ at each time-step.
One mathematically convenient basis is an IC POVM, followed by a set of IC preparation states which are independent of the measurement outcomes. However, this procedure requires fast projective control in device hardware. Although some progress has been made on this front~\cite{Corcoles2021}, fast control, in practice, is often too noisy and leads to a poor reconstruction. An exception to this is the recent work on process characterisation with intermediate measurements completed by Xiang \emph{et al.} in~\cite{Xiang2021}. A different approach to this measure-and-prepare strategy could implement this entire basis set through an interaction between the system qubit and an ancilla, followed by projective measurement in which the outcome of the ancilla is recorded. This too is problematic in practice as system-ancilla interactions will generate an operation which effectively depends on the circuit.

\par
For a typical NISQ device, all intermediate operations are limited to unitary transformations, and a measurement is only allowed at the end -- or if mid-circuit measurement with feed-forward is possible, it is typically much slower than $SE$ dynamics. Nevertheless, it is possible to work within the experimental limitations and implement an informationally incomplete set of basis operations. This constructs what is known as a \emph{restricted} process tensor~\cite{PT-limited-control} and has full predictive power for any operation in a subspace of operations. That is, these objects are well-defined as maps over the span of their incomplete basis, but do not form positive operators and do not uniquely fix a process tensor's Choi state.\par

We expand on the notion of a restricted process tensor here by offering an analogous quantum state perspective. The measurement of $\rho_k$ subject to some sequence of operations $\mathbf{B}^{\vec{\mu}}_{k-1:0}$ is akin to measuring a many-body observable on $\Upsilon_{k:0}$, as per Equation~\eqref{eq:PToutput}. The reconstruction of $\Upsilon_{k:0}$ then lies on the hyperplane defined by
\begin{gather}
\text{span}(\{\Pi_i,\mathbf{B}^{\vec{\mu}}_{k-1:0}\}).
\end{gather}
When the set of operations is tomographically incomplete, $\Upsilon_{k:0}$ is non-uniquely fixed, thus termed `restricted'. Consider, for example, in the state case, if only $Z$ and $X$ measurements were performed on each subsystem. Then a consistent, non-unique state could be estimated with the correct $Z$ and $X$ expectations. However, the $Y$ expectations would be a free parameter (up to positivity of the state).\par 

In the special case of unitary control, the Choi states representing each intervention are rank one projections onto a maximally entangled state of dimension $d_S^2$. Expanding these entangled measurements into a Hermitian basis yields only non-zero coefficients on the non-local terms. That is, in the Pauli basis for example, we have $\text{Tr}(\mathcal{B}_l^{\mu_l}\cdot \mathbb{I}^{(\mathfrak{i}_{l+1})}\otimes P_j^{(\mathfrak{o}_l)}) = \text{Tr}(\mathcal{B}_l^{\mu_l}\cdot P_i^{(\mathfrak{i}_{l+1})}\otimes \mathbb{I}^{(\mathfrak{o}_l)}) = 0\:\forall \:i,j$. Unitaries are consequently fully orthogonal to the span of non-unital (where the maximally mixed state is mapped to something more pure) and trace-decreasing (stochastically applied) maps. Linear inversion reconstruction of the process tensor, then, omits these local expectation values. \par 

A restricted process tensor is not a model restriction as such, but an observational restriction. Its properties are fully consistent with the discussed facets of full process tensors, but the mapping is only valid across the span of observed data. This means, for example, that it will provide a recipe for complete control (under the restriction) which is fully inclusive of non-Markovianity. However, the actual strength of the memory can only be inferred rather than directly measured. For example, in the absence of measurement causal breaks, correlations between past and future measurement statistics cannot be established. Moreover, any measure relying on a full eigendecomposition of the state (such as quantum mutual information, for example) is similarly out of reach. The relevant analogy then is performing QST without measuring in all possible bases. The state will not be fully determined, but the information provided through Born's rule to predict the future will still be valid, so long as predictions are made within linear combinations of the measured bases.

Working within the constraints of NISQ devices, we need to account for $d_S^4 - 2d_S^2 + 2$ unitary operations at each time-step. For a qubit, this amounts to $N=10$ unitary gates per time step. However, any estimation procedure will come with sampling error, leading to both an incorrect and unphysical representation of the map. In a practical setting with finite sampling error, it is best to set up tomographic protocols without bias in the basis vectors~\cite{PhysRevLett.105.030406}. This is especially true in high-dimensional spaces where even small errors may become significantly magnified. In Ref.~\cite{White-NM-2020}, we found that a minimal single-qubit ($N_{\text{min}}=10$) unitary basis incurs substantial error in reconstructing the process. We thus resorted to an overcomplete basis of $N_{\text{oc}}=24$, leading to very high fidelity reconstruction for the process tensor, within shot-noise precision. However, this was very resource-demanding, since:
\begin{equation}
\mbox{number of experiments} \sim \mathcal{O}(N_{\text{oc}}^k)
\end{equation}
for $k$ time-steps. \par 
Our focus here is to reduce the requirements of PTT reconstruction while obtaining accurate estimates. To do so, in the next section we first integrate the maximum likelihood optimisation for PTT. This has the advantage that we will no longer need an overcomplete basis and thus reduce the base from 24 to 10. In order to do this, we ensure that the MLE accommodates the affine conditions of the process tensor; devise a numerical method to generate an approximately unbiased basis, which minimises reconstuction errors; and develop a projection method to ensure the physicality of the process tensor. Of course, the MLE alone is not enough if we also cannot reduce the exponent $k$. In Section~\ref{sec:MO}, we integrate the MLE tools with a more generic truncation method of Markov order. The idea here is to truncate small long-time memory to exponentially reduce the number of reconstruction circuits, while retaining high fidelities for the reconstruction. Along the way, we demonstrate that both the MLE and the Markov order methods are practically implementable by applying these ideas to superconducting quantum devices. \par

\subsection{SPAM errors in process tensor tomography}
It is key to know the limitations of any QCVV procedures, in particular how the estimate is affected by faulty control -- or SPAM errors. In the context of PTT, the usual notions of SPAM need to be broadened. For example, quantum channels are assumed to act on ``known" input states; when this assumption breaks it can be problematic for QPT. In contrast, the initial state is a marginal to the process tensor, and any error is naturally estimated. A measurement probe, on the other hand, is required to read out information. Errors which are insensitive to the POVM effect will absorb into the process. If they are common to all bases, the predictive capabilities will be unaffected. However, the estimated process tensor itself will then look slightly noisier. Accounting for either this or basis-specific noise can be straightforwardly achieved by using an estimate of the device POVM in the model. Estimates may be obtained with consistent detector tomography outputs from procedures such as gate set tomography (GST)~\cite{nielsen-gst}.

For PTT, it is the input quantum operations that are assumed to be known, replacing `state preparation' in QPT. In order of increasing consequence, violation of this assumption can occur in three ways: (i) with gate-independent error (such as decoherence), (ii) with gate-dependent coherent error, and finally, (iii) with significant $SE$ interaction during the finite time gate. 
Similar to the measurement case, PTT is insensitive to independent error for the purposes of control, since it does not change the linear relation between basis elements. The second consideration is more problematic because it can lead to an inconsistent characterisation. This can be resolved in two ways: by \emph{a priori} characterising the gates themselves through GST, or by using an overcomplete basis to average over the coherent error. Lastly, if a non-Markovian interaction occurs with coupling $\mathcal{O}(1/\tau_p)$ for control width $\tau_p$ then the process tensor model will break down. However, we expect this final possibility to be extremely rare for any functioning device -- but indeed could be circumvented with virtual gates, such as those described in Ref.~\cite{PhysRevA.96.022330}.

Because the input control must be high fidelity, we view PTT predominantly as a useful tool for quantum devices clean enough to be sensitive to non-Markovian dynamics. In this work, the demonstrated results focus on single qubit unitary gates, for which the error is $\mathcal{O}(10^{-4})$.
A future extension to PTT one might consider is a self-calibrating simultaneous estimation of both the process tensor and the input interventions, as in GST.
Though incorrect characterisations through gate errors are not implausible, single qubit gate errors for a typical NISQ device are already smaller than the expected $1/\sqrt{N_{\text{shots}}}$ sensitivity, and most of this error is represented by decoherence during the small finite pulse width. 
Moreover, the control aspect may be self-consistently checked by comparing predictions made from estimates of the process tensor with random gate sequences on the real device, offering certification to the characterisation. We explore this concept further in Section~\ref{se:ReFi}.


\section{Maximum likelihood process tensor tomography}
\label{sec:MLPTT}
\begin{figure}
    \centering
    \includegraphics[width=\linewidth]{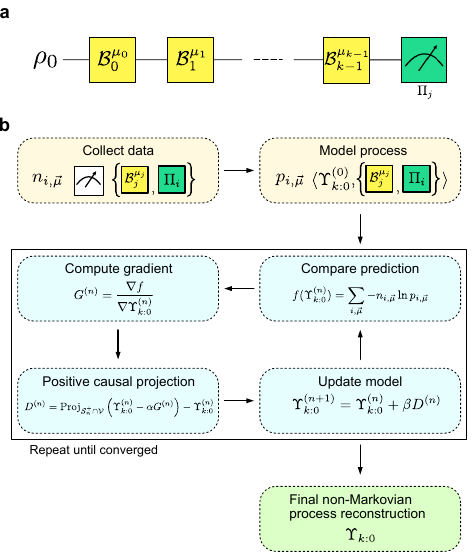}
    \caption{\textbf{a} Circuit structure for PTT. An arbitrary state is fed in, the experimenter acts with all combinations of different basis elements at different times, and a final measurement is recorded. \textbf{b} Logical flow of \texttt{pgdb} in the context of our MLE-PTT procedure. We maximise the likelihood of the model through iterative gradient descent and the physical projection of Section~\ref{ssec:projection} until some convergence condition is achieved. Full details of the algorithm is shown in Appendix~\ref{appendix:pt-maths}.}
    \label{fig:ml-flow}
\end{figure}

A major gap in the process tensor tomography toolkit is its lack of integration with standard tomography estimation tools like MLE, whose underlying principle is to find a physical model estimate that maximises the probability with the observed data. Due to the intricate affine conditions of causality, this integration is nontrivial in general. The complexity of the procedure further grows when applied to restricted process tensors, e.g. when control operations are restricted and/or when a finite Markov-order model is imposed. Our integration will naturally accommodate all of these variations.
We now close this gap and present a MLE construction for PTT, which helps to put this on the same footing as other tomographic techniques. The MLE procedure estimates the physical quantum map most consistent with the data, according to some desired measure, while respecting the constraints listed in Table~\ref{tab:tomography-summary}. The circuits required for tomography, depicted in Figure~\ref{fig:ml-flow}a, are the same as for linear inversion. That is, the scaling is the same. However, linear inversion typically requires an overcomplete basis to naturally average over inconsistencies. Meanwhile MLE treats the data such that a minimal tomographically complete basis suffices for accurate results.\par 
An estimate for the map is coupled with metric of goodness (the likelihood) which quantifies how consistent the map is with the data. The cost function is then minimised while enforcing the physicality of the map.
The stored data vector in PTT is the object $n_{i,\vec{\mu}}$, which contains the observed measurement probabilities for the $i$th effect of an IC-POVM, subject to a sequence of $k$ operations $\bigotimes_{j=0}^{k-1}\mathcal{B}_j^{\mu_j}$. As is typical in MLE tomography, this data is fit to a model for the process, $\Upsilon_{k:0}$, such that
\begin{equation}
\label{PT-vector-est}
    p_{i,\vec{\mu}} = \text{Tr}\left[(\Pi_i \otimes  \mathcal{B}_k^{\mu_{k-1}\text{T}}\otimes \cdots \otimes \mathcal{B}_0^{\mu_0\text{T}})\Upsilon_{k:0}\right].
\end{equation}
These predictions are then compared to the observed frequencies, $n_{i,\vec{\mu}}$. The `likelihood' of $\Upsilon_{k:0}$ subject to the data is given by $\mathcal{L} = \prod_{i,\vec{\mu}} (p_{i,\vec{\mu}})^{n_{i,\vec{\mu}}}$. The cost function of MLE algorithms is then the log-likelihood, i.e.,
\begin{equation}
\label{ML-cost}
    f(\Upsilon_{k:0}) = -\ln \mathcal{L} = \sum_{i,\vec{\mu}} -n_{i,\vec{\mu}}\ln p_{i,\vec{\mu}},
\end{equation}
whose minimisation is the maximiser of the likelihood. A key part of the appeal to MLE is that the cost function~\eqref{ML-cost} is convex. \par 
An extensive selection of different semi-definite program packages exist in the literature for the log-likelihood minimisation in QST and QPT under the appropriate constraints.
In our construction of the MLE-PTT procedure, we employ and adapt the algorithm from Ref.~\cite{QPT-projection}, used for QPT. This algorithm is termed `projected gradient descent with backtracking' (\texttt{pgdb}). We selected this both for its simplicity, and because it has been benchmarked as both faster and more accurate than other MLE-QPT algorithms. In this approach, the log-likelihood is minimised using conventional gradient descent, but at each iteration, a projection is made on the step direction to keep the map physical. The main steps are summarised in Figure~\ref{fig:ml-flow}b.

The relevant projection -- onto the intersection of the cone of CP maps with the affine space of TP channels  -- is performed using a procedure known as Dykstra's alternating projection algorithm~\cite{Birgin2005}.
We offer two key advancements here for PTT. First, we determine the exact affine space generated by causality conditions on process tensors, such that the physical constraints are mathematically elucidated. Then, we adapt and introduce a conic projection technique from optimisation literature in order to project onto the space of completely positive, causal processes \cite{conic-projection}. We find this projection method to far outperform Dykstra's alternating projection algorithm in the problem instances, a fact which may be of independent interest for QPT.
We detail each of these aspects in the following subsection. 
Finally, we benchmark the performance MLE-PTT on on superconducting quantum devices. These devices, as mentioned above, are limited to unitary control in the middle and a measurement at the end. This is insufficient to uniquely reconstruct the complete Choi state of a process. As such, our MLE procedure yields a operationally well-defined restricted process tensor which has been completed into a full process tensor. One might consider a `family' of process tensors generated by the intersection of positive causal matrices with the affine space of observed experimental data. This yields all possible process tensors whose restriction to unitary operations is consistent with the observed data. Therefore, further information about the full dynamics may be inferred even with limited data. We focus on the (non-unique) properties of the restricted process tensor family in Ref.~\cite{White2021}, and the performance of the restricted process tensor in the present work.



\subsection{Projecting onto the space of physical process tensors}
\label{ssec:projection}
Here, we describe in detail the physical conditions imposed on process tensors, as well as the approach used for projections onto the space of physical process tensors. Generally, this projection can be described as a problem of conic optimisation: finding the closest point lying on the intersection of a cone with an affine subspace. The affine constraints differ in each category: unit-trace for state tomography, trace-preservation for process tomography, and causality for the process tensor. 
Fundamentally, however, these techniques are applicable to all forms of quantum tomography, as shown in Table~\ref{tab:tomography-summary}.\par

Let $\Upsilon_{k:0}$ be the Choi form for a $k-$step process tensor (we will occasionally switch to $\Upsilon$ for brevity if the number of steps is not pertinent), and let $\kket{\Upsilon} := \text{vec}(\Upsilon)$, where we employ the row-vectorised convention \cite{gilchrist-vectorisation}. We discuss the mathematical demands of positivity and causality first, their individual projections, and then their simultaneous realisation. \par
Similar to a quantum channel, complete positivity of a process tensor is guaranteed by positivity of its Choi representation,
\begin{gather}
    \Upsilon_{k:0} \in \mathcal{S}_n^+,
\end{gather}
where $\mathcal{S}_n^+$ is the cone of $n\times n$ positive-semidefinite (PSD) matrices with complex entries. For $k$ time-steps, $n=2^{2k+1}$. The Euclidean projection is computed with a single eigendecomposition. Diagonalising $\Upsilon$ gives $\Upsilon = UDU^\dagger$ where $D = \text{diag}(\lambda_1, \lambda_2,\cdots,\lambda_n)$ is real. Then the projection onto $\mathcal{S}_n^+$ is:
\begin{gather}
    \label{PSD-proj}
    \text{Proj}_{\mathcal{S}_n^+}(\Upsilon) = U \text{diag}(\lambda^+_0,\cdots,\lambda^+_n)U^\dagger
\end{gather}
with $\lambda^+_j:=\max\{\lambda_j,0\}$. 

The Choi state must also obey causality, a generalisation of trace preservation. This is non-trivial to enforce, and ensures that future events should not influence past statistics. In the CJI picture of Figure~\ref{fig:PTT-explanation}b, there should be no correlations between the final input leg and the rest of the process when the final output leg is traced out. This is also a statement of containment of the process tensor: that the process over a subset of the total period is contained within the larger process tensor:
\begin{equation}
\label{containment}
    \text{Tr}_{\mathfrak{o}_k}\left[\Upsilon_{k':0}\right] = \mathbb{I}_{\mathfrak{i}_k}\otimes \Upsilon_{k'-1:0},
\end{equation}
iterated for all values of $k'$ from 1 to $k$. This statement is equivalent to causality in that the past stochastic process is unaffected by averaging over all future operations.\par 
We approach the problem of causality enforcement in the Pauli basis. Examining the Choi state, this condition places constraints on the values of these expectations. Let $\mathbf{P}:=\{\mathbb{I},X,Y,Z\}$ denote the single-qubit Pauli basis, $\mathbf{P}^n$ its $n-$qubit generalisation, and $\widetilde{\mathbf{P}}:=\{X,Y,Z\}$.
Focusing on the $\mathfrak{i}_{k'}$ subsystem, Equation~(\ref{containment}) can be enforced if all Pauli strings connecting the identity on the left subsystems with $\widetilde{\mathbf{P}}$ on the $\mathfrak{i}_{k'}$ subsystem have coefficients of zero. If this condition is imposed iteratively for all input legs on the process tensor, then Equation (\ref{containment}) will hold for all $k'$.
For example, in a two-step single qubit process (represented by a five-partite system), we have:
\begin{equation}
\label{containment-example}
\begin{split}
    &\langle \mathbb{I}_{\mathfrak{o}_2} P_{\mathfrak{i}_2}P_{\mathfrak{o}_1}P_{\mathfrak{i}_0}P_{\mathfrak{o}_0}\rangle = 0\: \forall \: P_{\mathfrak{i}_2}\in \widetilde{\mathbf{P}}; \: P_{\mathfrak{o}_1},P_{\mathfrak{i}_0},P_{\mathfrak{o}_0} \in \mathbf{P}\\ 
    &\langle \mathbb{I}_{\mathfrak{o}_2}\mathbb{I}_{\mathfrak{i}_2}\mathbb{I}_{\mathfrak{o}_1}P_{\mathfrak{i}_1}P_{\mathfrak{o}_0} \rangle = 0 \: \forall\: P_{\mathfrak{i}_0}\in\widetilde{\mathbf{P}}; P_{\mathfrak{o}_0}\in\mathbf{P}.
\end{split}
\end{equation}
\par 
A simple way to enforce this condition is with the projection of Pauli coefficients. 
In particular, let $\mathcal{P}$ be the elements of $\mathbf{P}^{2k+1}$ whose expectations must be zero from equations~\eqref{containment} and~\eqref{containment-example}.
We can write this as a single affine constraint in the matrix equation:
\begin{gather}
\label{affine-constraint}
    \begin{pmatrix} \langle\!\bra{\mathcal{P}_0} \\
    \langle\!\bra{\mathcal{P}_1}\\
    \vdots \\
    \langle\!\bra{\mathcal{P}_{m-2}}\\
    \langle\!\bra{\mathbb{I}}
    \end{pmatrix} \cdot \ket{\Upsilon}\!\rangle = \begin{pmatrix} 0\\0\\\vdots\\0\\d\end{pmatrix},
\end{gather}
where $d$ is the normalisation chosen for the Choi matrix (in this work, $d=1$). \par 

Letting this set the context for our discussion of the projection routine, consider a full rank constraint matrix $A\in\mathbb{C}^{m\times n^2}$, variable vector $\upsilon$, and fixed right hand side coefficient vector $b$.
Let $\mathcal{V}$ be the affine space:
\begin{gather}
    \mathcal{V} = \{ \upsilon\in \mathbb{C}^{n^2} | A\upsilon = b\}
\end{gather}
The projection onto $\mathcal{V}$ is given by
\begin{gather}
\label{affine-proj}
    \text{Proj}_{\mathcal{V}}(\upsilon_0) = \left[\mathbb{I} - A^\dagger (AA^\dagger)^{-1} A\right]\upsilon_0 + A^\dagger(AA^\dagger)^{-1}b.
\end{gather}

In general, however, a projection onto $\mathcal{S}_n^+$ and a projection onto $\mathcal{V}$ is not a projection onto $\mathcal{S}_n^+\cap \mathcal{V}$. The conic and affine constraints are difficult to simultaneously realise. 
One approach to this is to use Dykstra's alternating projection algorithm, as performed in \cite{QPT-projection} for quantum process tomography. This applies a select iterative sequence of~\eqref{affine-proj} and~\eqref{PSD-proj}. Although this method is straightforward and has guaranteed convergence, we find it unsuitable for larger-scale problems. For large gradient steps the convergence can take unreasonably many steps. More importantly, however, each step of the gradient descent requires many thousands of applications of \eqref{affine-proj}. Although much of this expression can be pre-computed, the complexity grows strictly with $n$, rather than the number of constraints. Moreover, the matrix inverse requirement can reduce much of the advantage of having a sparse $A$.\par
\begin{figure}[t!]
\includegraphics[width=0.85\linewidth]{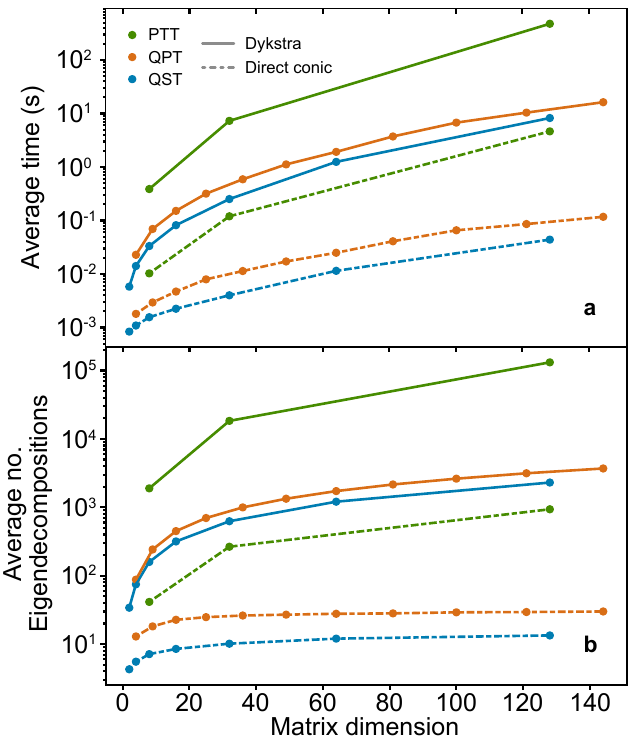}
\caption{A comparison between projection methods imposing physical conditions on 500 normally distributed random matrices. \textbf{a} Average time taken for a single projection for both Dykstra's alternating projection algorithm, and the direct conic projection. We compare conditions set by QST, QPT, and PTT. \textbf{b} Average number of eigendecompositions for each of the above. This dominates the runtime of each method.}
\label{fig:projection_comparison}
\end{figure}

  
In our MLE-PTT, instead of Dykstra's alternating projection algorithm, we integrate a variant of the technique introduced in~\cite{conic-projection} and discussed further in~\cite{conic-handbook}.  This method regularises the projection into a single unconstrained minimisation, such that only eigendecompositions and matrix-vector multiplications by $A$ are necessary, avoiding the need for~\eqref{affine-proj}. Note that to guarantee uniqueness of the projection as well as convergence of projected gradient descent in general, the closest physical process tensor at each step is found in terms of Euclidean distance. For further detail on this, see Refs.~\cite{hauswirth2016projected,conic-projection} Because the projection is the only component of \texttt{pgdb} that we change with respect to \cite{QPT-projection}, we explicitly walk through the steps in the following paragraph. We also benchmark this direct conic projection routine on normally distributed random matrices with respect to Dykstra's alternating projection algorithm for the case of QST, QPT, and PTT in Figure \ref{fig:projection_comparison}. The scaling for each method is similar (dominated by the cost of eigendecompositions), but the absolute savings are of two orders of magnitude. \par 
In each respective regime of tomography, the increased number of constraints increases the amount of time, on average, for the projection to complete. However, we find substantial improvements in both the run-time and in the number of eigendecomposition calls between the direct conic projection in comparison to Dykstra's. This is especially necessary for the fitting of process tensors where the difference between the two can be the difference between a run-time of days, or of fractions of a minute. We include QST here for completeness, however, note that the fixed projection of eigenvalues onto the canonical simplex with a single diagonalisation is more appropriate \cite{simplex-projection}.\par
We now explicitly step through the direct conic projection method. For a given $\upsilon_0$, we wish to find the closest (in Euclidean terms) $\upsilon \in \mathcal{S}_n^+ \cap \mathcal{V}$. That is, to compute 
\begin{gather}
    \label{primal}
    \argmin_{\upsilon \in \mathcal{S}_n^+ \cap \mathcal{V}} \|\upsilon - \upsilon_0\|^2.
\end{gather}
Note that when we talk about the vector $\upsilon$ being PSD, we mean that its matrix reshape is PSD. The dual approach introduces the Lagrangian, which is a function of the primal variable $\upsilon \in \mathcal{S}_n^+$ and dual variable $\lambda\in \mathbb{R}^m$ (for $m$ affine constraints): 
\begin{gather}
    \label{lagrangian}
    \mathcal{L}(\upsilon;\lambda) = \|\upsilon - \upsilon_0\|^2 - \lambda^\dagger (A\upsilon - b).
\end{gather}
Since $\upsilon$ is PSD, it is Hermitian, meaning that the matrix-vector product $A\upsilon$ is always real. This avoids the need for recasting the complex problem into real and imaginary pairs. \par 
The vector $\upsilon$ which minimises $\mathcal{L}$ for a given $\lambda$ provides a lower bound to the solution to the primal problem. We introduce the dual concave function
\begin{gather}
    \label{dual-func}
    \theta(\lambda) := \min_{\upsilon \in \mathcal{S}_n^+} \mathcal{L}(\upsilon;\lambda)
\end{gather}
whose maximum is exactly the solution to~\eqref{primal}. It is shown in Ref.~\cite{PSD-dual} that the minimum~\eqref{dual-func} is uniquely attained by $\upsilon(\lambda) = \text{Proj}_{\mathcal{S}_n^+}(\upsilon_0 + A^\dagger \lambda)$ \cite{PSD-dual}, and can hence be recast (up to a constant) as
\begin{gather}
    \label{dual-func-simplified}
    \theta(\lambda) = -\|\upsilon(\lambda)\|^2 + b^\dagger \lambda.
\end{gather}
It can further be shown that \eqref{dual-func-simplified} is differentiable on $\mathbb{R}^m$ with gradient
\begin{gather}
    \label{dual-func-grad}
    \nabla \theta(\lambda) = -A\upsilon(\lambda) + b.
\end{gather}
Thus, the solution to the projection problem \eqref{primal} becomes an unconstrained minimisation problem of \eqref{dual-func-simplified} with respect to $\lambda$, opening the door to a wealth of tested optimisation packages to be applied. The solution to the projection is then $\text{Proj}_{S_n^+}(\upsilon_0 + A^\dagger \lambda_{\text{min}})$. Specifically, in this work we select the L-BFGS algorithm to perform this minimisation, as we found it to give the fastest and most reliable solution \cite{lbfgs-alg}. Although we did not implement it here, it is also possible to compute the Clarke-generalised Jacobian of~\eqref{dual-func-grad}, allowing for exact second order optimisation techniques to be used \cite{projection-clarke-jacobian, spectral-derivatives}.
Note also that the difficulty of this minimisation is sensitive to the condition number of $A$. Thus, we find the best approach to be to always frame affine constraints in the Pauli basis to ensure a uniform spectrum.\par 
Another favourable reason to apply this conic projection method is in the arbitrary application of affine constraints. In Ref.~\cite{White2021}, we show how this can be used for searching (and thus bounding quantities of) manifolds of states consistent with an incomplete set of data. Introducing a feature matrix as part of the affine constraint with observed probabilities permits this exploration. Without the faster method, we found that this was infeasible to perform.\par
Using this modification of \texttt{pgdb} with an updated projection routine, we are able to implement PTT both in simulation and in real data. 
We have thus formalised MLE-PTT, and are now in position to benchmark the performance of MLE-PTT tomography against the linear inversion method and look at applications. The results, summarised in the following sections, suggest full characterisation of quantum non-Markovian dynamics in an object which is both mathematically minimal, and which obeys all of the physical constraints of a quantum stochastic process.

\subsection{Reconstruction Fidelity}
\label{se:ReFi}
In the linear inversion regime, the process tensor's action on basis sequences will result in the experimentally observed density matrices by construction. Note that by `experimentally observed', we mean the density matrices as reconstructed by QST. Since a process tensor is a linear operator, its action on linear combinations of basis sequences should be exactly the linear combinations of observed basis actions. This idea is expressed in Figure \ref{fig:trajectory}c: the $SE$ evolution between each operation is the same for all intermediate operations. In a linear combination, these arbitrarily strong dynamics are entirely accounted for. By tracing over the input space, we have the following relationship between the state conditioned on an arbitrary sequence of operations $\mathbf{A}_{k-1:0}$ and the states after each measured basis sequence:

\begin{equation}
    \label{eq:PT-response}
    \rho_k(\mathbf{A}_{k-1:0}) = \sum_{\vec{\mu}}\alpha^{\vec{\mu}} \rho_k(\mathbf{B}_{k-1:0}^{\vec{\mu}}),
\end{equation}
which is equivalent to Equation~\eqref{eq:PToutput}.
Hence, we may (in principle) determine the system's exact response to any sequence of operations in the presence of non-Markovian interaction.
We use this as the figure of merit for the quality of characterisation. That is, we compute the fidelity over random sequences of operation between the state predicted by the process tensor -- Equation~\eqref{eq:PToutput} -- and what is realised on the device. 
In~\cite{White-NM-2020}, we introduced this as the concept of reconstruction fidelity.

\begin{figure}
    \centering
    \includegraphics[width=\linewidth]{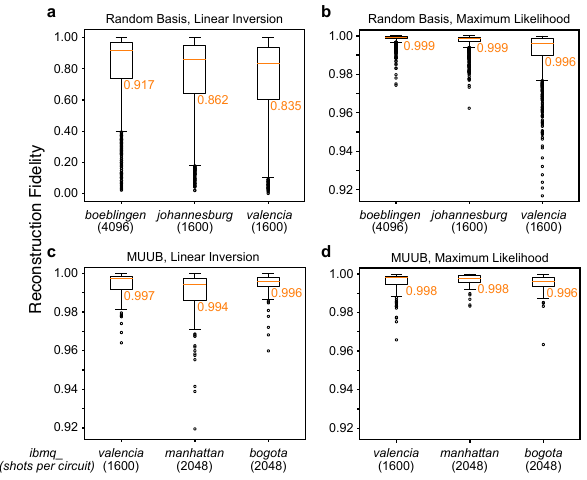}
    \caption{Reconstruction fidelity of various three-step process tensor procedures when using a minimal complete basis. Each data point represents a different randomly generated unitary sequence. The top and bottom of the boxplots are 75 and 25 percentiles, orange line is the median (figure also printed), whiskers are 1.5 times the interquartile range, and any remaining data points are outliers beyond this.} We compare reconstructions with \textbf{a} a randomly generated basis and linear inversion, \textbf{b} a randomly generated basis processed by maximum likelihood, \textbf{c} a MUUB processed with linear inversion, and \textbf{d} a MUUB processed with maximum likelihood.
    The results showcase high-fidelity, physical process tensors with minimal resources, in contrast with \cite{White-NM-2020} where the random basis data was taken.
    \label{fig:RF_boxplot}
\end{figure}

Formally, let the fidelity $F$ between two process tensors $\mathcal{T}^{(1)}_{k:0}$ and $\mathcal{T}^{(2)}_{k:0}$ for a given sequence of interventions $\mathbf{A}_{k-1:0}$ is given by
\begin{align}
&    F_{(1,2)}\left[\mathbf{A}_{k-1:0}\right]= F\left(\mathcal{T}^{(1)}_{k:0}[\mathbf{A}_{k-1:0}], \mathcal{T}^{(2)}_{k:0}[\mathbf{A}_{k-1:0}]\right),\notag \\
& \text{where}\quad F(\rho,\sigma) = \text{Tr}\left[\sqrt{\sqrt{\rho}\sigma\sqrt{\rho}}\right]^2.
\end{align}
Here, $\mathcal{T}^{(1)}$ is taken to be the reconstructed process and $\mathcal{T}^{(2)}$ to be the real process, i.e., the experimental outputs. Then, the average reconstruction fidelity is an estimate of
\begin{equation}
    \mathcal{F} := \int \text{d}\mathbf{A}_{k-1:0} \  F_{(1,2)}[\mathbf{A}_{k-1:0}].
\end{equation}
We can use this to estimate the quality of our reconstruction. The outputs to a real process is simply the state reconstruction conditioned on the sequence of gates $\{\mathcal{A}_0, \cdots, \mathcal{A}_{k-1}\}$. This integral can be estimated by performing sequences of randomly chosen operations and comparing the fidelity of the predictions made by $\mathcal{T}_{k:0}$ with the actual outcomes measured. We use this as a metric for the accuracy with which a process has been characterised. \par
As well as MLE-PTT, we also improve upon LI-PTT through particular choice of a basis. Sampling error typically averages out to zero on a circuit-by-circuit basis, however this noise can be both biased and amplified if certain regions of superoperator space are overrepresented -- i.e. if some elements of the basis overlap more than others. For this reason, we build upon the idea of a mutually unbiased basis (MUB) in conventional tomography. Unfortunately, mutually unbiased \textit{unitary} bases (MUUBs) do not exist in ten dimensions~\cite{Nasir2020}. However, in Appendix~\ref{appendix:muub} we numerically find the best approximation to a MUUB. \par
We probe each of these characteristics by looking at three-step PTT on IBM Quantum devices. Using combinations of basis choice, and LI/MLE post-processing, we compute the reconstruction fidelities for random sequences of unitaries. The boxplots showing these distributions are shown in Figure~\ref{fig:RF_boxplot}. Each data point constitutes a different sequence of random unitary gates. The number is then the fidelity between the density matrix predicted through action of the reconstruction process tensor on these unitary mappings, and the actual density matrix reconstructed through QST on the device after executing that specific unitary sequence. The use of MUUB alone finds substantial improvement in characterising the process. We continue to use these optimal parameter values as our unitary control basis for the remainder of this work. Reconstruction is improved further by MLE-PTT, in which we see not only an increase in median reconstruction fidelity, but there are far fewer outliers in the distribution. Compared to the random basis, linear inversion case, reconstruction fidelity increases greatly to within shot noise. This is essential for both validating process characterisation, and optimal control of the system.


\par 
In Ref.~\cite{White-NM-2020} (from which some of this device data is taken) much of the linear inversion characterisation noise was overcome with the use of an over-complete basis -- up to 24 unitaries. Since PTT is exponential in the size of the basis, the employment of maximum-likelihood methods as shown here can offer a significant reduction in experimental requirements. Thus, in addition to offering an algorithm imposing physicality constraints, we show how to make the technique more practical to implement. The scaling of MLE is therefore
\begin{equation}
\label{eq:mle-scaling}
    \mbox{number of experiments} \sim \mathcal{O}(N_{\text{mle}}^k),
\end{equation}
where $N_{\text{mle}}$ can now be 10 regardless of the specific basis choice.

As well as comparing processing methods, we also juxtapose our approximate MUUB with the minimal randomly-chosen unitary basis, where $N_{\text{muub}} = 10$. This, too, sees a drastic improvement of the method: though it is not guaranteed to produce a physical process tensor, we see that much higher quality predictions are possible without any additional effort in the linear inversion approach.

\section{Conditional Markov Order}
\label{sec:MO}
Even in the classical case, the price of characterising the joint statistics of a stochastic process in full generality is exponentially high. Often, however, this is unnecessary in practice as physical processes are sparse. This is because the memory must be carried by another physical system, whose size then bounds then the size of the memory. Often in practice, the \emph{necessary} complexity of a process characterisation only grows modestly with the size of its memory if, after a certain amount of time, the history and the future are independent from one another. 

In such cases, the joint statistics are no longer required between those points in time. This motivates the idea of Markov order~\cite{Rosvall2014, pollock-tomographic-equations}: the number of previous time steps in the process which are relevant to the present. Concretely, in classical theory, a stochastic process is described by the joint probability distribution of a sequence of events $\mathbb{P}(x_k, x_{k-1}, \dots, x_0)$, occurring at times $\{t_k, t_{k-1}, \dots, t_0\}$. A process with Markov order $\ell$ then conditionally separates the future $F_j=\{t_{j+\ell+1},\dots,t_{k}\}$ from the past $P_j=\{t_0,\dots,t_{j-1}\}$ given the knowledge of the state in the memory block $M_j=\{t_j, \dots, t_{j+\ell}\}$. That is, the above distribution takes the form
\begin{gather}\label{cl:cmo}
    \mathbb{P}(x_k, \dots, x_0) = 
    \sum_{M_j}
    \mathbb{P}(P_j|M_j) \
    \mathbb{P}(M_j) \
    \mathbb{P}(F_j|M_j).
\end{gather}
That is, in order know the probability of an event at a given time, we only need to look the past $\ell$ events. Anything beyond that will not affect the future. In the case where memory is indeed infinite but decays in time, we can turn the last equation into an approximate statement. Importantly, the complexity of the whole process goes as $d^k$, where $k$ is not bounded. While the complexity of a process with Markov order $\ell$ goes as $d^\ell$ with a fixed $\ell$. That is, the distribution in Eq.~\eqref{cl:cmo} is fully determine dy knowing $\mathbb{P}(M_j)$.

It is possible to extend the notion of Markov order to quantum stochastic processes by replacing $\mathbb{P}$ with its quantum counterpart $\Upsilon$, roughly speaking. We now apply this idea to MLE-PTT, and derive a resource-efficient way to characterise even processes with very large numbers of steps. Concretely, integrating MLE-PTT with a Markov order of $\ell$ would reduce the exponential scaling in Equation~\eqref{eq:mle-scaling} to
\begin{equation}
\label{eq:mo-mle-scaling}
    \mbox{number of experiments} \sim \mathcal{O}(k \cdot N_{\text{mle}}^\ell).
\end{equation}
To achieve this, we build upon the ideas established in Ref.~\cite{taranto1} and subsequently realised in Ref.~\cite{PhysRevLett.126.230401}.

A summary of our approach is to divide a $k-$step process up into a number of smaller, overlapping process tensors. These smaller process tensors are designed to account for a truncated number of past-time correlations. We then use MLE estimation to fit each of the memory process tensors according to a Markov order model chosen by the experimenter. The finite Markov order process tensor fitting, adaptive memory blocking, and action across sequences that we introduce here are all novel features of quantum Markov order, and offers a method by which non-Markovian behaviour on NISQ devices can be feasibly characterised and controlled. We note in passing that without MLE-PTT a Markov order integration would not be possible when working with restricted process tensors. This is because in this regime the partial traces of the process tensor are not well-defined, which makes it difficult to split the process into parts.\par 

\subsection{Structure of quantum Markov order}
In the quantum realm, the matter of Markov order is more nuanced than for classical processes. A quantum stochastic process has either Markov order one (the output at any leg is affected only by the previous input) or infinite Markov order (the memory persists indefinitely). That is to say, a Choi state may only be written as a product state, or there will exist correlations between all points in time (though, saying nothing about the strength of these correlations). For practical purposes however -- especially in the context of quantum computing -- there exists the useful concept of \emph{conditional} Markov order~\cite{taranto1}. 

To properly explain this statement about conditioning, we first re-emphasise that a process tensor represents a quantum stochastic process. As a consequence, if we gain extra information about the past -- for example, what operation was applied by the experimenter -- then we may update our description of the process when conditioned on that choice of operation. This is akin to the quantum state picture: if a measurement is made on one qubit as part of a many-body system, then the remaining state can be updated based on the outcome of the measurement. For a dynamical process, this intervention can, in full generality, be a quantum instrument. A quantum instrument is a set of completely positive, trace non-increasing maps whose sum is a CPTP map. We employ this terminology, but for readers unfamiliar with the object, it suffices to interpret this as any way an experimenter might manipulate a system, including unitaries, measurements, and re-preparations. Interested readers may consult Ref.~\cite{wilde_2013}. The conditional state of the process may then exhibit past-future independence.\par
We now introduce Markov order for a quantum stochastic process, as well as the related notion of instrument-specific conditional Markov order. We then make clear that processes conditioned on generic operations may only exhibit \emph{approximate} conditional Markov order. Finally, we explicitly walk through our calculations of tomographically reconstructing processes with an approximate conditional Markov order ansatz.\par 
To begin, we describe a $k-$step process with Markov order $\ell$. When $\ell=1$, the only relevant information to the CPTP map $\hat{\mathcal{E}}_i$ is the state mapped at the output of the $\hat{\mathcal{E}}_{i-1}$ step, described by the $\mathfrak{o}_{i-1}$ leg of the process tensor. Consequently, there is no context to the previous gates. The choice of operation $\mathcal{A}_{i-1}$ is only relevant insofar as determining the output state for time $(i-1)$. This constitutes a Markovian process, and the dynamics are CP-divisible. Otherwise, it is non-Markovian with $\ell = \infty$~\cite{taranto1}. Intuitively one may think of this as the statement that there is no way to consistently write a generic quantum state with strictly limited correlations -- for example where each subsystem might have nearest-neighbour correlations but zero correlations with any subsystem outside of this.  \par
Although finite $\ell>1$ Markov order is well-defined for classical stochastic processes, where there is only one basis, there is no generic way to write a quantum state with correlations persisting to the last $\ell$ subsystems. However, future and past statistics may be independent of one another for quantum stochastic processes when conditioned on the choice of an intermediate instrument. This notion of conditional Markov order may be described as follows. Consider a process tensor $\Upsilon_{k:0}$, which we denote as $\Upsilon_{FMP}$ with the groupings for the past, the memory, and the future, respectively:
\begin{equation}
\begin{split}
P_j &=\{t_0,\cdots,t_{j-1}\},\\
M_j &=\{t_{j}, \cdots, t_{j+\ell}\},\\
F_j &= \{t_{j+\ell+1}, \cdots, t_k\}.
\end{split}
\end{equation}

Let a sequence of operations $\mathbf{C}_{j+\ell:j}$, with $k > j+\ell$ act on the memory block of the process. For the moment, while discussing the basic properties of Markov order in quantum processes, we omit the $j$, and $\ell$. However, these will become important when propagating processes with a Markov order assumption. Thus $\mathbf{C}_{j+\ell:j}$ will be expressed as $\mathbf{C}_{M}$ henceforth. Let $\{\mathbf{B}_{M}^{\vec{\mu}}\}$ be a minimal IC basis for these times, which includes $\mathbf{C}_{M}$, and let $\{\mathbf{\Delta}_{M}^{\vec{\mu}}\}$ be its dual set. The conditional process is given by
\begin{equation}
    \Upsilon_{FP}^{(\mathbf{C}_{M})} = \text{Tr}_{M}\left[\Upsilon_{FMP} \mathbf{C}_{M}^\text{T}\right],
\end{equation}
where $\Upsilon_X$ is the process tensor across the legs given by the set(s) $X$. If the past and the future are independent in this conditional process, then it can be written as
\begin{equation}
\label{eq:conditional-independence}
    \Upsilon_{FP}^{(\mathbf{C}_{M})} = \Upsilon_{F}^{(\mathbf{C}_{M})} \otimes \Upsilon_{P}^{(\mathbf{C}_{M})},
\end{equation}
where
\begin{equation}
        \Upsilon_{X}^{(\mathbf{C}_{M})} = \text{Tr}_{M\overline{X}}\left[\Upsilon_{FMP} \mathbf{C}_{M}^\text{T}\right] \quad X\in\{F,P\}
\end{equation}
Note that the condensed language used here is identical to the description used in Equation~\eqref{eq:PToutput}. If Equation \eqref{eq:conditional-independence} holds for all elements of $\{\mathbf{B}_{M}^{\vec{\mu}}\}$, then the process, by construction, can be written as
\begin{equation}
\label{eq:cmo-pt}
    \Upsilon_{FMP} = \sum_{\vec{\mu}} \Upsilon_{F}^{(\mathbf{B}_{M}^{\vec{\mu}})} \otimes \mathbf{\Delta}_{M}^{\vec{\mu}} \otimes \Upsilon_{P}^{(\mathbf{B}_{M}^{\vec{\mu}})}.
\end{equation}

A fact of practical importance is that for all sequences of operations $\mathbf{A}_{M}\not\in \{\mathbf{B}_{M}^{\vec{\mu}}\}$, Equation~\eqref{eq:conditional-independence} cannot hold. This is because $\{\mathbf{B}_{M}^{\vec{\mu}}\}$ is informationally complete, meaning that some operation sequence from outside the set can be expressed as a linear combination
\begin{equation}
    \mathbf{A}_{M} = \sum_{\vec{\nu}} \alpha_{\vec{\nu}}\mathbf{B}_{M}^{\vec{\nu}}.
\end{equation}
Contracting this operation into the process then yields
\begin{equation}
    \begin{split}
        \Upsilon_{FMP}^{(\mathbf{A}_{M})} &= \text{Tr}_{M}\left[\Upsilon_{FMP}  \mathbf{A}_{M}^\text{T} \right]\\
        &= \text{Tr}_{M}\left[\Upsilon_{FMP} \left(\sum_{\vec{\nu}} \alpha_{\vec{\nu}}\mathbf{B}_{M}^{\vec{\nu}\text{T}}\right)\right]\\
        &= \sum_{\vec{\nu}} \alpha_{\vec{\nu}}\Upsilon_{F}^{(\mathbf{B}_{M}^{\vec{\nu}})}\otimes \Upsilon_{P}^{(\mathbf{B}_{M}^{\vec{\nu}})},
    \end{split}
\end{equation}
which is no longer a product state, and thus the future and the past are \emph{separable}, but not completely uncorrelated. \par
The complexity of characterising a process grows exponentially in $\ell$; we would prefer to drop the instrument-specific component, and employ a generic conditional Markov order model. Explicitly, in this model, we truncate all conditional future-past correlations, treating the conditional state as a product. i.e.,
\begin{equation}
\label{eq:cmo-approx}
    \Upsilon_{FMP}^{(\mathbf{A}_{M})} \approx \Upsilon_{F}^{(\mathbf{A}_{M})} \otimes \Upsilon_{P}^{(\mathbf{A}_{M})}.
\end{equation}
The cost, or approximation, in doing so will be determined by the actual memory strength of the process over different times. One meaningful measure of this is the quantum mutual information (QMI) of the conditional state, the LHS of Equation~\eqref{eq:cmo-approx}. Of course, this information is inaccessible in our truncated characterisation. Instead, we continue to use the reconstruction fidelity, and experimentally estimate this model error in the ability of each Markov order to predict the behaviour of actual sequences of random unitaries~\cite{taranto3}.

\begin{figure*}[ht!]
\includegraphics[width=0.85\linewidth]{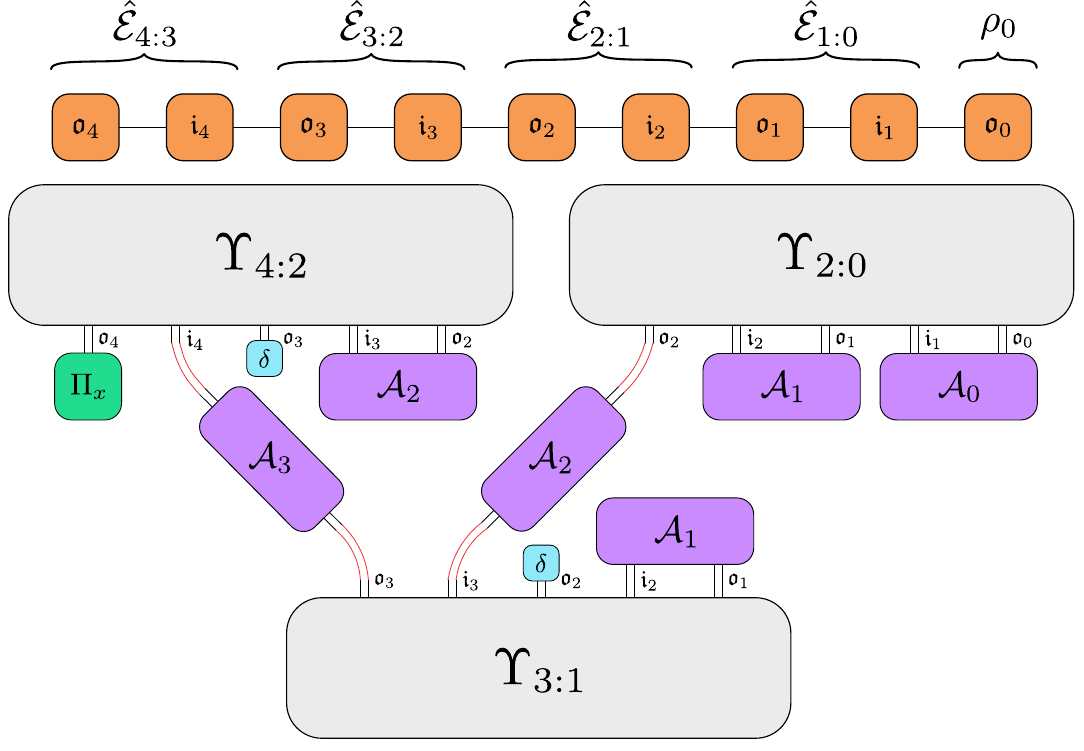}
\caption{A contraction strategy for mapping multi-time gate sequences using a conditional Markov order ansatz. Here, we show how a four-step process can be modelled by two-step memory process tensors. The memory process tensors are stitched together by first contracting the relevant operations to their times to account for correlations. After tracing over the state output (denoted by $\delta$), the conditionally independent parts can be treated as tensor products, and thus stitched together with the latest common operation, where the output of the earlier process tensor is mapped to the input of the later one.}
\label{fig:stitching-cmo}
\end{figure*}
\subsection{Stitching together finite Markov order processes}\label{se:MLPTTMO}
We turn now to our work in extending the concept of conditional quantum Markov order to a quantum circuit context, and tomographic characterisation. Here, we are interested not only in dividing up the process into a single past, memory, and future, but to do this for all times in the process. Then, for each time, the previous $\ell - 1$ operations are taken into account. This is performed by iterating through the above computation: at each time step, dividing the circuit up into past, memory and future. Correlations due to the memory are taken care of via contraction of the relevant operations, leaving the future and the past conditionally independent. This requires a tomographically reconstructed process tensor for the relevant memory steps. These memory process tensors are then stitched together by the overlapping operation's map of the earlier output state. 
Our goal is predict $\rho_k(\mathbf{A}_{k-1:0})$ by making use of the Markov order structure. At the first step, we have no past $P_0 = \{ \emptyset \}$, and the memory is given by the first $\ell$ operations. That is, we have to construct the full process tensor $\Upsilon_{M_0} = \Upsilon_{\ell:0}$, which contains all of the conditional states $\rho_j(\mathbf{A}_{j-1:0})$ in $M_0$, i.e. $j \le \ell$. To go beyond time $t_\ell$, we need the conditional state  $\rho_\ell(\mathbf{A}_{\ell-1:0})$ given by Equation~\eqref{eq:PToutput} (see also the first line of Equation~\eqref{conditional-PTs}): contracting $\Upsilon_{M_0}$ with $\mathbf{A}_{\ell-1:0}$. Importantly, this is the state propagated along with the sequences of operations. For this reason, the first memory process tensor is the only one for which the output state is not traced over, since we are not tracing over any alternative pasts. This means that it contracts one more local operation than the remainder. 

To get state $\rho_{\ell+1}(\mathbf{A}_{\ell:0})$, we move one step forward
with $P_1, \ M_1, \ F_{1}$. The relevant information is stored in $\Upsilon_{M_0}$ and $\Upsilon_{M_1} = \Upsilon_{\ell+1:1}$. For this process (and for all intermediate blocks in the process), there are three considerations: first, we must account for the memory through its action on the sequence $\mathbf{A}_{\ell-1:1}$ on $\Upsilon_{M_1}$ and trace over its output index at time $t_\ell$ (see the second line of Equation~\eqref{conditional-PTs}) since this state corresponds to a different, fixed past. Finally, the operation $\mathcal{A}_{\ell}$ connects $\Upsilon_{M_0}$ and $\Upsilon_{M_1}$ by mapping the state $\rho_{\ell}(\mathbf{A}_{\ell-1:0})$ to time $t_{\ell+1}$ since, as per our conditional Markov order assumption, once $M_1$ is accounted for, $F_1$ and $P_1$ are independent, i.e. their dynamics can be treated as a tensor product. See Figure~\ref{fig:stitching-cmo} for a graphical tensor network depiction. Note the distinction between here and Figure~\ref{fig:PTT-explanation}d. For a full four step process tensor, estimating a single expectation value involves contracting a tensor of matrix size $512\times 512$. With $\ell=2$ conditional Markov order, however, these requires only three tensors with matrix size $32\times 32$.\par 
Following this recipe, we proceed forward in single steps, generating blocks of $P_{j}, \ M_{j}, \ F_{j}$ until we reach time $t_k$ at which point the final state may be read out. For clarity, the sequence of conditional memory process tensor states is given by:
\begin{equation}
\label{conditional-PTs}
    \begin{split}
        &\rho_\ell(\mathbf{A}_{\ell-1:0}) = \text{Tr}_{\overline{\mathfrak{o}}_{\ell}}\left[\Upsilon_{M_0} \bigotimes_{i=0}^{\ell-1}\mathcal{A}_i^\text{T}\right],\\
        &\Upsilon_{j}^{(\mathbf{A}_{j-2:j-\ell})} := \text{Tr}_{\overline{j}}\left[\Upsilon_{M_{j-\ell}} \bigotimes_{i=j-\ell}^{j-2}\mathcal{A}_i^\text{T} \right].
    \end{split}
\end{equation}
The conditional state $\rho_\ell(\mathbf{A}_{\ell-1:0})$ has the free index $\mathfrak{o}_\ell$, corresponding to its output state. All others $\Upsilon_{j}^{(\mathbf{A}_{j-2:j-\ell})}$ have free indices $\mathfrak{i}_j$ and $\mathfrak{o}_j$ which, respectively, are contracted with the output and input legs of the operation $\mathcal{A}_j$ and $\mathcal{A}_{j+1}$, respectively. These are the operations which stitch together the different conditional memory process tensors, where the conditional independence means that the state can be mapped as though it were a tensor product. Finally, the last output leg $\mathfrak{o}_k$ is read out by some POVM. \par 
We condense this $k$-step Markov order $\ell$ process in the object $\mathbf{\Upsilon}_{k:0}^\ell:= \{\Upsilon_{M_{k-\ell}}, \Upsilon_{M_{k-\ell-1}},\cdots,\Upsilon_{M_0}\}$. That is, the final state is defined by the collective action of each $\Upsilon_{M_j}$ as

\begin{align}
    \label{CMO-action}
        &\rho_k(\mathbf{A}_{k-1:0}) \approx \mathbf{\Upsilon}_{k:0}^\ell \ast \mathbf{A}_{k-1:0} \\ \notag
        & \qquad := \text{Tr}_{\overline{\mathfrak{o}}_k}\left[ \Upsilon_{k}^{(\mathbf{A}_{k-2:k-\ell})} \left(\bigotimes_{j=\ell}^{k-1}\Upsilon_{j}^{(\mathbf{A}_{j-2:j-\ell})} \mathcal{\mathcal{A}}_j^\text{T}\right)\right].
\end{align}
Note that since the same control operation may contract into multiple different memory process tensors, this action is no longer linear in $\mathbf{A}_{k-1:0}$.\par 
Recalling our earlier depiction of the process tensor in Figure~\ref{fig:PTT-explanation}, the dynamics can be described as a collection of correlated CPTP maps $\{\hat{\mathcal{E}}_{j:j-1}\}$. In the CJI picture, past operations are equivalently seen as measurements on these earlier states. Thus, the $\Upsilon_{j}^{(\mathbf{A}_{j-2:j-\ell})}$ are exactly the conditional memory states $\hat{\mathcal{E}}_{j:j-1}^{(\mathbf{A}_{j-2:j-\ell})}$. With correlations accounted for, they can be treated locally in time. Any process may be written exactly as a sequence of conditional CPTP maps, but in full generality they depend on the whole past. Here, they only depend on the memory. The difference in complexity of characterisation is $\mathcal{O}(N^k)$ vs. $\mathcal{O}(N^\ell)$.
Putting it all together we have an equivalent form of Equation~\eqref{CMO-action}
\begin{align}\notag
    \rho_k(\mathbf{A}_{k-1:0}) \approx & \mathcal{E}_{k:k-1}^{(\mathbf{A}_{k-2:k-\ell})} \circ \mathcal{A}_{k-1} \circ \cdots \circ \mathcal{E}_{\ell+2:\ell+1}^{(\mathbf{A}_{\ell:2})} \\
    & \circ \mathcal{A}_{\ell+1} \circ  \notag
    \mathcal{E}_{\ell+1:\ell}^{(\mathbf{A}_{\ell-1:1})} \circ \mathcal{A}_{\ell} [\rho_{\ell}(\mathbf{A}_{\ell-1:0})] \\
    \mbox{with} \quad &\rho_{\ell}(\mathbf{A}_{\ell-1:0}) = \mathcal{T}_{M_0}[\mathbf{A}_{\ell-1:0}].
\end{align}

To summarise, the process with a conditional Markov order $\ell$ ansatz $\mathbf{\Upsilon}_{k:0}^{\ell}$ is represented by a collection of memory process tensors $\left\{\Upsilon_{k:k-\ell}, \Upsilon_{k-1:k-\ell-1}, \cdots, \Upsilon_{\ell+1:1}, \Upsilon_{\ell:0}\right\}$. As discussed, it cannot be represented generically by a quantum state, but this collection of memory process tensors defines its action on a sequence of $k$ operations.
The contraction strategy for a series of control operations (with $k=4$ and $\ell=2$) is shown in Figure~\ref{fig:stitching-cmo}. In short:
\begin{enumerate}
    \item Contract the first $\ell$ operations into $\Upsilon_{\ell:0}$, producing the output state at time $t_\ell$,
    \item Contract operations $2$ to $\ell$ into $\Upsilon_{\ell+1:1}$,
    \item Trace over the output index of $\Upsilon_{\ell+1:1}$ at time $t_\ell$ (since this is not representative of the actual state of the system subject to all operations),
    \item Taking the $(\ell+1)$th operation to be conditionally independent of the first, this can be applied across the tensor product of the two process tensors into the indices for the output state at time $t_\ell$ and the input for time $t_{\ell+1}$,
    \item Repeat this pattern for the next $k - \ell - 2$ process tensors,
    \item Read out the final state at time $t_k$.
\end{enumerate}
The full details of this computation can be found in Appendix \ref{appendix:mo}, where we fully describe an efficient tensor network contraction for the action of $\mathbf{\Upsilon}_{k:0}^\ell$.

\subsection{Circuits for $\Upsilon_{M_j}$}
\begin{figure}[ht]
    \centering
    \includegraphics[width=\linewidth]{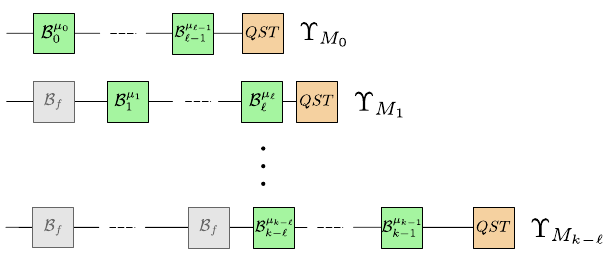}
    \caption{Circuits to construct a process tensor with conditional Markov order $\ell$. For a $k$-step process, $\mathbf{\Upsilon}_{k:0}^{\ell}$, there are $k-\ell+1$ memory process tensors that need constructing -- each circuit represents the estimation of each of these, by varying all combinations of all indices from $i_0$ to $i_{\ell-1}$. In the other gate positions, a fixed operation $\mathcal{B}_f$ is applied.}
    \label{fig:cmo_circs}
\end{figure}

With a framework established for constructing and operating a finite Markov order ansatz, we now explicitly detail how to tomographically reconstruct this model on a real device. This procedure does not deviate significantly from Section \ref{sec:MLPTT}. In order to estimate $\mathbf{\Upsilon}_{k:0}^\ell$, we must estimate each of the memory block process tensors. Recall that each $\Upsilon_{M_j} := \Upsilon_{\ell+j:j}$ is equivalent to $\Upsilon_{\ell+j:0}$ with the first $j$ times projected out onto some series of interventions. In order to experimentally reconstruct each $\Upsilon_{M_j}$, then, it suffices to fix the first $j$ operations in the circuit, and then perform a complete basis of operations in each position from $t_j$ to $t_{\ell+j}$ and estimate the associated $\ell$-step process tensor. The circuits required for each of these are illustrated in Figure~\ref{fig:cmo_circs}, with the fixed operation labelled $\mathcal{B}_f$. As well as sufficiently describing a Markov order $\ell$ model, these circuits contain all of the information required for any lower-order Markov model if it is a full process tensor. Under the unitary-only restriction, there will be a small number of extra experiments required for any smaller memory blocks terminating earlier than $t_{\ell}$. We note here also that the maximum likelihood procedure of Section~\ref{sec:MLPTT} is necessary for conditional Markov order models if the set of instruments is restricted to the unitaries. This is because Equation~\eqref{CMO-action} requires local partial traces, but unitary gates are equivalent to entangled measurements in the Choi picture. Hence, if a linear inversion restricted process tensor is constructed, the partial traces will not be well defined.
\par 
In the action of $\mathbf{\Upsilon}_{k:0}^\ell$, the state generated by each $\Upsilon_{M_j}$ is traced over for all $j>0$. Consequently, the fixed operations that precede $M_j$ in the reconstruction circuits should, in principle, not affect the final outcome $\rho_k$. Since, however, the CMO ansatz is an approximation to the true dynamics, then the fixed past operations will, in practice, affect this approximation. 
In the generic case, there is no reason to suspect any operation will put forth a better or worse approximation, hence we arbitrarily set this operation to be the first element of the basis set each time. This choice may require closer attention in practical situations.

With this, we have described how to adaptively characterise a process with quantum and classical requirements only as large as the complexity of the noise (or, up to the error the experimenter is willing to tolerate). Since a process characterisation may be verified through the reconstruction fidelity, the best approach to this is to progressively build up and verify a more complex model until the desired precision has been reached. This will depend on the intended applications of the characterisation. We believe that this is the first procedure to methodically characterise finite Markov order quantum processes. We expect this to be greatly useful in stemming the effects of both correlated and uncorrelated noise on NISQ devices, where the open dynamics is already clean enough so as to be mostly -- but not strongly -- non-Markovian. \par

\subsection{Estimating memory build-up}

\begin{figure}
    \centering
    \includegraphics[width=0.92\linewidth]{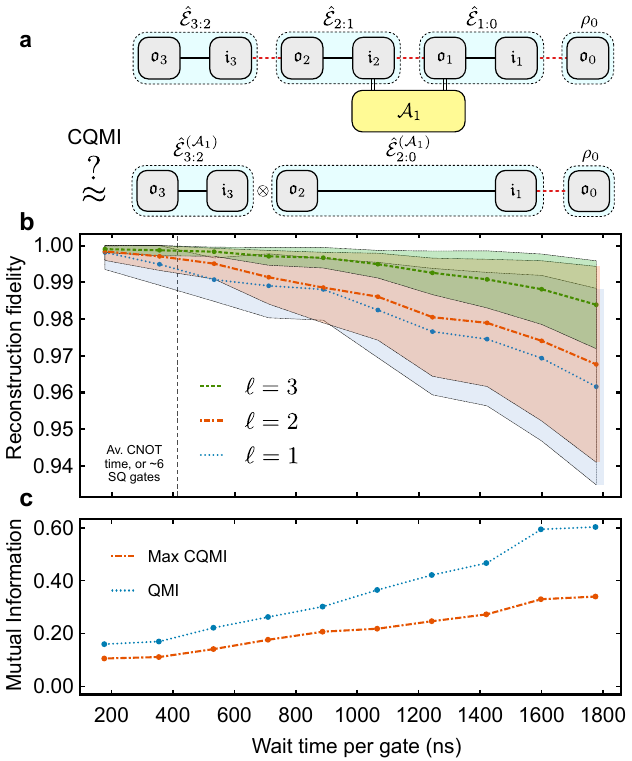}
    \caption{Build-up of temporal correlations with increasing interaction time. \textbf{a} The CQMI quantifies the error in truncating correlations beyond $\ell=2$ when conditioned on intermediate operations. \textbf{b} A four step process is considered with increasing wait time after each gate. Using Markov orders $\ell=1$, $\ell=2$, and $\ell=3$, we quantify how each model reconstructs 100 random unitary sequences. Points indicate average fidelity, shaded regions indicate standard deviation. \textbf{c} QMI and CQMI both increase with interaction time, respectively bounding the breakdown of $\ell=1$ and $\ell=2$ models.}
    \label{fig:cmo_reconstruction}
\end{figure}

We can combine some of the ideas introduced to give a more fine-grained measure of non-Markovian memory on quantum devices, both in terms of its length and its strength. Specifically, these simplified models of the process can be employed either to streamline control of the quantum system, or they may be used as a diagnostic tool by observing how well different restrictions describe the dynamics. This also validates our method of reconstructing processes with conditional Markov order in an efficient way.
To estimate non-Markovianity we construct increasingly complex models by accounting for increasing memory, and quantify how well they describe observed device dynamics under the measure of reconstruction fidelity. This measure is computationally convenient, scaling linearly in time-steps; has an immediately available interpretation; and may be performed up to acceptable approximation, or where costs become prohibitive. The breakdown of conditional Markov order is bounded by the maximum conditional quantum mutual information (CQMI) as described in Ref.~\cite{taranto3}. CQMI is taken for a three-step process tensor to be
\begin{equation}
    \max_{\mathcal{A}_1} S[\Upsilon_{3:0}^{(\mathcal{A}_1)} || \hat{\mathcal{E}}_{3:2}^{(\mathcal{A}_1)}\otimes \Upsilon_{2:0}^{(\mathcal{A}_1)}],
\end{equation}
where $S[\rho|| \sigma] := \text{Tr}[\rho(\log\rho - \log\sigma)]$ is the von Neumann relative entropy. The conditional Markov order approximation is illustrated in Figure~\ref{fig:cmo_reconstruction}a.

We performed this procedure as an example on \emph{ibmq\_guadalupe} to observe non-Markovianity as a function of time. For a four step process, we constructed process tensors $\mathbf{\Upsilon}_{4:0}^\ell$ for $\ell\in \{1,2,3\}$. With a $\ket{+}$ state neighbour, the duration of each step was varied across ten different times, ranging from 180 ns up to 1800 ns. For each value of $t$, we executed 100 sequences of random unitaries $\{\mathbf{U}_{3:0}^i\}$ followed by state reconstruction. Then the action of $\mathbf{\Upsilon}_{4:0}^{\ell}$ on $ \mathbf{U}_{3:0}^i$ is used to predict the resulting states. The fidelity between predicted state and actual state is computed, and the distribution for each data point shown in Figure~\ref{fig:cmo_reconstruction}. We see here the build-up of memory effects; the timescales constitute relatively short-depth effective circuits, meaning these temporal correlations are likely to accumulate across practical circuits. Interestingly, the $\ell=3$ model performs significantly better than the other two, whereas $\ell=2$ is only marginally better at predicting the dynamics than $\ell=1$ \footnote{$\ell = 2$ and $\ell=1$ models can be constructed from subsets of the $\ell=3$ data, however on their own they minimally require $3\times 300 = 900$ and $4\times 30 = 120$ circuits per time, respectively}. This suggests that most of the memory effects in these devices are higher order -- they persist across multiple times. This observation is substantiated by computing the QMI from Equation~\eqref{eq:qmi}, as well as finding the operation which maximises the conditional QMI for the three step process tensors. The ability to compute these measures comes from our compressed sensing approach to PTT, detailed in Ref.~\cite{White2021}. The increase of both of these non-Markovian measures is shown in Figure~\ref{fig:cmo_reconstruction}c. Because of the cumulative build-up, for long-time dynamical processes in real situations, mitigating the effects of these correlations would require either decoupling early, or fine-graining the process into many more time steps. These memory effects manifest themselves over a time frame of only a few CNOT gates, indicating that non-Markovian dynamics are likely a significant class of noise in regular circuits. 

Markovianity breakdown has been previously quantified in terms of model violation in gate sets, or in the loss of CP divisibility~\cite{RBK2017,white-POST,PhysRevA.83.052128}. However these approaches only coarsely diagnose temporal correlations, and are not generic to the process. We have presented a systematic framework by which different levels of finite conditional Markov order may be tested on quantum devices with both a rigorous foundation and practical interpretation. 
\section{Applications of multi-time characterisation}
\label{sec:applications}
The characterisation given in PTT can be useful for qualitatively different applications. Broadly speaking, these applications fall into two different camps: non-Markovian diagnostics, and non-Markovian optimal control. In the former, conventional many-body tools are applied to the Choi state to probe characteristics of the temporal correlations via correlations between the CPTP marginals $\hat{\mathcal{E}}_{j:j-1}$. These characteristics can reveal a great deal about the noise: its complexity, the probability of Markov model confusion, the size of the environment, as well as its quantum or classical nature, as some examples~\cite{Pollock2018, giarmatzi_quantum_2018, White2021}. Since the Choi state is, in general, non-uniquely defined for a restricted process tensor, we do not comment on these aspects here. In Ref.~\cite{White2021}, we focus on applying our methods to extract and bound information about these quantities on NISQ devices, with examples shown on IBM Quantum devices.\par 
Here, we focus on control. We present some examples of how a process tensor characterisation can straightforwardly yield superior circuit fidelities on real QIPs, and the extent to which a conditional Markov model can be used for this. Reconstruction fidelity validates the ability of the process tensor to accurately map a given sequence to its final state. This is especially applicable to near-term quantum devices whose control operations are high in fidelity but whose dynamics (non-Markovian or otherwise) are not under control. In the same way that a mathematical description of a quantum channel may be used to predict its behaviour on any input state, the process tensor can predict the output state of a process, subject to any sequence of input operations.
A mapping from unitary gates to outcomes allows an experimenter to ask `What is the optimal sequence of gates that best achieves this outcome?'. Two key features distinguishing this from regular quantum optimal control is firstly that after characterisation, all optimisation can be performed classically with confidence. Secondly, the process is fully inclusive of non-Markovian dynamics, allowing for the suppression of correlated errors.
Simply choose an objective function $\mathcal{L}$ of the final state. Then $\mathcal{L}\left(\mathcal{T}_{k:0}\left[\mathbf{A}_{k-1:0}]\right]\right)$ classically evaluates $\mathcal{L}$ conditioned on some operation sequence $\mathbf{A}_{k-1:0}$ using the process characterisation. This can be cast as a classical optimisation problem to find the sequence of gates which best results in the desired value of $\mathcal{L}$. This idea was preliminarily explored in Ref.~\cite{White-NM-2020}, and applied to pulse shaping in Ref.~\cite{PhysRevLett.126.200401}.\par 
\subsection{Optimising for states using a full process tensor}
\label{ssec:state-improvement}
\begin{figure}[b]
    \centering
    \includegraphics[width=\linewidth]{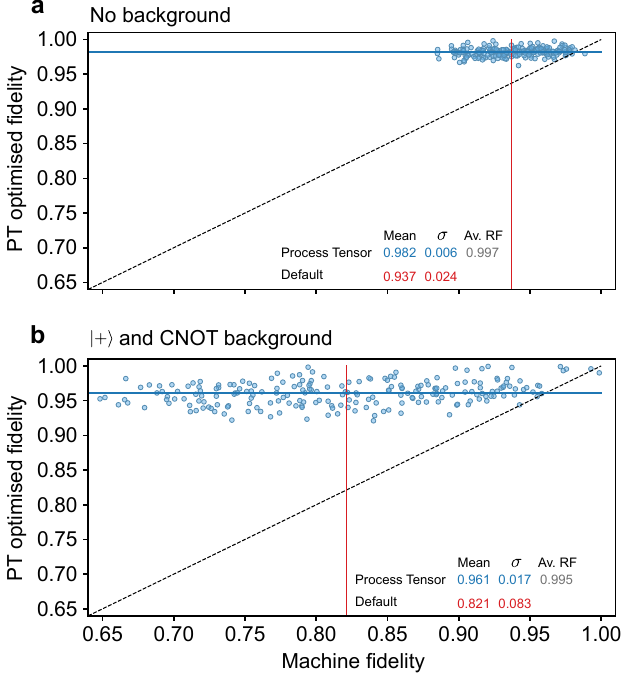}

    \caption{Results of using the process tensor to optimise multi-time circuits with different random single-qubit unitaries. The $x$-axis indicates the fidelity compared to ideal when the sequences are run on the device. The $y$-axis is the fidelity of the sequence when the process tensor is used to optimise to the ideal case. \textbf{a} Here, we have three sequential unitaries with a wait time interleaved similar to that of two CNOT gates (0.71 $\mu$s) on the \emph{ibmq\_manhattan}.  \textbf{b} A similar setup is considered on \emph{ibmq\_bogota}, but with the neighbouring qubit in a $\ket{+}$ state and subject to four CNOT gates per time-step (2.5 $\mu$s).}
    \label{fig:circuit_improvement}
\end{figure}
To demonstrate the utility of this idea, we use the process tensor to improve the fidelity of IBM Quantum devices over multi-time processes. Note that this characterisation overcomes both Markovian and non-Markovian errors.
In particular, we apply many sequences of random unitaries to a single qubit and measure the final state. Interleaved between each operation is a delay time roughly equivalent to the implementation duration of a CNOT gate. 
We then compare the fidelity of this output state to the ideal output subject to those unitaries. That is, generate a set of ideal outputs:
\begin{equation}
    \rho_{\text{ideal}}^{ijk} = \mathcal{A}_3^k\circ \mathcal{A}_2^j \circ \mathcal{A}_1^i [|0\rangle\langle 0|],
\end{equation}
with a set of values $\mathcal{F}(\rho_{\text{ideal}}, \rho_{\text{actual}})$.
Mirroring these dynamics, we construct a process tensor whose basis of inputs is at the same time as the target unitaries. We supplement the reconstruction by using GST to estimate the noisy device POVM. The estimated POVM is then used in the MLE processing, rather than the ideal projective measurements. The purpose of this is to avoid inflating any circuit improvement. For example, a relaxation process during the measurement operation would be absorbed into the process tensor estimate and could be artificially overcome by increasing the $\ket{1}$ population. By accounting for measurement errors in PTT and QST, we are considering only dynamics during the circuit as a more representative depiction of generic PTT capabilities.
Finally, using the PTT characterisation we determine which set of unitaries, $\mathcal{V}$ should be used (instead of the native ones, $\mathcal{A}$) in order to achieve the ideal output state. Let each unitary gate $V$ corresponding to the map $\mathcal{V}$ be parametrised in terms of $\theta$, $\phi$, and $\lambda$ as 
\begin{equation}
    V(\theta, \phi, \lambda) = \begin{pmatrix}
\cos(\theta/2) & -e^{i\lambda}\sin(\theta/2) \\
e^{i\phi}\sin(\theta/2) & e^{i\lambda+i\phi}\cos(\theta/2) 
\end{pmatrix}. \\
\end{equation}

The process prediction is then given with:

\begin{equation}
    \rho_{\text{predicted}}^{ijk}(\vec{\theta},\vec{\phi},\vec{\lambda}) = \mathcal{T}_{3:0}\left[\mathcal{V}^k_3, \mathcal{V}^j_2, \mathcal{V}^i_1\right].
\end{equation}
Then, for each combination $ijk$, we find
\begin{equation}
    \argmax_{\vec{\theta},\vec{\phi},\vec{\lambda}}\mathcal{F}(\rho_{\text{ideal}}^{ijk}, \rho_{\text{predicted}}^{ijk})
\end{equation}
and use these optimal values in sequences on the device.
The results across 216 random sequences are summarised in Figure~\ref{fig:circuit_improvement}a. The average observed improvement was 0.045, with a maximum of 0.10. In addition, the distribution of fidelities is much tighter, with the standard deviation of device-implemented unitaries at 0.0241, compared with our computed values at $0.00578$. We also repeated similar runs on \emph{ibmq\_bogota} intended to drive some characteristics of crosstalk: starting the neighbour in a $\ket{+}$ state followed by four sequential CNOTs to its other-side-nearest-neighbour between each unitary. Initialising the neighbour in a $\ket{+}$ state is intended to generate a passive entangling interaction between the two qubits due to the always-on $ZZ$ interaction found in superconducting transmons. We found that these native fidelities were much worse than on the \emph{ibmq\_manhattan}, despite possessing similar error rates -- implicating the effects of crosstalk. We emphasise that the noise encountered in all our results is naturally occurring from device fabrication, rather than a contrived environment. Nevertheless, the process-tensor-optimal fidelities in Figure~\ref{fig:circuit_improvement}b are nearly as high. This suggests a path forward whereby quantum devices may be characterised using PTT and circuits compiled according to the correlated noise of that device. An obvious drawback of this is the characterisation requirements. We now investigate carrying out a similar task with our conditional Markov order model.

\subsection{Optimising arbitrary circuits with finite Markov order process tensors}
\label{ssec:cmo-improvement}
Using a complete process tensor model to optimise circuit sections may be feasible for a small number of gates, and, indeed, may be necessary for highly correlated noise. However, it is not desirable in a generic sense to characterise redundant information. Moreover, it is impractical to optimise over specific circuits in a state-dependent way when inputs may be reduced subsystems of a larger register. Here, we address both of these points. We target longer circuits with larger values of $k$ by using a truncated Markov model. In doing so, we both validate our conditional Markov order methodology, and demonstrate the need for approaching the problem of NISQ noise with temporal correlations in mind. Further we also change our optimisation approach: instead of trying to create a specific state on a circuit-by-circuit basis, we numerically find the sequence of gates which most closely takes the effective process to be the identity channel. This allows for arbitrary addressing of non-Markovian noise without a priori knowing the input state.\par 
A five-step process is considered with delays of approximately 1.2$\mu$s after each gate. We characterise this process using conditional Markov order models of $\ell = 1$, $\ell=2$, and $\ell=3$ under three different cases: no operations on the background qubits, one nearest neighbour to the system initialised in a $\ket{+}$ state, and finally two nearest neighbours and one next-to-nearest neighbour in a $\ket{+}$ state. The first job took place on \emph{ibmq\_montreal} and the second and third on \emph{ibmq\_guadalupe}. The purpose of the latter two analyses is to encourage any (predominantly $ZZ$) interaction which realistically might occur between qubits in an algorithm. We then generate 100 sequences of 5 random unitary gates, followed by QST. These sequences serve two purposes: first, we evaluate the reconstruction fidelity for the different Markov order models, and secondly we use these as our benchmark for adaptively improving the native fidelity of the device. \par 
\begin{figure}
    \centering
    \includegraphics[width=\linewidth]{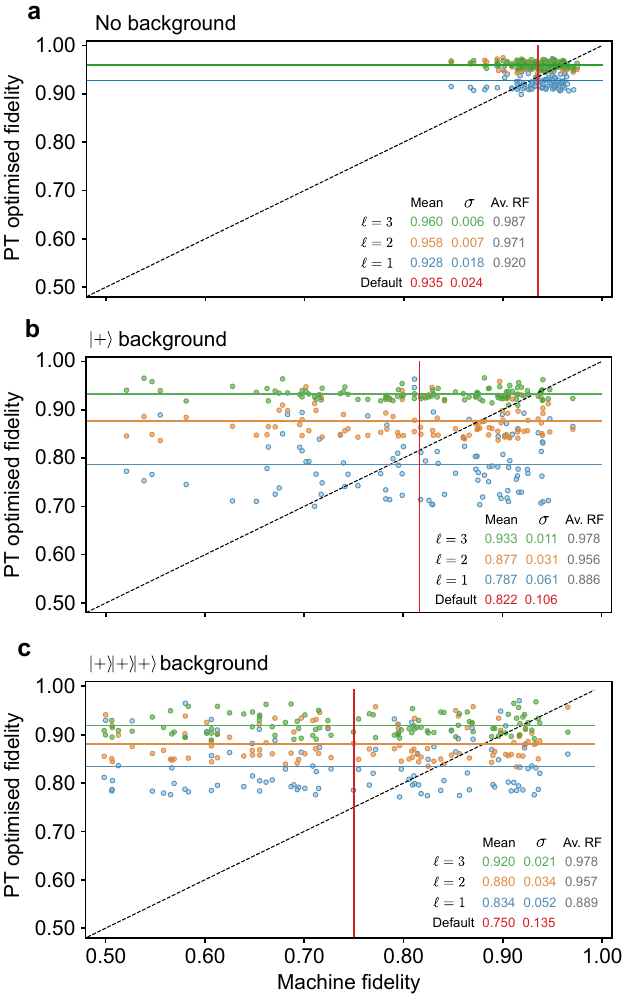}
    \caption{Results of using conditional Markov models to improve the fidelities of a five-step circuit on \emph{ibmq\_montreal} and \emph{ibmq\_guadalupe}. We construct conditional Markov order models for $\ell=1$, $\ell=2$, and $\ell=3$ across five steps under a variety of background conditions. For randomised inputs, these models are used to optimise the next four operations. The four optimal operations are then applied to each input and the results compared to machine fidelity. The mean and standard deviation for each set of circuits is listed, as well as average reconstruction fidelity of the models. \textbf{a} No background operations are applied. \textbf{b} The nearest neighbour to the system is initialised in a $\ket{+}$ state. \textbf{c} Two nearest neighbours and one next-to-nearest neighbour from the system are initialised in a $\ket{+}$ state.}
    \label{fig:circuit_improvement_CMO}
\end{figure}
We select an IC set of unitary gates $\{\mathcal{A}^i\}$ to be applied in the first circuit position, generating a set of ideal states $\{\rho_{\text{ideal}}\} := \{\mathcal{A}[\ket{0}\!\bra{0}\}$. That is, $i$ indexes a set of random states. We then parametrise the next four gates, again, in terms of $\vec{\theta}$, $\vec{\phi}$, and $\vec{\lambda}$. However, this time, the gates are the same for each input. Finally, using each $\mathbf{\Upsilon}_{5:0}^\ell$, we compute:
\begin{equation}
\label{CMO-opt}
    \argmax_{\vec{\theta},\vec{\phi},\vec{\lambda}} \sum_i \left[\mathcal{F}(\rho_{\text{ideal}}^i,\rho_{\text{predicted}}^i)\right]^2.
\end{equation}
In plain words, we are finding the four gates which simultaneously best preserve all of our random input states. After running this optimisation for each Markov order model and each background, we then aimed to create the ideal output from the 100 random sequences. First, by creating the state with the first gate, then applying the four gates found from Equation~\eqref{CMO-opt}. Following QST at the end, we compute the fidelity of each final state with respect to the ideal.
The purpose of this routine was two-fold: to determine whether active circuit improvements (akin to dynamical decoupling) could be systematically found, even in the presence of non-Markovian noise, and to ascertain how the inclusion of higher order temporal correlations in the model could help achieve this task. Without randomising over the inputs, we found that the $\ell=1$ model would hide each state in a decoherence-free subspace until the last gate, which is not a generalisable strategy.
The results of these runs are shown in Figure~\ref{fig:circuit_improvement_CMO} for each sequence and each Markov model with both the mean circuit fidelities and reconstruction fidelities printed. 
With no activity on neighbouring qubits, we find a moderate amount of non-Markovian noise at this time scale. Interestingly, $\ell=3$ predicts the dynamics moderately better than $\ell=2$, signalling the presence of higher order correlations in the dynamics. However, the optimal interventions improve the average circuit fidelity to a similar level for each. The generic correctability for given circuit structures may saturate, regardless of the completeness of characterisation.
For the second and third situations, the dynamics are more complex and we see a clear separation between the different Markov orders. By accounting for these higher order temporal correlations, we are able to more substantially increase circuit fidelities, both in terms of the mean value, and in terms of the tightness of the distribution. Only in the last case do we find that a Markov model $\ell=1$ is able to achieve an improvement, further highlighting the need for our multi-time process characterisation on NISQ devices.

\section{Discussion}
In NISQ devices, circuits performances are not solely determined by the simple composition of high-fidelity gates and measurements, but exhibit complex non-Markovian effects. It is therefore unavoidable to pivot the focus of characterisation techniques to the emergent, holistic behaviour of multi-time processes. 
In this paper, we have formally introduced a multi-time generalisation of quantum process tomography, in the form of estimating process tensor models. We have presented several key advancements that we believe will be valuable contributions to the community: we have shown how to obtain reliable, high-fidelity, minimal-resource estimations of quantum non-Markovian processes through our fast MLE procedure; derived a method by which low memory ans\"atze can be implemented; and shown how to use our tools to improve the performance of NISQ devices.
This is, to the best of our knowledge, the first development of a maximum-likelihood technique for reconstructing multi-time processes. Moreover, it permits a modular description of non-Markovian memory. These facets are important not just for quantum information processing, but in the study of multi-time correlations that naturally occur in out-of-equilibrium quantum stochastic phenomena, such in cold atoms, condensed matter physics, and quantum biology~\cite{PhysRevLett.124.043603,PhysRevLett.115.043601,PhysRevA.84.031602,PhysRevE.88.062719,nitzan2003electron,lambert2013quantum}.
\par 
Our technique requires computational resources which scale only linearly in time, while being exponential in the Markov order. The required Markov order plays two important roles: it limits the computational requirements, and benchmarks the degree of non-Markovianity in the device by answering ``how many previous time-steps are relevant to our current description of the dynamics?".
The next step is to further compress the process description by employing the myriad techniques for efficient tomography which are well described in the context of QST and QPT~\cite{flammia2012quantum,cramer2010efficient,arXiv:2006.02424,guochu2020,PhysRevLett.126.100402}.
Alternatively, if one were interested in an informative snapshot of the temporal correlations, shadow tomography could efficiently estimate linear functions of the Choi state, supposing that an informationally complete basis of operations were available~\cite{huang-shadow}. \par

Applications of the process tensor have been promising for improving circuit fidelities. However, much more work needs to be accomplished to render this practical: what is the minimal characterisation required to realise this superior control? When can it be re-used on other qubits? Can the characterisation be used to improve generic (possibly unknown) circuits? The practical and theoretical tools developed here pave the pathway for answering many of these questions.

One important aspect of the results so far, is that they allow for optimisations which mitigate highly correlated noise. These suggest that current devices may be closer to fault tolerance than presently realised. For example, the native error-per-gate suggested in Figure~\ref{fig:circuit_improvement_CMO}c is $\approx0.02$ in the presence of crosstalk, whereas applying an optimisation based on non-Markovian error characterisation reduces this to $\approx0.005$. But this is not the only consideration; not only do our methods allow for the device to be cleaned up in an absolute sense, but also the reduction of correlated errors has the effect of lowering the stringency of quantum error correcting thresholds (when compared to a correlated error model). Thus, a non-Markovian characterisation can both raise the performance while lowering the bar. We strongly believe that characterisation techniques and software will play a large role in the eventual realisation of a fault tolerant quantum computer.

\section*{Acknowledgments}
This work was supported by the University of Melbourne through the establishment of an IBM Quantum Network Hub at the University.
G.A.L.W. is supported by an Australian Government Research Training Program Scholarship. 
C.D.H. is supported through a Laby Foundation grant at The University of Melbourne. 
K.M. is supported through Australian Research Council Future Fellowship FT160100073.
K.M. and C.D.H. acknowledge the support of Australian Research Council's Discovery Project DP210100597.
K.M. and C.D.H. were recipients of the International Quantum U Tech Accelerator award by the US Air Force Research Laboratory.

\section*{References}


\begin{thebibliography}{93}%
\makeatletter
\providecommand \@ifxundefined [1]{%
 \@ifx{#1\undefined}
}%
\providecommand \@ifnum [1]{%
 \ifnum #1\expandafter \@firstoftwo
 \else \expandafter \@secondoftwo
 \fi
}%
\providecommand \@ifx [1]{%
 \ifx #1\expandafter \@firstoftwo
 \else \expandafter \@secondoftwo
 \fi
}%
\providecommand \natexlab [1]{#1}%
\providecommand \enquote  [1]{``#1''}%
\providecommand \bibnamefont  [1]{#1}%
\providecommand \bibfnamefont [1]{#1}%
\providecommand \citenamefont [1]{#1}%
\providecommand \href@noop [0]{\@secondoftwo}%
\providecommand \href [0]{\begingroup \@sanitize@url \@href}%
\providecommand \@href[1]{\@@startlink{#1}\@@href}%
\providecommand \@@href[1]{\endgroup#1\@@endlink}%
\providecommand \@sanitize@url [0]{\catcode `\\12\catcode `\$12\catcode
  `\&12\catcode `\#12\catcode `\^12\catcode `\_12\catcode `\%12\relax}%
\providecommand \@@startlink[1]{}%
\providecommand \@@endlink[0]{}%
\providecommand \url  [0]{\begingroup\@sanitize@url \@url }%
\providecommand \@url [1]{\endgroup\@href {#1}{\urlprefix }}%
\providecommand \urlprefix  [0]{URL }%
\providecommand \Eprint [0]{\href }%
\providecommand \doibase [0]{http://dx.doi.org/}%
\providecommand \selectlanguage [0]{\@gobble}%
\providecommand \bibinfo  [0]{\@secondoftwo}%
\providecommand \bibfield  [0]{\@secondoftwo}%
\providecommand \translation [1]{[#1]}%
\providecommand \BibitemOpen [0]{}%
\providecommand \bibitemStop [0]{}%
\providecommand \bibitemNoStop [0]{.\EOS\space}%
\providecommand \EOS [0]{\spacefactor3000\relax}%
\providecommand \BibitemShut  [1]{\csname bibitem#1\endcsname}%
\let\auto@bib@innerbib\@empty
\bibitem [{\citenamefont {Eisert}\ \emph {et~al.}(2020)\citenamefont {Eisert},
  \citenamefont {Hangleiter}, \citenamefont {Walk}, \citenamefont {Roth},
  \citenamefont {Markham}, \citenamefont {Parekh}, \citenamefont {Chabaud},\
  and\ \citenamefont {Kashefi}}]{Eisert2020}%
  \BibitemOpen
  \bibfield  {author} {\bibinfo {author} {\bibfnamefont {Jens}\ \bibnamefont
  {Eisert}}, \bibinfo {author} {\bibfnamefont {Dominik}\ \bibnamefont
  {Hangleiter}}, \bibinfo {author} {\bibfnamefont {Nathan}\ \bibnamefont
  {Walk}}, \bibinfo {author} {\bibfnamefont {Ingo}\ \bibnamefont {Roth}},
  \bibinfo {author} {\bibfnamefont {Damian}\ \bibnamefont {Markham}}, \bibinfo
  {author} {\bibfnamefont {Rhea}\ \bibnamefont {Parekh}}, \bibinfo {author}
  {\bibfnamefont {Ulysse}\ \bibnamefont {Chabaud}}, \ and\ \bibinfo {author}
  {\bibfnamefont {Elham}\ \bibnamefont {Kashefi}},\ }\bibfield  {title}
  {\enquote {\bibinfo {title} {{Quantum certification and benchmarking}},}\
  }\href {\doibase 10.1038/s42254-020-0186-4} {\bibfield  {journal} {\bibinfo
  {journal} {Nature Reviews Physics}\ }\textbf {\bibinfo {volume} {2}},\
  \bibinfo {pages} {382--390} (\bibinfo {year} {2020})},\ \Eprint
  {http://arxiv.org/abs/1910.06343} {arXiv:1910.06343} \BibitemShut {NoStop}%
\bibitem [{\citenamefont {Endo}\ \emph {et~al.}(2018)\citenamefont {Endo},
  \citenamefont {Benjamin},\ and\ \citenamefont {Li}}]{Endo2018}%
  \BibitemOpen
  \bibfield  {author} {\bibinfo {author} {\bibfnamefont {Suguru}\ \bibnamefont
  {Endo}}, \bibinfo {author} {\bibfnamefont {Simon~C.}\ \bibnamefont
  {Benjamin}}, \ and\ \bibinfo {author} {\bibfnamefont {Ying}\ \bibnamefont
  {Li}},\ }\bibfield  {title} {\enquote {\bibinfo {title} {{Practical Quantum
  Error Mitigation for Near-Future Applications}},}\ }\href {\doibase
  10.1103/PhysRevX.8.031027} {\bibfield  {journal} {\bibinfo  {journal}
  {Physical Review X}\ }\textbf {\bibinfo {volume} {8}},\ \bibinfo {pages}
  {31027} (\bibinfo {year} {2018})}\BibitemShut {NoStop}%
\bibitem [{\citenamefont {Ferracin}\ \emph {et~al.}(2019)\citenamefont
  {Ferracin}, \citenamefont {Kapourniotis},\ and\ \citenamefont
  {Datta}}]{Ferracin2019}%
  \BibitemOpen
  \bibfield  {author} {\bibinfo {author} {\bibfnamefont {Samuele}\ \bibnamefont
  {Ferracin}}, \bibinfo {author} {\bibfnamefont {Theodoros}\ \bibnamefont
  {Kapourniotis}}, \ and\ \bibinfo {author} {\bibfnamefont {Animesh}\
  \bibnamefont {Datta}},\ }\bibfield  {title} {\enquote {\bibinfo {title}
  {Accrediting outputs of noisy intermediate-scale quantum computing
  devices},}\ }\href {\doibase 10.1088/1367-2630/ab4fd6} {\bibfield  {journal}
  {\bibinfo  {journal} {New Journal of Physics}\ }\textbf {\bibinfo {volume}
  {21}},\ \bibinfo {pages} {113038} (\bibinfo {year} {2019})}\BibitemShut
  {NoStop}%
\bibitem [{\citenamefont {White}\ \emph
  {et~al.}(2021{\natexlab{a}})\citenamefont {White}, \citenamefont {Hill},\
  and\ \citenamefont {Hollenberg}}]{white-POST}%
  \BibitemOpen
  \bibfield  {author} {\bibinfo {author} {\bibfnamefont {G.~A.~L.}\
  \bibnamefont {White}}, \bibinfo {author} {\bibfnamefont {C.~D.}\ \bibnamefont
  {Hill}}, \ and\ \bibinfo {author} {\bibfnamefont {L.~C.~L.}\ \bibnamefont
  {Hollenberg}},\ }\href {\doibase 10.1103/PhysRevApplied.15.014023} {\bibfield
   {journal} {\bibinfo  {journal} {Physical Review Applied}\ }\textbf {\bibinfo
  {volume} {15}},\ \bibinfo {pages} {014023} (\bibinfo {year}
  {2021}{\natexlab{a}})},\ \Eprint {http://arxiv.org/abs/1911.12096}
  {arXiv:1911.12096} \BibitemShut {NoStop}%
\bibitem [{\citenamefont {Harper}\ \emph {et~al.}(2020)\citenamefont {Harper},
  \citenamefont {Flammia},\ and\ \citenamefont {Wallman}}]{Harper2020}%
  \BibitemOpen
  \bibfield  {author} {\bibinfo {author} {\bibfnamefont {Robin}\ \bibnamefont
  {Harper}}, \bibinfo {author} {\bibfnamefont {Steven~T.}\ \bibnamefont
  {Flammia}}, \ and\ \bibinfo {author} {\bibfnamefont {Joel~J.}\ \bibnamefont
  {Wallman}},\ }\bibfield  {title} {\enquote {\bibinfo {title} {{Efficient
  learning of quantum noise}},}\ }\href {\doibase 10.1038/s41567-020-0992-8}
  {\bibfield  {journal} {\bibinfo  {journal} {Nature Physics}\ }\textbf
  {\bibinfo {volume} {16}},\ \bibinfo {pages} {1184--1188} (\bibinfo {year}
  {2020})},\ \Eprint {http://arxiv.org/abs/1907.13022} {arXiv:1907.13022}
  \BibitemShut {NoStop}%
\bibitem [{\citenamefont {Jurcevic}\ \emph {et~al.}(2021)\citenamefont
  {Jurcevic}, \citenamefont {Javadi-Abhari}, \citenamefont {Bishop},
  \citenamefont {Lauer}, \citenamefont {Bogorin}, \citenamefont {Brink},
  \citenamefont {Capelluto}, \citenamefont {Günlük}, \citenamefont {Itoko},
  \citenamefont {Kanazawa}, \citenamefont {Kandala} \emph
  {et~al.}}]{Jurcevic2021}%
  \BibitemOpen
  \bibfield  {author} {\bibinfo {author} {\bibfnamefont {Petar}\ \bibnamefont
  {Jurcevic}}, \bibinfo {author} {\bibfnamefont {Ali}\ \bibnamefont
  {Javadi-Abhari}}, \bibinfo {author} {\bibfnamefont {Lev~S}\ \bibnamefont
  {Bishop}}, \bibinfo {author} {\bibfnamefont {Isaac}\ \bibnamefont {Lauer}},
  \bibinfo {author} {\bibfnamefont {Daniela~F}\ \bibnamefont {Bogorin}},
  \bibinfo {author} {\bibfnamefont {Markus}\ \bibnamefont {Brink}}, \bibinfo
  {author} {\bibfnamefont {Lauren}\ \bibnamefont {Capelluto}}, \bibinfo
  {author} {\bibfnamefont {Oktay}\ \bibnamefont {Günlük}}, \bibinfo {author}
  {\bibfnamefont {Toshinari}\ \bibnamefont {Itoko}}, \bibinfo {author}
  {\bibfnamefont {Naoki}\ \bibnamefont {Kanazawa}}, \bibinfo {author}
  {\bibfnamefont {Abhinav}\ \bibnamefont {Kandala}},  \emph {et~al.},\
  }\bibfield  {title} {\enquote {\bibinfo {title} {Demonstration of quantum
  volume 64 on a superconducting quantum computing system},}\ }\href {\doibase
  10.1088/2058-9565/abe519} {\bibfield  {journal} {\bibinfo  {journal} {Quantum
  Science and Technology}\ }\textbf {\bibinfo {volume} {6}},\ \bibinfo {pages}
  {025020} (\bibinfo {year} {2021})}\BibitemShut {NoStop}%
\bibitem [{\citenamefont {Blume-Kohout}\ \emph {et~al.}(2017)\citenamefont
  {Blume-Kohout}, \citenamefont {Gamble}, \citenamefont {Nielsen},
  \citenamefont {Rudinger}, \citenamefont {Mizrahi}, \citenamefont {Fortier},\
  and\ \citenamefont {Maunz}}]{RBK2017}%
  \BibitemOpen
  \bibfield  {author} {\bibinfo {author} {\bibfnamefont {Robin}\ \bibnamefont
  {Blume-Kohout}}, \bibinfo {author} {\bibfnamefont {John~King}\ \bibnamefont
  {Gamble}}, \bibinfo {author} {\bibfnamefont {Erik}\ \bibnamefont {Nielsen}},
  \bibinfo {author} {\bibfnamefont {Kenneth}\ \bibnamefont {Rudinger}},
  \bibinfo {author} {\bibfnamefont {Jonathan}\ \bibnamefont {Mizrahi}},
  \bibinfo {author} {\bibfnamefont {Kevin}\ \bibnamefont {Fortier}}, \ and\
  \bibinfo {author} {\bibfnamefont {Peter}\ \bibnamefont {Maunz}},\ }\bibfield
  {title} {\enquote {\bibinfo {title} {{Demonstration of qubit operations below
  a rigorous fault tolerance threshold with gate set tomography}},}\ }\href
  {\doibase 10.1038/ncomms14485} {\bibfield  {journal} {\bibinfo  {journal}
  {Nature Communications}\ }\textbf {\bibinfo {volume} {8}},\ \bibinfo {pages}
  {14485} (\bibinfo {year} {2017})},\ \Eprint {http://arxiv.org/abs/1605.07674}
  {arXiv:1605.07674} \BibitemShut {NoStop}%
\bibitem [{\citenamefont {Milz}\ \emph {et~al.}(2017)\citenamefont {Milz},
  \citenamefont {Pollock},\ and\ \citenamefont {Modi}}]{Milz2017}%
  \BibitemOpen
  \bibfield  {author} {\bibinfo {author} {\bibfnamefont {Simon}\ \bibnamefont
  {Milz}}, \bibinfo {author} {\bibfnamefont {Felix~A.}\ \bibnamefont
  {Pollock}}, \ and\ \bibinfo {author} {\bibfnamefont {Kavan}\ \bibnamefont
  {Modi}},\ }\bibfield  {title} {\enquote {\bibinfo {title} {{An introduction
  to operational quantum dynamics}},}\ }\href {\doibase
  10.1142/S1230161217400169} {\bibfield  {journal} {\bibinfo  {journal} {Open
  Syst. Inf. Dyn.}\ }\textbf {\bibinfo {volume} {24}},\ \bibinfo {pages}
  {1740016} (\bibinfo {year} {2017})}\BibitemShut {NoStop}%
\bibitem [{\citenamefont {Milz}\ and\ \citenamefont {Modi}(2021)}]{Milz2021}%
  \BibitemOpen
  \bibfield  {author} {\bibinfo {author} {\bibfnamefont {Simon}\ \bibnamefont
  {Milz}}\ and\ \bibinfo {author} {\bibfnamefont {Kavan}\ \bibnamefont
  {Modi}},\ }\bibfield  {title} {\enquote {\bibinfo {title} {Quantum stochastic
  processes and quantum non-markovian phenomena},}\ }\href {\doibase
  10.1103/PRXQuantum.2.030201} {\bibfield  {journal} {\bibinfo  {journal} {PRX
  Quantum}\ }\textbf {\bibinfo {volume} {2}},\ \bibinfo {pages} {030201}
  (\bibinfo {year} {2021})},\ \Eprint {http://arxiv.org/abs/2012.01894}
  {arXiv:2012.01894} \BibitemShut {NoStop}%
\bibitem [{\citenamefont {Li}\ \emph {et~al.}(2018)\citenamefont {Li},
  \citenamefont {Hall},\ and\ \citenamefont {Wiseman}}]{Li2018}%
  \BibitemOpen
  \bibfield  {author} {\bibinfo {author} {\bibfnamefont {Li}~\bibnamefont
  {Li}}, \bibinfo {author} {\bibfnamefont {Michael~J.W.}\ \bibnamefont {Hall}},
  \ and\ \bibinfo {author} {\bibfnamefont {Howard~M.}\ \bibnamefont
  {Wiseman}},\ }\bibfield  {title} {\enquote {\bibinfo {title} {{Concepts of
  quantum non-Markovianity: A hierarchy}},}\ }\href {\doibase
  10.1016/j.physrep.2018.07.001} {\bibfield  {journal} {\bibinfo  {journal}
  {Physics Reports}\ }\textbf {\bibinfo {volume} {759}},\ \bibinfo {pages}
  {1--51} (\bibinfo {year} {2018})},\ \Eprint {http://arxiv.org/abs/1712.08879}
  {arXiv:1712.08879} \BibitemShut {NoStop}%
\bibitem [{\citenamefont {Breuer}\ \emph {et~al.}(2016)\citenamefont {Breuer},
  \citenamefont {Laine}, \citenamefont {Piilo},\ and\ \citenamefont
  {Vacchini}}]{breuer2016}%
  \BibitemOpen
  \bibfield  {author} {\bibinfo {author} {\bibfnamefont {Heinz~Peter}\
  \bibnamefont {Breuer}}, \bibinfo {author} {\bibfnamefont {Elsi~Mari}\
  \bibnamefont {Laine}}, \bibinfo {author} {\bibfnamefont {Jyrki}\ \bibnamefont
  {Piilo}}, \ and\ \bibinfo {author} {\bibfnamefont {Bassano}\ \bibnamefont
  {Vacchini}},\ }\bibfield  {title} {\enquote {\bibinfo {title} {{Colloquium:
  Non-Markovian dynamics in open quantum systems}},}\ }\href {\doibase
  10.1103/RevModPhys.88.021002} {\bibfield  {journal} {\bibinfo  {journal}
  {Reviews of Modern Physics}\ }\textbf {\bibinfo {volume} {88}},\ \bibinfo
  {pages} {021002} (\bibinfo {year} {2016})}\BibitemShut {NoStop}%
\bibitem [{\citenamefont {de~Vega}\ and\ \citenamefont
  {Alonso}(2017)}]{deVega2017}%
  \BibitemOpen
  \bibfield  {author} {\bibinfo {author} {\bibfnamefont {In{\'e}s}\
  \bibnamefont {de~Vega}}\ and\ \bibinfo {author} {\bibfnamefont {Daniel}\
  \bibnamefont {Alonso}},\ }\bibfield  {title} {\enquote {\bibinfo {title}
  {Dynamics of non-markovian open quantum systems},}\ }\href {\doibase
  10.1103/RevModPhys.89.015001} {\bibfield  {journal} {\bibinfo  {journal}
  {Rev. Mod. Phys.}\ }\textbf {\bibinfo {volume} {89}},\ \bibinfo {pages}
  {015001} (\bibinfo {year} {2017})}\BibitemShut {NoStop}%
\bibitem [{\citenamefont {Rivas}\ \emph {et~al.}(2014)\citenamefont {Rivas},
  \citenamefont {Huelga},\ and\ \citenamefont {Plenio}}]{rivas-NM-review}%
  \BibitemOpen
  \bibfield  {author} {\bibinfo {author} {\bibfnamefont {{\'{A}}ngel}\
  \bibnamefont {Rivas}}, \bibinfo {author} {\bibfnamefont {Susana~F.}\
  \bibnamefont {Huelga}}, \ and\ \bibinfo {author} {\bibfnamefont {Martin~B.}\
  \bibnamefont {Plenio}},\ }\bibfield  {title} {\enquote {\bibinfo {title}
  {{Quantum non-Markovianity: Characterization, quantification and
  detection}},}\ }\href {\doibase 10.1088/0034-4885/77/9/094001} {\bibfield
  {journal} {\bibinfo  {journal} {Reports on Progress in Physics}\ }\textbf
  {\bibinfo {volume} {77}},\ \bibinfo {pages} {094001} (\bibinfo {year}
  {2014})},\ \Eprint {http://arxiv.org/abs/1405.0303} {arXiv:1405.0303}
  \BibitemShut {NoStop}%
\bibitem [{\citenamefont {Cross}\ \emph {et~al.}(2019)\citenamefont {Cross},
  \citenamefont {Bishop}, \citenamefont {Sheldon}, \citenamefont {Nation},\
  and\ \citenamefont {Gambetta}}]{Cross-QV}%
  \BibitemOpen
  \bibfield  {author} {\bibinfo {author} {\bibfnamefont {Andrew~W.}\
  \bibnamefont {Cross}}, \bibinfo {author} {\bibfnamefont {Lev~S.}\
  \bibnamefont {Bishop}}, \bibinfo {author} {\bibfnamefont {Sarah}\
  \bibnamefont {Sheldon}}, \bibinfo {author} {\bibfnamefont {Paul~D.}\
  \bibnamefont {Nation}}, \ and\ \bibinfo {author} {\bibfnamefont {Jay~M.}\
  \bibnamefont {Gambetta}},\ }\bibfield  {title} {\enquote {\bibinfo {title}
  {{Validating quantum computers using randomized model circuits}},}\ }\href
  {\doibase 10.1103/PhysRevA.100.032328} {\bibfield  {journal} {\bibinfo
  {journal} {Physical Review A}\ }\textbf {\bibinfo {volume} {100}},\ \bibinfo
  {pages} {032328} (\bibinfo {year} {2019})},\ \Eprint
  {http://arxiv.org/abs/1811.12926} {arXiv:1811.12926} \BibitemShut {NoStop}%
\bibitem [{\citenamefont {Arute}\ \emph {et~al.}(2019)\citenamefont {Arute},
  \citenamefont {Arya}, \citenamefont {Babbush}, \citenamefont {Bacon},
  \citenamefont {Bardin}, \citenamefont {Barends}, \citenamefont {Biswas},
  \citenamefont {Boixo}, \citenamefont {Brandao}, \citenamefont {Buell} \emph
  {et~al.}}]{Arute2019}%
  \BibitemOpen
  \bibfield  {author} {\bibinfo {author} {\bibfnamefont {Frank}\ \bibnamefont
  {Arute}}, \bibinfo {author} {\bibfnamefont {Kunal}\ \bibnamefont {Arya}},
  \bibinfo {author} {\bibfnamefont {Ryan}\ \bibnamefont {Babbush}}, \bibinfo
  {author} {\bibfnamefont {Dave}\ \bibnamefont {Bacon}}, \bibinfo {author}
  {\bibfnamefont {Joseph~C}\ \bibnamefont {Bardin}}, \bibinfo {author}
  {\bibfnamefont {Rami}\ \bibnamefont {Barends}}, \bibinfo {author}
  {\bibfnamefont {Rupak}\ \bibnamefont {Biswas}}, \bibinfo {author}
  {\bibfnamefont {Sergio}\ \bibnamefont {Boixo}}, \bibinfo {author}
  {\bibfnamefont {Fernando~GSL}\ \bibnamefont {Brandao}}, \bibinfo {author}
  {\bibfnamefont {David~A}\ \bibnamefont {Buell}},  \emph {et~al.},\ }\bibfield
   {title} {\enquote {\bibinfo {title} {Quantum supremacy using a programmable
  superconducting processor},}\ }\href {\doibase 10.1038/s41586-019-1666-5}
  {\bibfield  {journal} {\bibinfo  {journal} {Nature}\ }\textbf {\bibinfo
  {volume} {574}},\ \bibinfo {pages} {505--510} (\bibinfo {year} {2019})},\
  \Eprint {http://arxiv.org/abs/1910.11333} {arXiv:1910.11333} \BibitemShut
  {NoStop}%
\bibitem [{\citenamefont {Pogorelov}\ \emph {et~al.}(2021)\citenamefont
  {Pogorelov}, \citenamefont {Feldker}, \citenamefont {Marciniak},
  \citenamefont {Postler}, \citenamefont {Jacob}, \citenamefont
  {Krieglsteiner}, \citenamefont {Podlesnic}, \citenamefont {Meth},
  \citenamefont {Negnevitsky}, \citenamefont {Stadler} \emph
  {et~al.}}]{Pogorelov2021}%
  \BibitemOpen
  \bibfield  {author} {\bibinfo {author} {\bibfnamefont {Ivan}\ \bibnamefont
  {Pogorelov}}, \bibinfo {author} {\bibfnamefont {Thomas}\ \bibnamefont
  {Feldker}}, \bibinfo {author} {\bibfnamefont {Christian~D.}\ \bibnamefont
  {Marciniak}}, \bibinfo {author} {\bibfnamefont {Lukas}\ \bibnamefont
  {Postler}}, \bibinfo {author} {\bibfnamefont {Georg}\ \bibnamefont {Jacob}},
  \bibinfo {author} {\bibfnamefont {Oliver}\ \bibnamefont {Krieglsteiner}},
  \bibinfo {author} {\bibfnamefont {Verena}\ \bibnamefont {Podlesnic}},
  \bibinfo {author} {\bibfnamefont {Michael}\ \bibnamefont {Meth}}, \bibinfo
  {author} {\bibfnamefont {Vlad}\ \bibnamefont {Negnevitsky}}, \bibinfo
  {author} {\bibfnamefont {Martin}\ \bibnamefont {Stadler}},  \emph {et~al.},\
  }\bibfield  {title} {\enquote {\bibinfo {title} {{A compact ion-trap quantum
  computing demonstrator}},}\ }\href {http://arxiv.org/abs/2101.11390}
  {\bibfield  {journal} {\bibinfo  {journal} {arXiv:2101.11390}\ } (\bibinfo
  {year} {2021})}\BibitemShut {NoStop}%
\bibitem [{\citenamefont {Sung}\ \emph {et~al.}(2021)\citenamefont {Sung},
  \citenamefont {Ding}, \citenamefont {Braum\"uller}, \citenamefont
  {Veps\"al\"ainen}, \citenamefont {Kannan}, \citenamefont {Kjaergaard},
  \citenamefont {Greene}, \citenamefont {Samach}, \citenamefont {McNally},
  \citenamefont {Kim} \emph {et~al.}}]{PhysRevX.11.021058}%
  \BibitemOpen
  \bibfield  {author} {\bibinfo {author} {\bibfnamefont {Youngkyu}\
  \bibnamefont {Sung}}, \bibinfo {author} {\bibfnamefont {Leon}\ \bibnamefont
  {Ding}}, \bibinfo {author} {\bibfnamefont {Jochen}\ \bibnamefont
  {Braum\"uller}}, \bibinfo {author} {\bibfnamefont {Antti}\ \bibnamefont
  {Veps\"al\"ainen}}, \bibinfo {author} {\bibfnamefont {Bharath}\ \bibnamefont
  {Kannan}}, \bibinfo {author} {\bibfnamefont {Morten}\ \bibnamefont
  {Kjaergaard}}, \bibinfo {author} {\bibfnamefont {Ami}\ \bibnamefont
  {Greene}}, \bibinfo {author} {\bibfnamefont {Gabriel~O.}\ \bibnamefont
  {Samach}}, \bibinfo {author} {\bibfnamefont {Chris}\ \bibnamefont {McNally}},
  \bibinfo {author} {\bibfnamefont {David}\ \bibnamefont {Kim}},  \emph
  {et~al.},\ }\bibfield  {title} {\enquote {\bibinfo {title} {{Realization of
  High-Fidelity CZ and $ZZ$-Free iSWAP Gates with a Tunable Coupler}},}\ }\href
  {\doibase 10.1103/PhysRevX.11.021058} {\bibfield  {journal} {\bibinfo
  {journal} {Physical Review X}\ }\textbf {\bibinfo {volume} {11}},\ \bibinfo
  {pages} {021058} (\bibinfo {year} {2021})},\ \Eprint
  {http://arxiv.org/abs/2011.01261} {arXiv:2011.01261} \BibitemShut {NoStop}%
\bibitem [{\citenamefont {Proctor}\ \emph {et~al.}(2020)\citenamefont
  {Proctor}, \citenamefont {Rudinger}, \citenamefont {Young}, \citenamefont
  {Nielsen},\ and\ \citenamefont {Blume-Kohout}}]{mirror-benchmarking}%
  \BibitemOpen
  \bibfield  {author} {\bibinfo {author} {\bibfnamefont {Timothy}\ \bibnamefont
  {Proctor}}, \bibinfo {author} {\bibfnamefont {Kenneth}\ \bibnamefont
  {Rudinger}}, \bibinfo {author} {\bibfnamefont {Kevin}\ \bibnamefont {Young}},
  \bibinfo {author} {\bibfnamefont {Erik}\ \bibnamefont {Nielsen}}, \ and\
  \bibinfo {author} {\bibfnamefont {Robin}\ \bibnamefont {Blume-Kohout}},\
  }\bibfield  {title} {\enquote {\bibinfo {title} {{Measuring the Capabilities
  of Quantum Computers}},}\ }\href {http://arxiv.org/abs/2008.11294} {\bibfield
   {journal} {\bibinfo  {journal} {arXiv:2008.11294}\ } (\bibinfo {year}
  {2020})}\BibitemShut {NoStop}%
\bibitem [{\citenamefont {Clader}\ \emph {et~al.}(2021)\citenamefont {Clader},
  \citenamefont {Trout}, \citenamefont {Barnes}, \citenamefont {Schultz},
  \citenamefont {Quiroz},\ and\ \citenamefont {Titum}}]{Clader2021}%
  \BibitemOpen
  \bibfield  {author} {\bibinfo {author} {\bibfnamefont {B.~D.}\ \bibnamefont
  {Clader}}, \bibinfo {author} {\bibfnamefont {Colin~J.}\ \bibnamefont
  {Trout}}, \bibinfo {author} {\bibfnamefont {Jeff~P.}\ \bibnamefont {Barnes}},
  \bibinfo {author} {\bibfnamefont {Kevin}\ \bibnamefont {Schultz}}, \bibinfo
  {author} {\bibfnamefont {Gregory}\ \bibnamefont {Quiroz}}, \ and\ \bibinfo
  {author} {\bibfnamefont {Paraj}\ \bibnamefont {Titum}},\ }\bibfield  {title}
  {\enquote {\bibinfo {title} {{Impact of correlations and heavy tails on
  quantum error correction}},}\ }\href {\doibase 10.1103/PhysRevA.103.052428}
  {\bibfield  {journal} {\bibinfo  {journal} {Physical Review A}\ }\textbf
  {\bibinfo {volume} {103}},\ \bibinfo {pages} {052428} (\bibinfo {year}
  {2021})},\ \Eprint {http://arxiv.org/abs/2101.11631} {arXiv:2101.11631}
  \BibitemShut {NoStop}%
\bibitem [{\citenamefont {Nickerson}\ and\ \citenamefont
  {Brown}(2019)}]{correlated-qec}%
  \BibitemOpen
  \bibfield  {author} {\bibinfo {author} {\bibfnamefont {Naomi~H.}\
  \bibnamefont {Nickerson}}\ and\ \bibinfo {author} {\bibfnamefont
  {Benjamin~J.}\ \bibnamefont {Brown}},\ }\bibfield  {title} {\enquote
  {\bibinfo {title} {{Analysing correlated noise on the surface code using
  adaptive decoding algorithms}},}\ }\href {\doibase 10.22331/q-2019-04-08-131}
  {\bibfield  {journal} {\bibinfo  {journal} {Quantum}\ }\textbf {\bibinfo
  {volume} {3}},\ \bibinfo {pages} {131} (\bibinfo {year} {2019})},\ \Eprint
  {http://arxiv.org/abs/1712.00502} {arXiv:1712.00502} \BibitemShut {NoStop}%
\bibitem [{\citenamefont {White}\ \emph {et~al.}(2020)\citenamefont {White},
  \citenamefont {Hill}, \citenamefont {Pollock}, \citenamefont {Hollenberg},\
  and\ \citenamefont {Modi}}]{White-NM-2020}%
  \BibitemOpen
  \bibfield  {author} {\bibinfo {author} {\bibfnamefont {G.~A.~L.}\
  \bibnamefont {White}}, \bibinfo {author} {\bibfnamefont {C.~D.}\ \bibnamefont
  {Hill}}, \bibinfo {author} {\bibfnamefont {F.~A.}\ \bibnamefont {Pollock}},
  \bibinfo {author} {\bibfnamefont {L.~C.~L.}\ \bibnamefont {Hollenberg}}, \
  and\ \bibinfo {author} {\bibfnamefont {K.}~\bibnamefont {Modi}},\ }\bibfield
  {title} {\enquote {\bibinfo {title} {{Demonstration of non-Markovian process
  characterisation and control on a quantum processor}},}\ }\href {\doibase
  10.1038/s41467-020-20113-3} {\bibfield  {journal} {\bibinfo  {journal}
  {Nature Communications}\ }\textbf {\bibinfo {volume} {11}},\ \bibinfo {pages}
  {6301} (\bibinfo {year} {2020})},\ \Eprint {http://arxiv.org/abs/2004.14018}
  {arXiv:2004.14018} \BibitemShut {NoStop}%
\bibitem [{\citenamefont {Nielsen}\ \emph
  {et~al.}(2020{\natexlab{a}})\citenamefont {Nielsen}, \citenamefont {Gamble},
  \citenamefont {Rudinger}, \citenamefont {Scholten}, \citenamefont {Young},\
  and\ \citenamefont {Blume-Kohout}}]{nielsen-gst}%
  \BibitemOpen
  \bibfield  {author} {\bibinfo {author} {\bibfnamefont {Erik}\ \bibnamefont
  {Nielsen}}, \bibinfo {author} {\bibfnamefont {John~King}\ \bibnamefont
  {Gamble}}, \bibinfo {author} {\bibfnamefont {Kenneth}\ \bibnamefont
  {Rudinger}}, \bibinfo {author} {\bibfnamefont {Travis}\ \bibnamefont
  {Scholten}}, \bibinfo {author} {\bibfnamefont {Kevin}\ \bibnamefont {Young}},
  \ and\ \bibinfo {author} {\bibfnamefont {Robin}\ \bibnamefont
  {Blume-Kohout}},\ }\bibfield  {title} {\enquote {\bibinfo {title} {{Gate set
  tomography}},}\ }\href {http://arxiv.org/abs/2009.07301} {\bibfield
  {journal} {\bibinfo  {journal} {arXiv:2009.07301}\ } (\bibinfo {year}
  {2020}{\natexlab{a}})}\BibitemShut {NoStop}%
\bibitem [{\citenamefont {Sarovar}\ \emph {et~al.}(2020)\citenamefont
  {Sarovar}, \citenamefont {Proctor}, \citenamefont {Rudinger}, \citenamefont
  {Young}, \citenamefont {Nielsen},\ and\ \citenamefont
  {Blume-Kohout}}]{Sarovar2020detectingcrosstalk}%
  \BibitemOpen
  \bibfield  {author} {\bibinfo {author} {\bibfnamefont {Mohan}\ \bibnamefont
  {Sarovar}}, \bibinfo {author} {\bibfnamefont {Timothy}\ \bibnamefont
  {Proctor}}, \bibinfo {author} {\bibfnamefont {Kenneth}\ \bibnamefont
  {Rudinger}}, \bibinfo {author} {\bibfnamefont {Kevin}\ \bibnamefont {Young}},
  \bibinfo {author} {\bibfnamefont {Erik}\ \bibnamefont {Nielsen}}, \ and\
  \bibinfo {author} {\bibfnamefont {Robin}\ \bibnamefont {Blume-Kohout}},\
  }\bibfield  {title} {\enquote {\bibinfo {title} {Detecting crosstalk errors
  in quantum information processors},}\ }\href {\doibase
  10.22331/q-2020-09-11-321} {\bibfield  {journal} {\bibinfo  {journal}
  {{Quantum}}\ }\textbf {\bibinfo {volume} {4}},\ \bibinfo {pages} {321}
  (\bibinfo {year} {2020})}\BibitemShut {NoStop}%
\bibitem [{\citenamefont {Rudinger}\ \emph {et~al.}(2019)\citenamefont
  {Rudinger}, \citenamefont {Proctor}, \citenamefont {Langharst}, \citenamefont
  {Sarovar}, \citenamefont {Young},\ and\ \citenamefont
  {Blume-Kohout}}]{PhysRevX.9.021045}%
  \BibitemOpen
  \bibfield  {author} {\bibinfo {author} {\bibfnamefont {Kenneth}\ \bibnamefont
  {Rudinger}}, \bibinfo {author} {\bibfnamefont {Timothy}\ \bibnamefont
  {Proctor}}, \bibinfo {author} {\bibfnamefont {Dylan}\ \bibnamefont
  {Langharst}}, \bibinfo {author} {\bibfnamefont {Mohan}\ \bibnamefont
  {Sarovar}}, \bibinfo {author} {\bibfnamefont {Kevin}\ \bibnamefont {Young}},
  \ and\ \bibinfo {author} {\bibfnamefont {Robin}\ \bibnamefont
  {Blume-Kohout}},\ }\bibfield  {title} {\enquote {\bibinfo {title} {Probing
  context-dependent errors in quantum processors},}\ }\href {\doibase
  10.1103/PhysRevX.9.021045} {\bibfield  {journal} {\bibinfo  {journal} {Phys.
  Rev. X}\ }\textbf {\bibinfo {volume} {9}},\ \bibinfo {pages} {021045}
  (\bibinfo {year} {2019})}\BibitemShut {NoStop}%
\bibitem [{\citenamefont {Veitia}\ \emph {et~al.}(2020)\citenamefont {Veitia},
  \citenamefont {da~Silva}, \citenamefont {Blume-Kohout},\ and\ \citenamefont
  {van Enk}}]{veitia2020macroscopic}%
  \BibitemOpen
  \bibfield  {author} {\bibinfo {author} {\bibfnamefont {Andrzej}\ \bibnamefont
  {Veitia}}, \bibinfo {author} {\bibfnamefont {Marcus~P}\ \bibnamefont
  {da~Silva}}, \bibinfo {author} {\bibfnamefont {Robin}\ \bibnamefont
  {Blume-Kohout}}, \ and\ \bibinfo {author} {\bibfnamefont {Steven~J}\
  \bibnamefont {van Enk}},\ }\bibfield  {title} {\enquote {\bibinfo {title}
  {Macroscopic instructions vs microscopic operations in quantum circuits},}\
  }\href@noop {} {\bibfield  {journal} {\bibinfo  {journal} {Physics Letters
  A}\ }\textbf {\bibinfo {volume} {384}},\ \bibinfo {pages} {126131} (\bibinfo
  {year} {2020})}\BibitemShut {NoStop}%
\bibitem [{\citenamefont {Veitia}\ and\ \citenamefont {van
  Enk}(2018)}]{veitia2018testing}%
  \BibitemOpen
  \bibfield  {author} {\bibinfo {author} {\bibfnamefont {Andrzej}\ \bibnamefont
  {Veitia}}\ and\ \bibinfo {author} {\bibfnamefont {Steven~J}\ \bibnamefont
  {van Enk}},\ }\bibfield  {title} {\enquote {\bibinfo {title} {Testing the
  context-independence of quantum gates},}\ }\href@noop {} {\  (\bibinfo {year}
  {2018})},\ \Eprint {http://arxiv.org/abs/1810.05945} {arXiv:1810.05945}
  \BibitemShut {NoStop}%
\bibitem [{\citenamefont {Helsen}\ \emph {et~al.}(2019)\citenamefont {Helsen},
  \citenamefont {Battistel},\ and\ \citenamefont
  {Terhal}}]{helsen2019spectral}%
  \BibitemOpen
  \bibfield  {author} {\bibinfo {author} {\bibfnamefont {Jonas}\ \bibnamefont
  {Helsen}}, \bibinfo {author} {\bibfnamefont {Francesco}\ \bibnamefont
  {Battistel}}, \ and\ \bibinfo {author} {\bibfnamefont {Barbara~M}\
  \bibnamefont {Terhal}},\ }\bibfield  {title} {\enquote {\bibinfo {title}
  {Spectral quantum tomography},}\ }\href@noop {} {\bibfield  {journal}
  {\bibinfo  {journal} {npj Quantum Information}\ }\textbf {\bibinfo {volume}
  {5}},\ \bibinfo {pages} {74} (\bibinfo {year} {2019})}\BibitemShut {NoStop}%
\bibitem [{\citenamefont {Pollock}\ \emph
  {et~al.}(2018{\natexlab{a}})\citenamefont {Pollock}, \citenamefont
  {Rodr{\'{i}}guez-Rosario}, \citenamefont {Frauenheim}, \citenamefont
  {Paternostro},\ and\ \citenamefont {Modi}}]{Pollock2018a}%
  \BibitemOpen
  \bibfield  {author} {\bibinfo {author} {\bibfnamefont {Felix~A.}\
  \bibnamefont {Pollock}}, \bibinfo {author} {\bibfnamefont {C{\'{e}}sar}\
  \bibnamefont {Rodr{\'{i}}guez-Rosario}}, \bibinfo {author} {\bibfnamefont
  {Thomas}\ \bibnamefont {Frauenheim}}, \bibinfo {author} {\bibfnamefont
  {Mauro}\ \bibnamefont {Paternostro}}, \ and\ \bibinfo {author} {\bibfnamefont
  {Kavan}\ \bibnamefont {Modi}},\ }\bibfield  {title} {\enquote {\bibinfo
  {title} {{Non-Markovian quantum processes: Complete framework and efficient
  characterization}},}\ }\href {\doibase 10.1103/PhysRevA.97.012127} {\bibfield
   {journal} {\bibinfo  {journal} {Physical Review A}\ }\textbf {\bibinfo
  {volume} {97}},\ \bibinfo {pages} {012127} (\bibinfo {year}
  {2018}{\natexlab{a}})},\ \Eprint {http://arxiv.org/abs/1512.00589}
  {arXiv:1512.00589} \BibitemShut {NoStop}%
\bibitem [{\citenamefont {Taranto}\ \emph
  {et~al.}(2019{\natexlab{a}})\citenamefont {Taranto}, \citenamefont {Pollock},
  \citenamefont {Milz}, \citenamefont {Tomamichel},\ and\ \citenamefont
  {Modi}}]{taranto1}%
  \BibitemOpen
  \bibfield  {author} {\bibinfo {author} {\bibfnamefont {Philip}\ \bibnamefont
  {Taranto}}, \bibinfo {author} {\bibfnamefont {Felix~A}\ \bibnamefont
  {Pollock}}, \bibinfo {author} {\bibfnamefont {Simon}\ \bibnamefont {Milz}},
  \bibinfo {author} {\bibfnamefont {Marco}\ \bibnamefont {Tomamichel}}, \ and\
  \bibinfo {author} {\bibfnamefont {Kavan}\ \bibnamefont {Modi}},\ }\bibfield
  {title} {\enquote {\bibinfo {title} {{Quantum Markov Order}},}\ }\href
  {\doibase 10.1103/PhysRevLett.122.140401} {\bibfield  {journal} {\bibinfo
  {journal} {Physical Review Letters}\ }\textbf {\bibinfo {volume} {122}},\
  \bibinfo {pages} {140401} (\bibinfo {year} {2019}{\natexlab{a}})}\BibitemShut
  {NoStop}%
\bibitem [{\citenamefont {Cramer}\ \emph {et~al.}(2010)\citenamefont {Cramer},
  \citenamefont {Plenio}, \citenamefont {Flammia}, \citenamefont {Somma},
  \citenamefont {Gross}, \citenamefont {Bartlett}, \citenamefont
  {Landon-Cardinal}, \citenamefont {Poulin},\ and\ \citenamefont
  {Liu}}]{cramer2010efficient}%
  \BibitemOpen
  \bibfield  {author} {\bibinfo {author} {\bibfnamefont {Marcus}\ \bibnamefont
  {Cramer}}, \bibinfo {author} {\bibfnamefont {Martin~B}\ \bibnamefont
  {Plenio}}, \bibinfo {author} {\bibfnamefont {Steven~T}\ \bibnamefont
  {Flammia}}, \bibinfo {author} {\bibfnamefont {Rolando}\ \bibnamefont
  {Somma}}, \bibinfo {author} {\bibfnamefont {David}\ \bibnamefont {Gross}},
  \bibinfo {author} {\bibfnamefont {Stephen~D}\ \bibnamefont {Bartlett}},
  \bibinfo {author} {\bibfnamefont {Olivier}\ \bibnamefont {Landon-Cardinal}},
  \bibinfo {author} {\bibfnamefont {David}\ \bibnamefont {Poulin}}, \ and\
  \bibinfo {author} {\bibfnamefont {Yi-Kai}\ \bibnamefont {Liu}},\ }\bibfield
  {title} {\enquote {\bibinfo {title} {Efficient quantum state tomography},}\
  }\href {\doibase https://doi.org/10.1038/ncomms1147} {\bibfield  {journal}
  {\bibinfo  {journal} {Nature Communications}\ }\textbf {\bibinfo {volume}
  {1}},\ \bibinfo {pages} {149} (\bibinfo {year} {2010})},\ \Eprint
  {http://arxiv.org/abs/1101.4366} {arXiv:1101.4366} \BibitemShut {NoStop}%
\bibitem [{\citenamefont {Cygorek}\ \emph {et~al.}(2022)\citenamefont
  {Cygorek}, \citenamefont {Cosacchi}, \citenamefont {Vagov}, \citenamefont
  {Axt}, \citenamefont {Lovett}, \citenamefont {Keeling},\ and\ \citenamefont
  {Gauger}}]{cygorek2022simulation}%
  \BibitemOpen
  \bibfield  {author} {\bibinfo {author} {\bibfnamefont {Moritz}\ \bibnamefont
  {Cygorek}}, \bibinfo {author} {\bibfnamefont {Michael}\ \bibnamefont
  {Cosacchi}}, \bibinfo {author} {\bibfnamefont {Alexei}\ \bibnamefont
  {Vagov}}, \bibinfo {author} {\bibfnamefont {Vollrath~Martin}\ \bibnamefont
  {Axt}}, \bibinfo {author} {\bibfnamefont {Brendon~W}\ \bibnamefont {Lovett}},
  \bibinfo {author} {\bibfnamefont {Jonathan}\ \bibnamefont {Keeling}}, \ and\
  \bibinfo {author} {\bibfnamefont {Erik~M}\ \bibnamefont {Gauger}},\
  }\bibfield  {title} {\enquote {\bibinfo {title} {Simulation of open quantum
  systems by automated compression of arbitrary environments},}\ }\href@noop {}
  {\bibfield  {journal} {\bibinfo  {journal} {Nature Physics}\ } (\bibinfo
  {year} {2022})}\BibitemShut {NoStop}%
\bibitem [{\citenamefont {Dang}\ \emph {et~al.}(2021)\citenamefont {Dang},
  \citenamefont {White}, \citenamefont {Hollenberg},\ and\ \citenamefont
  {Hill}}]{dang2021process}%
  \BibitemOpen
  \bibfield  {author} {\bibinfo {author} {\bibfnamefont {Aidan}\ \bibnamefont
  {Dang}}, \bibinfo {author} {\bibfnamefont {Gregory~AL}\ \bibnamefont
  {White}}, \bibinfo {author} {\bibfnamefont {Lloyd~CL}\ \bibnamefont
  {Hollenberg}}, \ and\ \bibinfo {author} {\bibfnamefont {Charles~D}\
  \bibnamefont {Hill}},\ }\bibfield  {title} {\enquote {\bibinfo {title}
  {Process tomography on a 7-qubit quantum processor via tensor network
  contraction path finding},}\ }\href@noop {} {\bibfield  {journal} {\bibinfo
  {journal} {arXiv preprint arXiv:2112.06364}\ } (\bibinfo {year}
  {2021})}\BibitemShut {NoStop}%
\bibitem [{\citenamefont {Baumgratz}\ \emph {et~al.}(2013)\citenamefont
  {Baumgratz}, \citenamefont {Gross}, \citenamefont {Cramer},\ and\
  \citenamefont {Plenio}}]{PhysRevLett.111.020401}%
  \BibitemOpen
  \bibfield  {author} {\bibinfo {author} {\bibfnamefont {T.}~\bibnamefont
  {Baumgratz}}, \bibinfo {author} {\bibfnamefont {D.}~\bibnamefont {Gross}},
  \bibinfo {author} {\bibfnamefont {M.}~\bibnamefont {Cramer}}, \ and\ \bibinfo
  {author} {\bibfnamefont {M.~B.}\ \bibnamefont {Plenio}},\ }\bibfield  {title}
  {\enquote {\bibinfo {title} {Scalable reconstruction of density matrices},}\
  }\href {\doibase 10.1103/PhysRevLett.111.020401} {\bibfield  {journal}
  {\bibinfo  {journal} {Phys. Rev. Lett.}\ }\textbf {\bibinfo {volume} {111}},\
  \bibinfo {pages} {020401} (\bibinfo {year} {2013})}\BibitemShut {NoStop}%
\bibitem [{\citenamefont {Blume-Kohout}\ \emph {et~al.}(2020)\citenamefont
  {Blume-Kohout}, \citenamefont {Rudinger}, \citenamefont {Nielsen},
  \citenamefont {Proctor},\ and\ \citenamefont {Young}}]{RBH-wildcard}%
  \BibitemOpen
  \bibfield  {author} {\bibinfo {author} {\bibfnamefont {Robin}\ \bibnamefont
  {Blume-Kohout}}, \bibinfo {author} {\bibfnamefont {Kenneth}\ \bibnamefont
  {Rudinger}}, \bibinfo {author} {\bibfnamefont {Erik}\ \bibnamefont
  {Nielsen}}, \bibinfo {author} {\bibfnamefont {Timothy}\ \bibnamefont
  {Proctor}}, \ and\ \bibinfo {author} {\bibfnamefont {Kevin}\ \bibnamefont
  {Young}},\ }\bibfield  {title} {\enquote {\bibinfo {title} {{Wildcard error:
  Quantifying unmodeled errors in quantum processors}},}\ }\href
  {http://arxiv.org/abs/2012.12231} {\bibfield  {journal} {\bibinfo  {journal}
  {arxiv:2012.12231}\ } (\bibinfo {year} {2020})}\BibitemShut {NoStop}%
\bibitem [{\citenamefont {Knill}\ \emph {et~al.}(2008)\citenamefont {Knill},
  \citenamefont {Leibfried}, \citenamefont {Reichle}, \citenamefont {Britton},
  \citenamefont {Blakestad}, \citenamefont {Jost}, \citenamefont {Langer},
  \citenamefont {Ozeri}, \citenamefont {Seidelin},\ and\ \citenamefont
  {Wineland}}]{PhysRevA.77.012307}%
  \BibitemOpen
  \bibfield  {author} {\bibinfo {author} {\bibfnamefont {E.}~\bibnamefont
  {Knill}}, \bibinfo {author} {\bibfnamefont {D.}~\bibnamefont {Leibfried}},
  \bibinfo {author} {\bibfnamefont {R.}~\bibnamefont {Reichle}}, \bibinfo
  {author} {\bibfnamefont {J.}~\bibnamefont {Britton}}, \bibinfo {author}
  {\bibfnamefont {R.~B.}\ \bibnamefont {Blakestad}}, \bibinfo {author}
  {\bibfnamefont {J.~D.}\ \bibnamefont {Jost}}, \bibinfo {author}
  {\bibfnamefont {C.}~\bibnamefont {Langer}}, \bibinfo {author} {\bibfnamefont
  {R.}~\bibnamefont {Ozeri}}, \bibinfo {author} {\bibfnamefont
  {S.}~\bibnamefont {Seidelin}}, \ and\ \bibinfo {author} {\bibfnamefont
  {D.~J.}\ \bibnamefont {Wineland}},\ }\bibfield  {title} {\enquote {\bibinfo
  {title} {Randomized benchmarking of quantum gates},}\ }\href {\doibase
  10.1103/PhysRevA.77.012307} {\bibfield  {journal} {\bibinfo  {journal} {Phys.
  Rev. A}\ }\textbf {\bibinfo {volume} {77}},\ \bibinfo {pages} {012307}
  (\bibinfo {year} {2008})}\BibitemShut {NoStop}%
\bibitem [{\citenamefont {Zhang}\ and\ \citenamefont
  {Sarovar}(2014)}]{PhysRevLett.113.080401}%
  \BibitemOpen
  \bibfield  {author} {\bibinfo {author} {\bibfnamefont {Jun}\ \bibnamefont
  {Zhang}}\ and\ \bibinfo {author} {\bibfnamefont {Mohan}\ \bibnamefont
  {Sarovar}},\ }\bibfield  {title} {\enquote {\bibinfo {title} {Quantum
  hamiltonian identification from measurement time traces},}\ }\href {\doibase
  10.1103/PhysRevLett.113.080401} {\bibfield  {journal} {\bibinfo  {journal}
  {Phys. Rev. Lett.}\ }\textbf {\bibinfo {volume} {113}},\ \bibinfo {pages}
  {080401} (\bibinfo {year} {2014})}\BibitemShut {NoStop}%
\bibitem [{\citenamefont {Wang}\ \emph {et~al.}(2015)\citenamefont {Wang},
  \citenamefont {Deng},\ and\ \citenamefont {Duan}}]{Wang_2015}%
  \BibitemOpen
  \bibfield  {author} {\bibinfo {author} {\bibfnamefont {Sheng-Tao}\
  \bibnamefont {Wang}}, \bibinfo {author} {\bibfnamefont {Dong-Ling}\
  \bibnamefont {Deng}}, \ and\ \bibinfo {author} {\bibfnamefont {L-M}\
  \bibnamefont {Duan}},\ }\bibfield  {title} {\enquote {\bibinfo {title}
  {Hamiltonian tomography for quantum many-body systems with arbitrary
  couplings},}\ }\href {\doibase 10.1088/1367-2630/17/9/093017} {\bibfield
  {journal} {\bibinfo  {journal} {New Journal of Physics}\ }\textbf {\bibinfo
  {volume} {17}},\ \bibinfo {pages} {093017} (\bibinfo {year}
  {2015})}\BibitemShut {NoStop}%
\bibitem [{\citenamefont {Pollock}\ and\ \citenamefont
  {Modi}(2018)}]{pollock-tomographic-equations}%
  \BibitemOpen
  \bibfield  {author} {\bibinfo {author} {\bibfnamefont {Felix~A.}\
  \bibnamefont {Pollock}}\ and\ \bibinfo {author} {\bibfnamefont {Kavan}\
  \bibnamefont {Modi}},\ }\bibfield  {title} {\enquote {\bibinfo {title}
  {{Tomographically reconstructed master equations for any open quantum
  dynamics}},}\ }\href {\doibase 10.22331/q-2018-07-11-76} {\bibfield
  {journal} {\bibinfo  {journal} {Quantum}\ }\textbf {\bibinfo {volume} {2}},\
  \bibinfo {pages} {76} (\bibinfo {year} {2018})},\ \Eprint
  {http://arxiv.org/abs/1704.06204} {arXiv:1704.06204} \BibitemShut {NoStop}%
\bibitem [{\citenamefont {Lorenzo}\ \emph {et~al.}(2016)\citenamefont
  {Lorenzo}, \citenamefont {Ciccarello},\ and\ \citenamefont
  {Palma}}]{PhysRevA.93.052111}%
  \BibitemOpen
  \bibfield  {author} {\bibinfo {author} {\bibfnamefont {Salvatore}\
  \bibnamefont {Lorenzo}}, \bibinfo {author} {\bibfnamefont {Francesco}\
  \bibnamefont {Ciccarello}}, \ and\ \bibinfo {author} {\bibfnamefont
  {G.~Massimo}\ \bibnamefont {Palma}},\ }\bibfield  {title} {\enquote {\bibinfo
  {title} {Class of exact memory-kernel master equations},}\ }\href {\doibase
  10.1103/PhysRevA.93.052111} {\bibfield  {journal} {\bibinfo  {journal} {Phys.
  Rev. A}\ }\textbf {\bibinfo {volume} {93}},\ \bibinfo {pages} {052111}
  (\bibinfo {year} {2016})}\BibitemShut {NoStop}%
\bibitem [{\citenamefont {Flammia}\ \emph {et~al.}(2012)\citenamefont
  {Flammia}, \citenamefont {Gross}, \citenamefont {Liu},\ and\ \citenamefont
  {Eisert}}]{flammia2012quantum}%
  \BibitemOpen
  \bibfield  {author} {\bibinfo {author} {\bibfnamefont {Steven~T}\
  \bibnamefont {Flammia}}, \bibinfo {author} {\bibfnamefont {David}\
  \bibnamefont {Gross}}, \bibinfo {author} {\bibfnamefont {Yi-Kai}\
  \bibnamefont {Liu}}, \ and\ \bibinfo {author} {\bibfnamefont {Jens}\
  \bibnamefont {Eisert}},\ }\bibfield  {title} {\enquote {\bibinfo {title}
  {Quantum tomography via compressed sensing: error bounds, sample complexity
  and efficient estimators},}\ }\href {\doibase
  https://doi.org/10.1088/1367-2630/14/9/095022} {\bibfield  {journal}
  {\bibinfo  {journal} {New Journal of Physics}\ }\textbf {\bibinfo {volume}
  {14}},\ \bibinfo {pages} {095022} (\bibinfo {year} {2012})},\ \Eprint
  {http://arxiv.org/abs/1205.2300} {arXiv:1205.2300} \BibitemShut {NoStop}%
\bibitem [{\citenamefont {Blume-Kohout}(2010)}]{RBK2010}%
  \BibitemOpen
  \bibfield  {author} {\bibinfo {author} {\bibfnamefont {Robin}\ \bibnamefont
  {Blume-Kohout}},\ }\bibfield  {title} {\enquote {\bibinfo {title} {Optimal,
  reliable estimation of quantum states},}\ }\href {\doibase
  10.1088/1367-2630/12/4/043034} {\bibfield  {journal} {\bibinfo  {journal}
  {New Journal of Physics}\ }\textbf {\bibinfo {volume} {12}},\ \bibinfo
  {pages} {043034} (\bibinfo {year} {2010})},\ \Eprint
  {http://arxiv.org/abs/0611080} {arXiv:0611080} \BibitemShut {NoStop}%
\bibitem [{\citenamefont {Merkel}\ \emph {et~al.}(2013)\citenamefont {Merkel},
  \citenamefont {Gambetta}, \citenamefont {Smolin}, \citenamefont {Poletto},
  \citenamefont {C{\'{o}}rcoles}, \citenamefont {Johnson}, \citenamefont
  {Ryan},\ and\ \citenamefont {Steffen}}]{PhysRevA.87.062119}%
  \BibitemOpen
  \bibfield  {author} {\bibinfo {author} {\bibfnamefont {Seth~T}\ \bibnamefont
  {Merkel}}, \bibinfo {author} {\bibfnamefont {Jay~M}\ \bibnamefont
  {Gambetta}}, \bibinfo {author} {\bibfnamefont {John~A}\ \bibnamefont
  {Smolin}}, \bibinfo {author} {\bibfnamefont {Stefano}\ \bibnamefont
  {Poletto}}, \bibinfo {author} {\bibfnamefont {Antonio~D}\ \bibnamefont
  {C{\'{o}}rcoles}}, \bibinfo {author} {\bibfnamefont {Blake~R}\ \bibnamefont
  {Johnson}}, \bibinfo {author} {\bibfnamefont {Colm~A}\ \bibnamefont {Ryan}},
  \ and\ \bibinfo {author} {\bibfnamefont {Matthias}\ \bibnamefont {Steffen}},\
  }\bibfield  {title} {\enquote {\bibinfo {title} {{Self-consistent quantum
  process tomography}},}\ }\href {\doibase 10.1103/PhysRevA.87.062119}
  {\bibfield  {journal} {\bibinfo  {journal} {Physical Review A}\ }\textbf
  {\bibinfo {volume} {87}},\ \bibinfo {pages} {62119} (\bibinfo {year}
  {2013})}\BibitemShut {NoStop}%
\bibitem [{\citenamefont {Blume-Kohout}\ \emph {et~al.}(2013)\citenamefont
  {Blume-Kohout}, \citenamefont {Gamble}, \citenamefont {Nielsen},
  \citenamefont {Mizrahi}, \citenamefont {Sterk},\ and\ \citenamefont
  {Maunz}}]{gst-2013}%
  \BibitemOpen
  \bibfield  {author} {\bibinfo {author} {\bibfnamefont {Robin}\ \bibnamefont
  {Blume-Kohout}}, \bibinfo {author} {\bibfnamefont {John}\ \bibnamefont
  {Gamble}}, \bibinfo {author} {\bibfnamefont {Erik}\ \bibnamefont {Nielsen}},
  \bibinfo {author} {\bibfnamefont {Jonathan}\ \bibnamefont {Mizrahi}},
  \bibinfo {author} {\bibfnamefont {Jonathan}\ \bibnamefont {Sterk}}, \ and\
  \bibinfo {author} {\bibfnamefont {Peter}\ \bibnamefont {Maunz}},\ }\bibfield
  {title} {\enquote {\bibinfo {title} {{Robust, self-consistent, closed-form
  tomography of quantum logic gates on a trapped ion qubit}},}\ }\href
  {https://arxiv.org/abs/1310.4492} {\bibfield  {journal} {\bibinfo  {journal}
  {arXiv:1310.4492}\ } (\bibinfo {year} {2013})}\BibitemShut {NoStop}%
\bibitem [{\citenamefont {Greenbaum}(2015)}]{intro-GST}%
  \BibitemOpen
  \bibfield  {author} {\bibinfo {author} {\bibfnamefont {Daniel}\ \bibnamefont
  {Greenbaum}},\ }\bibfield  {title} {\enquote {\bibinfo {title} {{Introduction
  to Quantum Gate Set Tomography}},}\ }\href {http://arxiv.org/abs/1509.02921}
  {\bibfield  {journal} {\bibinfo  {journal} {arXiv:1509.02921}\ } (\bibinfo
  {year} {2015})}\BibitemShut {NoStop}%
\bibitem [{\citenamefont {Hradil}\ \emph {et~al.}(2004)\citenamefont {Hradil},
  \citenamefont {{\v{R}}eh{\'a}{\v{c}}ek}, \citenamefont {Fiur{\'a}{\v{s}}ek},\
  and\ \citenamefont {Je{\v{z}}ek}}]{Hradil2004}%
  \BibitemOpen
  \bibfield  {author} {\bibinfo {author} {\bibfnamefont {Zden{\v{e}}k}\
  \bibnamefont {Hradil}}, \bibinfo {author} {\bibfnamefont {Jaroslav}\
  \bibnamefont {{\v{R}}eh{\'a}{\v{c}}ek}}, \bibinfo {author} {\bibfnamefont
  {Jarom{\'i}r}\ \bibnamefont {Fiur{\'a}{\v{s}}ek}}, \ and\ \bibinfo {author}
  {\bibfnamefont {Miroslav}\ \bibnamefont {Je{\v{z}}ek}},\ }\enquote {\bibinfo
  {title} {Maximum-likelihood methods in quantum mechanics},}\ in\ \href
  {\doibase 10.1007/978-3-540-44481-7_3} {\emph {\bibinfo {booktitle} {Quantum
  State Estimation}}},\ \bibinfo {editor} {edited by\ \bibinfo {editor}
  {\bibfnamefont {Matteo}\ \bibnamefont {Paris}}\ and\ \bibinfo {editor}
  {\bibfnamefont {Jaroslav}\ \bibnamefont {{\v{R}}eh{\'a}{\v{c}}ek}}}\
  (\bibinfo  {publisher} {Springer Berlin Heidelberg},\ \bibinfo {address}
  {Berlin, Heidelberg},\ \bibinfo {year} {2004})\ pp.\ \bibinfo {pages}
  {59--112}\BibitemShut {NoStop}%
\bibitem [{\citenamefont {Costa}\ and\ \citenamefont
  {Shrapnel}(2016)}]{1367-2630-18-6-063032}%
  \BibitemOpen
  \bibfield  {author} {\bibinfo {author} {\bibfnamefont {Fabio}\ \bibnamefont
  {Costa}}\ and\ \bibinfo {author} {\bibfnamefont {Sally}\ \bibnamefont
  {Shrapnel}},\ }\bibfield  {title} {\enquote {\bibinfo {title} {Quantum causal
  modelling},}\ }\href {http://stacks.iop.org/1367-2630/18/i=6/a=063032}
  {\bibfield  {journal} {\bibinfo  {journal} {New Journal of Physics}\ }\textbf
  {\bibinfo {volume} {18}},\ \bibinfo {pages} {063032} (\bibinfo {year}
  {2016})}\BibitemShut {NoStop}%
\bibitem [{\citenamefont {Milz}\ \emph {et~al.}(2020)\citenamefont {Milz},
  \citenamefont {Sakuldee}, \citenamefont {Pollock},\ and\ \citenamefont
  {Modi}}]{Milz2020}%
  \BibitemOpen
  \bibfield  {author} {\bibinfo {author} {\bibfnamefont {Simon}\ \bibnamefont
  {Milz}}, \bibinfo {author} {\bibfnamefont {Fattah}\ \bibnamefont {Sakuldee}},
  \bibinfo {author} {\bibfnamefont {Felix~A.}\ \bibnamefont {Pollock}}, \ and\
  \bibinfo {author} {\bibfnamefont {Kavan}\ \bibnamefont {Modi}},\ }\bibfield
  {title} {\enquote {\bibinfo {title} {{Kolmogorov extension theorem for
  (quantum) causal modelling and general probabilistic theories}},}\ }\href
  {\doibase 10.22331/q-2020-04-20-255} {\bibfield  {journal} {\bibinfo
  {journal} {Quantum}\ }\textbf {\bibinfo {volume} {4}},\ \bibinfo {pages}
  {255} (\bibinfo {year} {2020})},\ \Eprint {http://arxiv.org/abs/1712.02589}
  {arXiv:1712.02589} \BibitemShut {NoStop}%
\bibitem [{\citenamefont {Chiribella}\ \emph {et~al.}(2008)\citenamefont
  {Chiribella}, \citenamefont {D'Ariano},\ and\ \citenamefont
  {Perinotti}}]{chiribella_memory_2008}%
  \BibitemOpen
  \bibfield  {author} {\bibinfo {author} {\bibfnamefont {Giulio}\ \bibnamefont
  {Chiribella}}, \bibinfo {author} {\bibfnamefont {Giacomo~M.}\ \bibnamefont
  {D'Ariano}}, \ and\ \bibinfo {author} {\bibfnamefont {Paolo}\ \bibnamefont
  {Perinotti}},\ }\bibfield  {title} {\enquote {\bibinfo {title} {Memory
  {Effects} in {Quantum} {Channel} {Discrimination}},}\ }\href {\doibase
  10.1103/PhysRevLett.101.180501} {\bibfield  {journal} {\bibinfo  {journal}
  {Physical Review Letters}\ }\textbf {\bibinfo {volume} {101}},\ \bibinfo
  {pages} {180501} (\bibinfo {year} {2008})}\BibitemShut {NoStop}%
\bibitem [{\citenamefont {Shrapnel}\ \emph {et~al.}(2018)\citenamefont
  {Shrapnel}, \citenamefont {Costa},\ and\ \citenamefont
  {Milburn}}]{Shrapnel_2018}%
  \BibitemOpen
  \bibfield  {author} {\bibinfo {author} {\bibfnamefont {Sally}\ \bibnamefont
  {Shrapnel}}, \bibinfo {author} {\bibfnamefont {Fabio}\ \bibnamefont {Costa}},
  \ and\ \bibinfo {author} {\bibfnamefont {Gerard}\ \bibnamefont {Milburn}},\
  }\bibfield  {title} {\enquote {\bibinfo {title} {Updating the born rule},}\
  }\href {\doibase 10.1088/1367-2630/aabe12} {\bibfield  {journal} {\bibinfo
  {journal} {New Journal of Physics}\ }\textbf {\bibinfo {volume} {20}},\
  \bibinfo {pages} {053010} (\bibinfo {year} {2018})}\BibitemShut {NoStop}%
\bibitem [{\citenamefont {Wolf}\ \emph {et~al.}(2008)\citenamefont {Wolf},
  \citenamefont {Eisert}, \citenamefont {Cubitt},\ and\ \citenamefont
  {Cirac}}]{PhysRevLett.101.150402}%
  \BibitemOpen
  \bibfield  {author} {\bibinfo {author} {\bibfnamefont {M.~M.}\ \bibnamefont
  {Wolf}}, \bibinfo {author} {\bibfnamefont {J.}~\bibnamefont {Eisert}},
  \bibinfo {author} {\bibfnamefont {T.~S.}\ \bibnamefont {Cubitt}}, \ and\
  \bibinfo {author} {\bibfnamefont {J.~I.}\ \bibnamefont {Cirac}},\ }\bibfield
  {title} {\enquote {\bibinfo {title} {Assessing non-{Markovian} quantum
  dynamics},}\ }\href {\doibase 10.1103/PhysRevLett.101.150402} {\bibfield
  {journal} {\bibinfo  {journal} {Phys. Rev. Lett.}\ }\textbf {\bibinfo
  {volume} {101}},\ \bibinfo {pages} {150402} (\bibinfo {year}
  {2008})}\BibitemShut {NoStop}%
\bibitem [{\citenamefont {Breuer}\ \emph {et~al.}(2009)\citenamefont {Breuer},
  \citenamefont {Laine},\ and\ \citenamefont {Piilo}}]{PhysRevLett.103.210401}%
  \BibitemOpen
  \bibfield  {author} {\bibinfo {author} {\bibfnamefont {Heinz-Peter}\
  \bibnamefont {Breuer}}, \bibinfo {author} {\bibfnamefont {Elsi-Mari}\
  \bibnamefont {Laine}}, \ and\ \bibinfo {author} {\bibfnamefont {Jyrki}\
  \bibnamefont {Piilo}},\ }\bibfield  {title} {\enquote {\bibinfo {title}
  {{Measure for the Degree of Non-Markovian Behavior of Quantum Processes in
  Open Systems}},}\ }\href {\doibase 10.1103/PhysRevLett.103.210401} {\bibfield
   {journal} {\bibinfo  {journal} {Phys. Rev. Lett.}\ }\textbf {\bibinfo
  {volume} {103}},\ \bibinfo {pages} {210401} (\bibinfo {year}
  {2009})}\BibitemShut {NoStop}%
\bibitem [{\citenamefont {Rivas}\ \emph {et~al.}(2010)\citenamefont {Rivas},
  \citenamefont {Huelga},\ and\ \citenamefont
  {Plenio}}]{PhysRevLett.105.050403}%
  \BibitemOpen
  \bibfield  {author} {\bibinfo {author} {\bibfnamefont {\'Angel}\ \bibnamefont
  {Rivas}}, \bibinfo {author} {\bibfnamefont {Susana~F.}\ \bibnamefont
  {Huelga}}, \ and\ \bibinfo {author} {\bibfnamefont {Martin~B.}\ \bibnamefont
  {Plenio}},\ }\bibfield  {title} {\enquote {\bibinfo {title} {Entanglement and
  non-markovianity of quantum evolutions},}\ }\href {\doibase
  10.1103/PhysRevLett.105.050403} {\bibfield  {journal} {\bibinfo  {journal}
  {Phys. Rev. Lett.}\ }\textbf {\bibinfo {volume} {105}},\ \bibinfo {pages}
  {050403} (\bibinfo {year} {2010})}\BibitemShut {NoStop}%
\bibitem [{\citenamefont {Chru\ifmmode \acute{s}\else
  \'{s}\fi{}ci\ifmmode~\acute{n}\else \'{n}\fi{}ski}\ \emph
  {et~al.}(2011)\citenamefont {Chru\ifmmode \acute{s}\else
  \'{s}\fi{}ci\ifmmode~\acute{n}\else \'{n}\fi{}ski}, \citenamefont
  {Kossakowski},\ and\ \citenamefont {Rivas}}]{PhysRevA.83.052128}%
  \BibitemOpen
  \bibfield  {author} {\bibinfo {author} {\bibfnamefont {Dariusz}\ \bibnamefont
  {Chru\ifmmode \acute{s}\else \'{s}\fi{}ci\ifmmode~\acute{n}\else
  \'{n}\fi{}ski}}, \bibinfo {author} {\bibfnamefont {Andrzej}\ \bibnamefont
  {Kossakowski}}, \ and\ \bibinfo {author} {\bibfnamefont {\'Angel}\
  \bibnamefont {Rivas}},\ }\bibfield  {title} {\enquote {\bibinfo {title}
  {Measures of non-markovianity: Divisibility versus backflow of
  information},}\ }\href {\doibase 10.1103/PhysRevA.83.052128} {\bibfield
  {journal} {\bibinfo  {journal} {Phys. Rev. A}\ }\textbf {\bibinfo {volume}
  {83}},\ \bibinfo {pages} {052128} (\bibinfo {year} {2011})}\BibitemShut
  {NoStop}%
\bibitem [{\citenamefont {Vacchini}(2013)}]{vacchini_non-markovian_2013}%
  \BibitemOpen
  \bibfield  {author} {\bibinfo {author} {\bibfnamefont {Bassano}\ \bibnamefont
  {Vacchini}},\ }\bibfield  {title} {\enquote {\bibinfo {title}
  {Non-{Markovian} master equations from piecewise dynamics},}\ }\href
  {\doibase 10.1103/PhysRevA.87.030101} {\bibfield  {journal} {\bibinfo
  {journal} {Phys. Rev. A}\ }\textbf {\bibinfo {volume} {87}},\ \bibinfo
  {pages} {030101} (\bibinfo {year} {2013})}\BibitemShut {NoStop}%
\bibitem [{\citenamefont {Corcoles}\ \emph {et~al.}(2021)\citenamefont
  {Corcoles}, \citenamefont {Takita}, \citenamefont {Inoue}, \citenamefont
  {Lekuch}, \citenamefont {Minev}, \citenamefont {Chow},\ and\ \citenamefont
  {Gambetta}}]{Corcoles2021}%
  \BibitemOpen
  \bibfield  {author} {\bibinfo {author} {\bibfnamefont {Antonio~D.}\
  \bibnamefont {Corcoles}}, \bibinfo {author} {\bibfnamefont {Maika}\
  \bibnamefont {Takita}}, \bibinfo {author} {\bibfnamefont {Ken}\ \bibnamefont
  {Inoue}}, \bibinfo {author} {\bibfnamefont {Scott}\ \bibnamefont {Lekuch}},
  \bibinfo {author} {\bibfnamefont {Zlatko~K.}\ \bibnamefont {Minev}}, \bibinfo
  {author} {\bibfnamefont {Jerry~M.}\ \bibnamefont {Chow}}, \ and\ \bibinfo
  {author} {\bibfnamefont {Jay~M.}\ \bibnamefont {Gambetta}},\ }\bibfield
  {title} {\enquote {\bibinfo {title} {{Exploiting dynamic quantum circuits in
  a quantum algorithm with superconducting qubits}},}\ }\href
  {http://arxiv.org/abs/2102.01682} {\bibfield  {journal} {\bibinfo  {journal}
  {arxiv:2102.01682}\ } (\bibinfo {year} {2021})}\BibitemShut {NoStop}%
\bibitem [{\citenamefont {Xiang}\ \emph {et~al.}(2021)\citenamefont {Xiang},
  \citenamefont {Zong}, \citenamefont {Zhan}, \citenamefont {Fei},
  \citenamefont {Run}, \citenamefont {Wu}, \citenamefont {Jin}, \citenamefont
  {Xiao}, \citenamefont {Jia}, \citenamefont {Duan}, \citenamefont {Wu},
  \citenamefont {Yin},\ and\ \citenamefont {Guo}}]{Xiang2021}%
  \BibitemOpen
  \bibfield  {author} {\bibinfo {author} {\bibfnamefont {Liang}\ \bibnamefont
  {Xiang}}, \bibinfo {author} {\bibfnamefont {Zhiwen}\ \bibnamefont {Zong}},
  \bibinfo {author} {\bibfnamefont {Ze}~\bibnamefont {Zhan}}, \bibinfo {author}
  {\bibfnamefont {Ying}\ \bibnamefont {Fei}}, \bibinfo {author} {\bibfnamefont
  {Chongxin}\ \bibnamefont {Run}}, \bibinfo {author} {\bibfnamefont {Yaozu}\
  \bibnamefont {Wu}}, \bibinfo {author} {\bibfnamefont {Wenyan}\ \bibnamefont
  {Jin}}, \bibinfo {author} {\bibfnamefont {Cong}\ \bibnamefont {Xiao}},
  \bibinfo {author} {\bibfnamefont {Zhilong}\ \bibnamefont {Jia}}, \bibinfo
  {author} {\bibfnamefont {Peng}\ \bibnamefont {Duan}}, \bibinfo {author}
  {\bibfnamefont {Jianlan}\ \bibnamefont {Wu}}, \bibinfo {author}
  {\bibfnamefont {Yi}~\bibnamefont {Yin}}, \ and\ \bibinfo {author}
  {\bibfnamefont {Guoping}\ \bibnamefont {Guo}},\ }\bibfield  {title} {\enquote
  {\bibinfo {title} {{Quantify the Non-Markovian Process with Intermediate
  Projections in a Superconducting Processor}},}\ }\href
  {http://arxiv.org/abs/2105.03333} {\bibfield  {journal} {\bibinfo  {journal}
  {arXiv:2105.03333}\ } (\bibinfo {year} {2021})}\BibitemShut {NoStop}%
\bibitem [{\citenamefont {Milz}\ \emph {et~al.}(2018)\citenamefont {Milz},
  \citenamefont {Pollock},\ and\ \citenamefont {Modi}}]{PT-limited-control}%
  \BibitemOpen
  \bibfield  {author} {\bibinfo {author} {\bibfnamefont {Simon}\ \bibnamefont
  {Milz}}, \bibinfo {author} {\bibfnamefont {Felix~A.}\ \bibnamefont
  {Pollock}}, \ and\ \bibinfo {author} {\bibfnamefont {Kavan}\ \bibnamefont
  {Modi}},\ }\bibfield  {title} {\enquote {\bibinfo {title} {{Reconstructing
  non-Markovian quantum dynamics with limited control}},}\ }\href {\doibase
  10.1103/PhysRevA.98.012108} {\bibfield  {journal} {\bibinfo  {journal}
  {Physical Review A}\ }\textbf {\bibinfo {volume} {98}},\ \bibinfo {pages}
  {012108} (\bibinfo {year} {2018})},\ \Eprint
  {http://arxiv.org/abs/1610.02152} {arXiv:1610.02152} \BibitemShut {NoStop}%
\bibitem [{\citenamefont {Adamson}\ and\ \citenamefont
  {Steinberg}(2010)}]{PhysRevLett.105.030406}%
  \BibitemOpen
  \bibfield  {author} {\bibinfo {author} {\bibfnamefont {R.~B.~A.}\
  \bibnamefont {Adamson}}\ and\ \bibinfo {author} {\bibfnamefont {A.~M.}\
  \bibnamefont {Steinberg}},\ }\bibfield  {title} {\enquote {\bibinfo {title}
  {{Improving Quantum State Estimation with Mutually Unbiased Bases}},}\ }\href
  {\doibase 10.1103/PhysRevLett.105.030406} {\bibfield  {journal} {\bibinfo
  {journal} {Physical Review Letters}\ }\textbf {\bibinfo {volume} {105}},\
  \bibinfo {pages} {030406} (\bibinfo {year} {2010})}\BibitemShut {NoStop}%
\bibitem [{\citenamefont {McKay}\ \emph {et~al.}(2017)\citenamefont {McKay},
  \citenamefont {Wood}, \citenamefont {Sheldon}, \citenamefont {Chow},\ and\
  \citenamefont {Gambetta}}]{PhysRevA.96.022330}%
  \BibitemOpen
  \bibfield  {author} {\bibinfo {author} {\bibfnamefont {David~C}\ \bibnamefont
  {McKay}}, \bibinfo {author} {\bibfnamefont {Christopher~J}\ \bibnamefont
  {Wood}}, \bibinfo {author} {\bibfnamefont {Sarah}\ \bibnamefont {Sheldon}},
  \bibinfo {author} {\bibfnamefont {Jerry~M}\ \bibnamefont {Chow}}, \ and\
  \bibinfo {author} {\bibfnamefont {Jay~M}\ \bibnamefont {Gambetta}},\
  }\bibfield  {title} {\enquote {\bibinfo {title} {{Efficient {\$}Z{\$} gates
  for quantum computing}},}\ }\href {\doibase 10.1103/PhysRevA.96.022330}
  {\bibfield  {journal} {\bibinfo  {journal} {Phys. Rev. A}\ }\textbf {\bibinfo
  {volume} {96}},\ \bibinfo {pages} {22330} (\bibinfo {year} {2017})},\ \Eprint
  {http://arxiv.org/abs/1612.00858} {arXiv:1612.00858} \BibitemShut {NoStop}%
\bibitem [{\citenamefont {Knee}\ \emph {et~al.}(2018)\citenamefont {Knee},
  \citenamefont {Bolduc}, \citenamefont {Leach},\ and\ \citenamefont
  {Gauger}}]{QPT-projection}%
  \BibitemOpen
  \bibfield  {author} {\bibinfo {author} {\bibfnamefont {George~C.}\
  \bibnamefont {Knee}}, \bibinfo {author} {\bibfnamefont {Eliot}\ \bibnamefont
  {Bolduc}}, \bibinfo {author} {\bibfnamefont {Jonathan}\ \bibnamefont
  {Leach}}, \ and\ \bibinfo {author} {\bibfnamefont {Erik~M.}\ \bibnamefont
  {Gauger}},\ }\bibfield  {title} {\enquote {\bibinfo {title} {{Quantum process
  tomography via completely positive and trace-preserving projection}},}\
  }\href {\doibase 10.1103/PhysRevA.98.062336} {\bibfield  {journal} {\bibinfo
  {journal} {Physical Review A}\ }\textbf {\bibinfo {volume} {98}},\ \bibinfo
  {pages} {062336} (\bibinfo {year} {2018})},\ \Eprint
  {http://arxiv.org/abs/1803.10062} {arXiv:1803.10062} \BibitemShut {NoStop}%
\bibitem [{\citenamefont {Birgin}\ and\ \citenamefont
  {Raydan}(2005)}]{Birgin2005}%
  \BibitemOpen
  \bibfield  {author} {\bibinfo {author} {\bibfnamefont {Ernesto}\ \bibnamefont
  {Birgin}}\ and\ \bibinfo {author} {\bibfnamefont {Marcos}\ \bibnamefont
  {Raydan}},\ }\bibfield  {title} {\enquote {\bibinfo {title} {{Robust Stopping
  Criteria for Dykstra's Algorithm}},}\ }\href {\doibase 10.1137/03060062X}
  {\bibfield  {journal} {\bibinfo  {journal} {SIAM J. Scientific Computing}\
  }\textbf {\bibinfo {volume} {26}},\ \bibinfo {pages} {1405--1414} (\bibinfo
  {year} {2005})}\BibitemShut {NoStop}%
\bibitem [{\citenamefont {Henrion}\ and\ \citenamefont
  {Malick}(2011)}]{conic-projection}%
  \BibitemOpen
  \bibfield  {author} {\bibinfo {author} {\bibfnamefont {Didier}\ \bibnamefont
  {Henrion}}\ and\ \bibinfo {author} {\bibfnamefont {J{\'{e}}r{\^{o}}me}\
  \bibnamefont {Malick}},\ }\bibfield  {title} {\enquote {\bibinfo {title}
  {{Projection methods for conic feasibility problems: Applications to
  polynomial sum-of-squares decompositions}},}\ }\href {\doibase
  10.1080/10556780903191165} {\bibfield  {journal} {\bibinfo  {journal}
  {Optimization Methods and Software}\ }\textbf {\bibinfo {volume} {26}},\
  \bibinfo {pages} {23--46} (\bibinfo {year} {2011})}\BibitemShut {NoStop}%
\bibitem [{\citenamefont {White}\ \emph
  {et~al.}(2021{\natexlab{b}})\citenamefont {White}, \citenamefont {Pollock},
  \citenamefont {Hollenberg}, \citenamefont {Hill},\ and\ \citenamefont
  {Modi}}]{White2021}%
  \BibitemOpen
  \bibfield  {author} {\bibinfo {author} {\bibfnamefont {Gregory A.~L.}\
  \bibnamefont {White}}, \bibinfo {author} {\bibfnamefont {Felix~A.}\
  \bibnamefont {Pollock}}, \bibinfo {author} {\bibfnamefont {Lloyd C.~L.}\
  \bibnamefont {Hollenberg}}, \bibinfo {author} {\bibfnamefont {Charles~D.}\
  \bibnamefont {Hill}}, \ and\ \bibinfo {author} {\bibfnamefont {Kavan}\
  \bibnamefont {Modi}},\ }\bibfield  {title} {\enquote {\bibinfo {title}
  {{Diagnosing temporal quantum correlations: compressed non-Markovian
  calipers}},}\ }\href {http://arxiv.org/abs/2107.13934} {\bibfield  {journal}
  {\bibinfo  {journal} {arXiv:2107.13934}\ } (\bibinfo {year}
  {2021}{\natexlab{b}})}\BibitemShut {NoStop}%
\bibitem [{\citenamefont {Gilchrist}\ \emph {et~al.}(2009)\citenamefont
  {Gilchrist}, \citenamefont {Terno},\ and\ \citenamefont
  {Wood}}]{gilchrist-vectorisation}%
  \BibitemOpen
  \bibfield  {author} {\bibinfo {author} {\bibfnamefont {Alexei}\ \bibnamefont
  {Gilchrist}}, \bibinfo {author} {\bibfnamefont {Daniel~R.}\ \bibnamefont
  {Terno}}, \ and\ \bibinfo {author} {\bibfnamefont {Christopher~J.}\
  \bibnamefont {Wood}},\ }\bibfield  {title} {\enquote {\bibinfo {title}
  {{Vectorization of quantum operations and its use}},}\ }\href
  {http://arxiv.org/abs/0911.2539} {\bibfield  {journal} {\bibinfo  {journal}
  {arXiv:0911.2539}\ } (\bibinfo {year} {2009})}\BibitemShut {NoStop}%
\bibitem [{\citenamefont {Anjos}\ and\ \citenamefont
  {Lasserre}(2012)}]{conic-handbook}%
  \BibitemOpen
  \bibfield  {author} {\bibinfo {author} {\bibfnamefont {Miguel~F.}\
  \bibnamefont {Anjos}}\ and\ \bibinfo {author} {\bibfnamefont {Jean~B.}\
  \bibnamefont {Lasserre}},\ }\href {\doibase 10.1007/978-1-4614-0769-0} {\emph
  {\bibinfo {title} {International Series in Operations Research and Management
  Science}}},\ Vol.\ \bibinfo {volume} {166}\ (\bibinfo  {publisher} {Springer
  US},\ \bibinfo {year} {2012})\ Chap.~\bibinfo {chapter} {20}, pp.\ \bibinfo
  {pages} {XI, 960}\BibitemShut {NoStop}%
\bibitem [{\citenamefont {Hauswirth}\ \emph {et~al.}(2016)\citenamefont
  {Hauswirth}, \citenamefont {Bolognani}, \citenamefont {Hug},\ and\
  \citenamefont {D{\"o}rfler}}]{hauswirth2016projected}%
  \BibitemOpen
  \bibfield  {author} {\bibinfo {author} {\bibfnamefont {Adrian}\ \bibnamefont
  {Hauswirth}}, \bibinfo {author} {\bibfnamefont {Saverio}\ \bibnamefont
  {Bolognani}}, \bibinfo {author} {\bibfnamefont {Gabriela}\ \bibnamefont
  {Hug}}, \ and\ \bibinfo {author} {\bibfnamefont {Florian}\ \bibnamefont
  {D{\"o}rfler}},\ }\bibfield  {title} {\enquote {\bibinfo {title} {Projected
  gradient descent on riemannian manifolds with applications to online power
  system optimization},}\ }in\ \href@noop {} {\emph {\bibinfo {booktitle} {2016
  54th Annual Allerton Conference on Communication, Control, and Computing
  (Allerton)}}}\ (\bibinfo {organization} {IEEE},\ \bibinfo {year} {2016})\
  pp.\ \bibinfo {pages} {225--232}\BibitemShut {NoStop}%
\bibitem [{\citenamefont {Michelot}(1986)}]{simplex-projection}%
  \BibitemOpen
  \bibfield  {author} {\bibinfo {author} {\bibfnamefont {C.}~\bibnamefont
  {Michelot}},\ }\bibfield  {title} {\enquote {\bibinfo {title} {{A finite
  algorithm for finding the projection of a point onto the canonical simplex of
  $\alpha_n$}},}\ }\href {\doibase 10.1007/BF00938486} {\bibfield  {journal}
  {\bibinfo  {journal} {Journal of Optimization Theory and Applications}\
  }\textbf {\bibinfo {volume} {50}},\ \bibinfo {pages} {195--200} (\bibinfo
  {year} {1986})}\BibitemShut {NoStop}%
\bibitem [{\citenamefont {Malick}(2004)}]{PSD-dual}%
  \BibitemOpen
  \bibfield  {author} {\bibinfo {author} {\bibfnamefont {J{\'{e}}r{\^{o}}me}\
  \bibnamefont {Malick}},\ }\bibfield  {title} {\enquote {\bibinfo {title} {{A
  Dual Approach to Semidefinite Least-Squares Problems}},}\ }\href {\doibase
  10.1137/S0895479802413856} {\bibfield  {journal} {\bibinfo  {journal} {SIAM
  J. Matrix Analysis Applications}\ }\textbf {\bibinfo {volume} {26}},\
  \bibinfo {pages} {272--284} (\bibinfo {year} {2004})}\BibitemShut {NoStop}%
\bibitem [{\citenamefont {{Dong C. Liu}}\ and\ \citenamefont {{Jorge
  Nocedal}}(1989)}]{lbfgs-alg}%
  \BibitemOpen
  \bibfield  {author} {\bibinfo {author} {\bibnamefont {{Dong C. Liu}}}\ and\
  \bibinfo {author} {\bibnamefont {{Jorge Nocedal}}},\ }\bibfield  {title}
  {\enquote {\bibinfo {title} {{On the limited memory BFGS method for large
  scale optimization}},}\ }\href
  {https://link.springer.com/content/pdf/10.1007{\%}2FBF01589116.pdf}
  {\bibfield  {journal} {\bibinfo  {journal} {Mathematical Programming}\
  }\textbf {\bibinfo {volume} {45}},\ \bibinfo {pages} {503--528} (\bibinfo
  {year} {1989})}\BibitemShut {NoStop}%
\bibitem [{\citenamefont {Malick}\ and\ \citenamefont
  {Sendov}(2006)}]{projection-clarke-jacobian}%
  \BibitemOpen
  \bibfield  {author} {\bibinfo {author} {\bibfnamefont {J{\'{e}}r{\^{o}}me}\
  \bibnamefont {Malick}}\ and\ \bibinfo {author} {\bibfnamefont {Hristo~S.}\
  \bibnamefont {Sendov}},\ }\bibfield  {title} {\enquote {\bibinfo {title}
  {{Clarke generalized Jacobian of the projection onto the cone of positive
  semidefinite matrices}},}\ }\href {\doibase 10.1007/s11228-005-0005-1}
  {\bibfield  {journal} {\bibinfo  {journal} {Set-Valued Analysis}\ }\textbf
  {\bibinfo {volume} {14}},\ \bibinfo {pages} {273--293} (\bibinfo {year}
  {2006})}\BibitemShut {NoStop}%
\bibitem [{\citenamefont {Sendov}(2006)}]{spectral-derivatives}%
  \BibitemOpen
  \bibfield  {author} {\bibinfo {author} {\bibfnamefont {Hristo~S.}\
  \bibnamefont {Sendov}},\ }\bibfield  {title} {\enquote {\bibinfo {title}
  {{Generalized Hadamard product and the derivatives of spectral functions}},}\
  }\href {\doibase 10.1137/050623206} {\bibfield  {journal} {\bibinfo
  {journal} {SIAM Journal on Matrix Analysis and Applications}\ }\textbf
  {\bibinfo {volume} {28}},\ \bibinfo {pages} {667--681} (\bibinfo {year}
  {2006})},\ \Eprint {http://arxiv.org/abs/0404347} {arXiv:0404347 [math]}
  \BibitemShut {NoStop}%
\bibitem [{\citenamefont {Nasir}\ \emph {et~al.}(2020)\citenamefont {Nasir},
  \citenamefont {Shaari},\ and\ \citenamefont {Mancini}}]{Nasir2020}%
  \BibitemOpen
  \bibfield  {author} {\bibinfo {author} {\bibfnamefont {Rinie~N.M.}\
  \bibnamefont {Nasir}}, \bibinfo {author} {\bibfnamefont {Jesni~Shamsul}\
  \bibnamefont {Shaari}}, \ and\ \bibinfo {author} {\bibfnamefont {Stefano}\
  \bibnamefont {Mancini}},\ }\bibfield  {title} {\enquote {\bibinfo {title}
  {{Mutually unbiased unitary bases of operators on $d$-dimensional Hilbert
  space}},}\ }\href {\doibase 10.1142/S0219749919410260} {\bibfield  {journal}
  {\bibinfo  {journal} {International Journal of Quantum Information}\ ,\
  \bibinfo {pages} {1941026}} (\bibinfo {year} {2020})},\ \Eprint
  {http://arxiv.org/abs/2003.12201} {arXiv:2003.12201} \BibitemShut {NoStop}%
\bibitem [{\citenamefont {Rosvall}\ \emph {et~al.}(2014)\citenamefont
  {Rosvall}, \citenamefont {Esquivel}, \citenamefont {Lancichinetti},
  \citenamefont {West},\ and\ \citenamefont {Lambiotte}}]{Rosvall2014}%
  \BibitemOpen
  \bibfield  {author} {\bibinfo {author} {\bibfnamefont {Martin}\ \bibnamefont
  {Rosvall}}, \bibinfo {author} {\bibfnamefont {Alcides~V}\ \bibnamefont
  {Esquivel}}, \bibinfo {author} {\bibfnamefont {Andrea}\ \bibnamefont
  {Lancichinetti}}, \bibinfo {author} {\bibfnamefont {Jevin~D}\ \bibnamefont
  {West}}, \ and\ \bibinfo {author} {\bibfnamefont {Renaud}\ \bibnamefont
  {Lambiotte}},\ }\bibfield  {title} {\enquote {\bibinfo {title} {Memory in
  network flows and its effects on spreading dynamics and community
  detection},}\ }\href {\doibase https://doi.org/10.1038/ncomms5630} {\bibfield
   {journal} {\bibinfo  {journal} {Nature Communications}\ }\textbf {\bibinfo
  {volume} {5}},\ \bibinfo {pages} {4630} (\bibinfo {year} {2014})},\ \Eprint
  {http://arxiv.org/abs/1305.4807} {arXiv:1305.4807} \BibitemShut {NoStop}%
\bibitem [{\citenamefont {Guo}\ \emph {et~al.}(2021)\citenamefont {Guo},
  \citenamefont {Taranto}, \citenamefont {Liu}, \citenamefont {Hu},
  \citenamefont {Huang}, \citenamefont {Li},\ and\ \citenamefont
  {Guo}}]{PhysRevLett.126.230401}%
  \BibitemOpen
  \bibfield  {author} {\bibinfo {author} {\bibfnamefont {Yu}~\bibnamefont
  {Guo}}, \bibinfo {author} {\bibfnamefont {Philip}\ \bibnamefont {Taranto}},
  \bibinfo {author} {\bibfnamefont {Bi-Heng}\ \bibnamefont {Liu}}, \bibinfo
  {author} {\bibfnamefont {Xiao-Min}\ \bibnamefont {Hu}}, \bibinfo {author}
  {\bibfnamefont {Yun-Feng}\ \bibnamefont {Huang}}, \bibinfo {author}
  {\bibfnamefont {Chuan-Feng}\ \bibnamefont {Li}}, \ and\ \bibinfo {author}
  {\bibfnamefont {Guang-Can}\ \bibnamefont {Guo}},\ }\bibfield  {title}
  {\enquote {\bibinfo {title} {{Experimental Demonstration of
  Instrument-Specific Quantum Memory Effects and Non-Markovian Process Recovery
  for Common-Cause Processes}},}\ }\href {\doibase
  10.1103/PhysRevLett.126.230401} {\bibfield  {journal} {\bibinfo  {journal}
  {Physical Review Letters}\ }\textbf {\bibinfo {volume} {126}},\ \bibinfo
  {pages} {230401} (\bibinfo {year} {2021})}\BibitemShut {NoStop}%
\bibitem [{\citenamefont {Wilde}(2013)}]{wilde_2013}%
  \BibitemOpen
  \bibfield  {author} {\bibinfo {author} {\bibfnamefont {Mark~M.}\ \bibnamefont
  {Wilde}},\ }\href {\doibase 10.1017/CBO9781139525343} {\emph {\bibinfo
  {title} {Quantum Information Theory}}}\ (\bibinfo  {publisher} {Cambridge
  University Press},\ \bibinfo {year} {2013})\BibitemShut {NoStop}%
\bibitem [{\citenamefont {Taranto}\ \emph
  {et~al.}(2019{\natexlab{b}})\citenamefont {Taranto}, \citenamefont
  {Pollock},\ and\ \citenamefont {Modi}}]{taranto3}%
  \BibitemOpen
  \bibfield  {author} {\bibinfo {author} {\bibfnamefont {Philip}\ \bibnamefont
  {Taranto}}, \bibinfo {author} {\bibfnamefont {Felix~A}\ \bibnamefont
  {Pollock}}, \ and\ \bibinfo {author} {\bibfnamefont {Kavan}\ \bibnamefont
  {Modi}},\ }\bibfield  {title} {\enquote {\bibinfo {title} {{Memory Strength
  and Recoverability of Non-Markovian Quantum Stochastic Processes}},}\ }\href
  {http://arxiv.org/abs/1907.12583} {\  (\bibinfo {year}
  {2019}{\natexlab{b}})},\ \Eprint {http://arxiv.org/abs/1907.12583}
  {arXiv:1907.12583} \BibitemShut {NoStop}%
\bibitem [{Note1()}]{Note1}%
  \BibitemOpen
  \bibinfo {note} {$\ell = 2$ and $\ell =1$ models can be constructed from
  subsets of the $\ell =3$ data, however on their own they minimally require
  $3\times 300 = 900$ and $4\times 30 = 120$ circuits per time,
  respectively}\BibitemShut {NoStop}%
\bibitem [{\citenamefont {Pollock}\ \emph
  {et~al.}(2018{\natexlab{b}})\citenamefont {Pollock}, \citenamefont
  {Rodr{\'{i}}guez-Rosario}, \citenamefont {Frauenheim}, \citenamefont
  {Paternostro},\ and\ \citenamefont {Modi}}]{Pollock2018}%
  \BibitemOpen
  \bibfield  {author} {\bibinfo {author} {\bibfnamefont {Felix~A.}\
  \bibnamefont {Pollock}}, \bibinfo {author} {\bibfnamefont {C{\'{e}}sar}\
  \bibnamefont {Rodr{\'{i}}guez-Rosario}}, \bibinfo {author} {\bibfnamefont
  {Thomas}\ \bibnamefont {Frauenheim}}, \bibinfo {author} {\bibfnamefont
  {Mauro}\ \bibnamefont {Paternostro}}, \ and\ \bibinfo {author} {\bibfnamefont
  {Kavan}\ \bibnamefont {Modi}},\ }\bibfield  {title} {\enquote {\bibinfo
  {title} {{Operational Markov Condition for Quantum Processes}},}\ }\href
  {\doibase 10.1103/PhysRevLett.120.040405} {\bibfield  {journal} {\bibinfo
  {journal} {Physical Review Letters}\ }\textbf {\bibinfo {volume} {120}},\
  \bibinfo {pages} {040405} (\bibinfo {year} {2018}{\natexlab{b}})},\ \Eprint
  {http://arxiv.org/abs/1801.09811} {arXiv:1801.09811} \BibitemShut {NoStop}%
\bibitem [{\citenamefont {Giarmatzi}\ and\ \citenamefont
  {Costa}(2018)}]{giarmatzi_quantum_2018}%
  \BibitemOpen
  \bibfield  {author} {\bibinfo {author} {\bibfnamefont {Christina}\
  \bibnamefont {Giarmatzi}}\ and\ \bibinfo {author} {\bibfnamefont {Fabio}\
  \bibnamefont {Costa}},\ }\bibfield  {title} {\enquote {\bibinfo {title} {A
  quantum causal discovery algorithm},}\ }\href {\doibase
  10.1038/s41534-018-0062-6} {\bibfield  {journal} {\bibinfo  {journal} {npj
  Quantum Information}\ }\textbf {\bibinfo {volume} {4}},\ \bibinfo {pages}
  {17} (\bibinfo {year} {2018})}\BibitemShut {NoStop}%
\bibitem [{\citenamefont {Fux}\ \emph {et~al.}(2021)\citenamefont {Fux},
  \citenamefont {Butler}, \citenamefont {Eastham}, \citenamefont {Lovett},\
  and\ \citenamefont {Keeling}}]{PhysRevLett.126.200401}%
  \BibitemOpen
  \bibfield  {author} {\bibinfo {author} {\bibfnamefont {Gerald~E.}\
  \bibnamefont {Fux}}, \bibinfo {author} {\bibfnamefont {Eoin~P.}\ \bibnamefont
  {Butler}}, \bibinfo {author} {\bibfnamefont {Paul~R.}\ \bibnamefont
  {Eastham}}, \bibinfo {author} {\bibfnamefont {Brendon~W.}\ \bibnamefont
  {Lovett}}, \ and\ \bibinfo {author} {\bibfnamefont {Jonathan}\ \bibnamefont
  {Keeling}},\ }\bibfield  {title} {\enquote {\bibinfo {title} {{Efficient
  Exploration of Hamiltonian Parameter Space for Optimal Control of
  Non-Markovian Open Quantum Systems}},}\ }\href {\doibase
  10.1103/PhysRevLett.126.200401} {\bibfield  {journal} {\bibinfo  {journal}
  {Physical Review Letters}\ }\textbf {\bibinfo {volume} {126}},\ \bibinfo
  {pages} {200401} (\bibinfo {year} {2021})}\BibitemShut {NoStop}%
\bibitem [{\citenamefont {Sinha}\ \emph {et~al.}(2020)\citenamefont {Sinha},
  \citenamefont {Meystre}, \citenamefont {Goldschmidt}, \citenamefont {Fatemi},
  \citenamefont {Rolston},\ and\ \citenamefont
  {Solano}}]{PhysRevLett.124.043603}%
  \BibitemOpen
  \bibfield  {author} {\bibinfo {author} {\bibfnamefont {Kanupriya}\
  \bibnamefont {Sinha}}, \bibinfo {author} {\bibfnamefont {Pierre}\
  \bibnamefont {Meystre}}, \bibinfo {author} {\bibfnamefont {Elizabeth~A.}\
  \bibnamefont {Goldschmidt}}, \bibinfo {author} {\bibfnamefont {Fredrik~K.}\
  \bibnamefont {Fatemi}}, \bibinfo {author} {\bibfnamefont {S.~L.}\
  \bibnamefont {Rolston}}, \ and\ \bibinfo {author} {\bibfnamefont {Pablo}\
  \bibnamefont {Solano}},\ }\bibfield  {title} {\enquote {\bibinfo {title}
  {Non-markovian collective emission from macroscopically separated
  emitters},}\ }\href {\doibase 10.1103/PhysRevLett.124.043603} {\bibfield
  {journal} {\bibinfo  {journal} {Phys. Rev. Lett.}\ }\textbf {\bibinfo
  {volume} {124}},\ \bibinfo {pages} {043603} (\bibinfo {year}
  {2020})}\BibitemShut {NoStop}%
\bibitem [{\citenamefont {Nagy}\ and\ \citenamefont
  {Domokos}(2015)}]{PhysRevLett.115.043601}%
  \BibitemOpen
  \bibfield  {author} {\bibinfo {author} {\bibfnamefont {D.}~\bibnamefont
  {Nagy}}\ and\ \bibinfo {author} {\bibfnamefont {P.}~\bibnamefont {Domokos}},\
  }\bibfield  {title} {\enquote {\bibinfo {title} {Nonequilibrium quantum
  criticality and non-markovian environment: Critical exponent of a quantum
  phase transition},}\ }\href {\doibase 10.1103/PhysRevLett.115.043601}
  {\bibfield  {journal} {\bibinfo  {journal} {Phys. Rev. Lett.}\ }\textbf
  {\bibinfo {volume} {115}},\ \bibinfo {pages} {043601} (\bibinfo {year}
  {2015})}\BibitemShut {NoStop}%
\bibitem [{\citenamefont {Haikka}\ \emph {et~al.}(2011)\citenamefont {Haikka},
  \citenamefont {McEndoo}, \citenamefont {De~Chiara}, \citenamefont {Palma},\
  and\ \citenamefont {Maniscalco}}]{PhysRevA.84.031602}%
  \BibitemOpen
  \bibfield  {author} {\bibinfo {author} {\bibfnamefont {P.}~\bibnamefont
  {Haikka}}, \bibinfo {author} {\bibfnamefont {S.}~\bibnamefont {McEndoo}},
  \bibinfo {author} {\bibfnamefont {G.}~\bibnamefont {De~Chiara}}, \bibinfo
  {author} {\bibfnamefont {G.~M.}\ \bibnamefont {Palma}}, \ and\ \bibinfo
  {author} {\bibfnamefont {S.}~\bibnamefont {Maniscalco}},\ }\bibfield  {title}
  {\enquote {\bibinfo {title} {Quantifying, characterizing, and controlling
  information flow in ultracold atomic gases},}\ }\href {\doibase
  10.1103/PhysRevA.84.031602} {\bibfield  {journal} {\bibinfo  {journal} {Phys.
  Rev. A}\ }\textbf {\bibinfo {volume} {84}},\ \bibinfo {pages} {031602}
  (\bibinfo {year} {2011})}\BibitemShut {NoStop}%
\bibitem [{\citenamefont {Mujica-Martinez}\ \emph {et~al.}(2013)\citenamefont
  {Mujica-Martinez}, \citenamefont {Nalbach},\ and\ \citenamefont
  {Thorwart}}]{PhysRevE.88.062719}%
  \BibitemOpen
  \bibfield  {author} {\bibinfo {author} {\bibfnamefont {C.~A.}\ \bibnamefont
  {Mujica-Martinez}}, \bibinfo {author} {\bibfnamefont {P.}~\bibnamefont
  {Nalbach}}, \ and\ \bibinfo {author} {\bibfnamefont {M.}~\bibnamefont
  {Thorwart}},\ }\bibfield  {title} {\enquote {\bibinfo {title}
  {{Quantification of non-Markovian effects in the Fenna-Matthews-Olson
  complex}},}\ }\href {\doibase 10.1103/PhysRevE.88.062719} {\bibfield
  {journal} {\bibinfo  {journal} {Phys. Rev. E}\ }\textbf {\bibinfo {volume}
  {88}},\ \bibinfo {pages} {062719} (\bibinfo {year} {2013})}\BibitemShut
  {NoStop}%
\bibitem [{\citenamefont {Nitzan}\ and\ \citenamefont
  {Ratner}(2003)}]{nitzan2003electron}%
  \BibitemOpen
  \bibfield  {author} {\bibinfo {author} {\bibfnamefont {Abraham}\ \bibnamefont
  {Nitzan}}\ and\ \bibinfo {author} {\bibfnamefont {Mark~A}\ \bibnamefont
  {Ratner}},\ }\bibfield  {title} {\enquote {\bibinfo {title} {Electron
  transport in molecular wire junctions},}\ }\href {\doibase
  10.1126/science.1081572} {\bibfield  {journal} {\bibinfo  {journal}
  {Science}\ }\textbf {\bibinfo {volume} {300}},\ \bibinfo {pages} {1384--1389}
  (\bibinfo {year} {2003})}\BibitemShut {NoStop}%
\bibitem [{\citenamefont {Lambert}\ \emph {et~al.}(2013)\citenamefont
  {Lambert}, \citenamefont {Chen}, \citenamefont {Cheng}, \citenamefont {Li},
  \citenamefont {Chen},\ and\ \citenamefont {Nori}}]{lambert2013quantum}%
  \BibitemOpen
  \bibfield  {author} {\bibinfo {author} {\bibfnamefont {Neill}\ \bibnamefont
  {Lambert}}, \bibinfo {author} {\bibfnamefont {Yueh-Nan}\ \bibnamefont
  {Chen}}, \bibinfo {author} {\bibfnamefont {Yuan-Chung}\ \bibnamefont
  {Cheng}}, \bibinfo {author} {\bibfnamefont {Che-Ming}\ \bibnamefont {Li}},
  \bibinfo {author} {\bibfnamefont {Guang-Yin}\ \bibnamefont {Chen}}, \ and\
  \bibinfo {author} {\bibfnamefont {Franco}\ \bibnamefont {Nori}},\ }\bibfield
  {title} {\enquote {\bibinfo {title} {Quantum biology},}\ }\href {\doibase
  https://doi.org/10.1038/nphys2474} {\bibfield  {journal} {\bibinfo  {journal}
  {Nature Physics}\ }\textbf {\bibinfo {volume} {9}},\ \bibinfo {pages}
  {10--18} (\bibinfo {year} {2013})}\BibitemShut {NoStop}%
\bibitem [{\citenamefont {Torlai}\ \emph {et~al.}(2020)\citenamefont {Torlai},
  \citenamefont {Wood}, \citenamefont {Acharya}, \citenamefont {Carleo},
  \citenamefont {Carrasquilla},\ and\ \citenamefont
  {Aolita}}]{arXiv:2006.02424}%
  \BibitemOpen
  \bibfield  {author} {\bibinfo {author} {\bibfnamefont {Giacomo}\ \bibnamefont
  {Torlai}}, \bibinfo {author} {\bibfnamefont {Christopher~J}\ \bibnamefont
  {Wood}}, \bibinfo {author} {\bibfnamefont {Atithi}\ \bibnamefont {Acharya}},
  \bibinfo {author} {\bibfnamefont {Giuseppe}\ \bibnamefont {Carleo}}, \bibinfo
  {author} {\bibfnamefont {Juan}\ \bibnamefont {Carrasquilla}}, \ and\ \bibinfo
  {author} {\bibfnamefont {Leandro}\ \bibnamefont {Aolita}},\ }\bibfield
  {title} {\enquote {\bibinfo {title} {Quantum process tomography with
  unsupervised learning and tensor networks},}\ }\href
  {https://www.arxiv.org/abs/2006.02424} {\bibfield  {journal} {\bibinfo
  {journal} {arXiv:2006.02424}\ } (\bibinfo {year} {2020})}\BibitemShut
  {NoStop}%
\bibitem [{\citenamefont {Guo}\ \emph {et~al.}(2020)\citenamefont {Guo},
  \citenamefont {Modi},\ and\ \citenamefont {Poletti}}]{guochu2020}%
  \BibitemOpen
  \bibfield  {author} {\bibinfo {author} {\bibfnamefont {Chu}\ \bibnamefont
  {Guo}}, \bibinfo {author} {\bibfnamefont {Kavan}\ \bibnamefont {Modi}}, \
  and\ \bibinfo {author} {\bibfnamefont {Dario}\ \bibnamefont {Poletti}},\
  }\bibfield  {title} {\enquote {\bibinfo {title} {{Tensor-network-based
  machine learning of non-Markovian quantum processes}},}\ }\href {\doibase
  10.1103/PhysRevA.102.062414} {\bibfield  {journal} {\bibinfo  {journal}
  {Physical Review A}\ }\textbf {\bibinfo {volume} {102}},\ \bibinfo {pages}
  {062414} (\bibinfo {year} {2020})}\BibitemShut {NoStop}%
\bibitem [{\citenamefont {Rambach}\ \emph {et~al.}(2021)\citenamefont
  {Rambach}, \citenamefont {Qaryan}, \citenamefont {Kewming}, \citenamefont
  {Ferrie}, \citenamefont {White},\ and\ \citenamefont
  {Romero}}]{PhysRevLett.126.100402}%
  \BibitemOpen
  \bibfield  {author} {\bibinfo {author} {\bibfnamefont {Markus}\ \bibnamefont
  {Rambach}}, \bibinfo {author} {\bibfnamefont {Mahdi}\ \bibnamefont {Qaryan}},
  \bibinfo {author} {\bibfnamefont {Michael}\ \bibnamefont {Kewming}}, \bibinfo
  {author} {\bibfnamefont {Christopher}\ \bibnamefont {Ferrie}}, \bibinfo
  {author} {\bibfnamefont {Andrew~G.}\ \bibnamefont {White}}, \ and\ \bibinfo
  {author} {\bibfnamefont {Jacquiline}\ \bibnamefont {Romero}},\ }\bibfield
  {title} {\enquote {\bibinfo {title} {Robust and efficient high-dimensional
  quantum state tomography},}\ }\href {\doibase 10.1103/PhysRevLett.126.100402}
  {\bibfield  {journal} {\bibinfo  {journal} {Phys. Rev. Lett.}\ }\textbf
  {\bibinfo {volume} {126}},\ \bibinfo {pages} {100402} (\bibinfo {year}
  {2021})},\ \Eprint {http://arxiv.org/abs/2010.00632} {arXiv:2010.00632}
  \BibitemShut {NoStop}%
\bibitem [{\citenamefont {Huang}\ \emph {et~al.}(2020)\citenamefont {Huang},
  \citenamefont {Kueng},\ and\ \citenamefont {Preskill}}]{huang-shadow}%
  \BibitemOpen
  \bibfield  {author} {\bibinfo {author} {\bibfnamefont {Hsin-Yuan}\
  \bibnamefont {Huang}}, \bibinfo {author} {\bibfnamefont {Richard}\
  \bibnamefont {Kueng}}, \ and\ \bibinfo {author} {\bibfnamefont {John}\
  \bibnamefont {Preskill}},\ }\bibfield  {title} {\enquote {\bibinfo {title}
  {{Predicting many properties of a quantum system from very few
  measurements}},}\ }\href {\doibase https://doi.org/10.1038/s41567-020-0932-7}
  {\bibfield  {journal} {\bibinfo  {journal} {Nature Physics}\ }\textbf
  {\bibinfo {volume} {16}},\ \bibinfo {pages} {1050--1057} (\bibinfo {year}
  {2020})},\ \Eprint {http://arxiv.org/abs/2002.08953} {arXiv:2002.08953}
  \BibitemShut {NoStop}%
\bibitem [{\citenamefont {Gray}(2018)}]{quimb}%
  \BibitemOpen
  \bibfield  {author} {\bibinfo {author} {\bibfnamefont {Johnnie}\ \bibnamefont
  {Gray}},\ }\bibfield  {title} {\enquote {\bibinfo {title} {{quimb: A python
  package for quantum information and many-body calculations}},}\ }\href
  {\doibase 10.21105/joss.00819} {\bibfield  {journal} {\bibinfo  {journal}
  {Journal of Open Source Software}\ }\textbf {\bibinfo {volume} {3}},\
  \bibinfo {pages} {819} (\bibinfo {year} {2018})}\BibitemShut {NoStop}%
\bibitem [{\citenamefont {\ifmmode \check{R}\else
  \v{R}\fi{}eh\'a\ifmmode~\check{c}\else \v{c}\fi{}ek}\ \emph
  {et~al.}(2004)\citenamefont {\ifmmode \check{R}\else
  \v{R}\fi{}eh\'a\ifmmode~\check{c}\else \v{c}\fi{}ek}, \citenamefont
  {Englert},\ and\ \citenamefont {Kaszlikowski}}]{PhysRevA.70.052321}%
  \BibitemOpen
  \bibfield  {author} {\bibinfo {author} {\bibfnamefont {Jaroslav}\
  \bibnamefont {\ifmmode \check{R}\else \v{R}\fi{}eh\'a\ifmmode~\check{c}\else
  \v{c}\fi{}ek}}, \bibinfo {author} {\bibfnamefont {Berthold-Georg}\
  \bibnamefont {Englert}}, \ and\ \bibinfo {author} {\bibfnamefont {Dagomir}\
  \bibnamefont {Kaszlikowski}},\ }\bibfield  {title} {\enquote {\bibinfo
  {title} {Minimal qubit tomography},}\ }\href {\doibase
  10.1103/PhysRevA.70.052321} {\bibfield  {journal} {\bibinfo  {journal}
  {Physical Review A}\ }\textbf {\bibinfo {volume} {70}},\ \bibinfo {pages}
  {052321} (\bibinfo {year} {2004})}\BibitemShut {NoStop}%
\bibitem [{\citenamefont {Nielsen}\ \emph
  {et~al.}(2020{\natexlab{b}})\citenamefont {Nielsen}, \citenamefont
  {Rudinger}, \citenamefont {Proctor}, \citenamefont {Russo}, \citenamefont
  {Young},\ and\ \citenamefont {Blume-Kohout}}]{pygsti}%
  \BibitemOpen
  \bibfield  {author} {\bibinfo {author} {\bibfnamefont {Erik}\ \bibnamefont
  {Nielsen}}, \bibinfo {author} {\bibfnamefont {Kenneth}\ \bibnamefont
  {Rudinger}}, \bibinfo {author} {\bibfnamefont {Timothy}\ \bibnamefont
  {Proctor}}, \bibinfo {author} {\bibfnamefont {Antonio}\ \bibnamefont
  {Russo}}, \bibinfo {author} {\bibfnamefont {Kevin}\ \bibnamefont {Young}}, \
  and\ \bibinfo {author} {\bibfnamefont {Robin}\ \bibnamefont {Blume-Kohout}},\
  }\bibfield  {title} {\enquote {\bibinfo {title} {Probing quantum processor
  performance with {pyGSTi}},}\ }\href {\doibase 10.1088/2058-9565/ab8aa4}
  {\bibfield  {journal} {\bibinfo  {journal} {Quantum Science and Technology}\
  }\textbf {\bibinfo {volume} {5}},\ \bibinfo {pages} {044002} (\bibinfo {year}
  {2020}{\natexlab{b}})}\BibitemShut {NoStop}%
\end{thebibliography}
%

\clearpage
\onecolumngrid

\section*{Appendix}
\appendix
\section{Process tensor contraction and maximum likelihood}
\label{appendix:pt-maths}
Throughout the main text we omitted some of the lengthier computations and descriptions relevant to both linear inversion and maximum likelihood PTT. We include these as follows for completeness, as well as an outline of our algorithmic implementation of PTT.
\subsection*{Construction of a dual set}
The procedure to construct the dual operators is as follows: we compile an IC operation set $\{\mathcal{B}_i\}$ into a single matrix $\mathfrak{B}$.
Write each $\mathcal{B}_i = \sum_j b_{ij}\Gamma_j$, where $\{\Gamma_j\}$ form a Hermitian, self-dual, linearly-independent basis satisfying $\text{tr}[\Gamma_j\Gamma_k]=\delta_{jk}$. 
In our case, we select $\{\Gamma_j\}$ to be the standard basis, meaning that the $k$th column of the matrix $\mathfrak{B} = \sum_{ij}b_{ij}\ket{i}\!\bra{j}$ is $\mathcal{B}_k$ flattened into a $1$D vector. 
Because the $\{\mathcal{B}_i\}$ are linearly independent, $\mathfrak{B}$ is invertible. 
Let the matrix $\mathfrak{F}^\dagger = \mathfrak{B}^{-1}$ such that $\mathfrak{B}\cdot \mathfrak{F}^\dagger = \mathbb{I}$.
This means that the rows of $\mathfrak{F}^\dagger$ are orthogonal to the rows of $\mathfrak{B}$.
The dual matrices can then be defined as $\Delta_i = \sum_j f_{ij}\Gamma_j$, ensuring that $\text{tr}[\mathcal{B}_i\Delta_j] = \delta_{ij}$.
Note that in this work, our basis is restricted to the sub-manifold of unitary matrices.
This means that the dimension $d$ of the space is less than the order $n$ of the matrices. 
Therefore we construct $\mathfrak{F}^\dagger$ as the Moore-Penrose or the right inverse of $\mathfrak{B}$. If a set of duals is with respect to an overcomplete basis, the same strategy may also be used.
Here, we relax the duality condition $\text{tr}[\mathcal{B}_i\Delta_j]=\delta_{ij}$, but retain $\sum_i\Delta_i = \mathbb{I}$ to ensure that the expansion of any operation within the basis is complete. 
\par 

\subsection{Action of Choi states}
When written in terms of its dual construction, it becomes apparent that the action of a quantum channel through its Choi representation is a linear expansion in terms of its action on an IC basis of inputs. Here, we step through this computation and then through the same computation for the process tensor in order to emphasise their parallels. Explicitly, consider $\sigma = \sum_{i=1}^n \alpha_i \rho_i$. The action of $\mathcal{E}$ on $\sigma$ is given by:
\begin{equation}
\begin{split}
    &\text{Tr}_{\text{in}}\left[(\mathbb{I}_{\text{out}}\otimes \sigma^T)\hat{\mathcal{E}}\right]\\
    &= \text{Tr}_{\text{in}}\left[(\mathbb{I}_{\text{out}}\otimes \sigma^T)\sum_{i=1}^n \rho'_i \otimes \omega_i^T\right]\\
    &=\text{Tr}_{\text{in}}\left[(\mathbb{I}_{\text{out}}\otimes \sum_{j=1}^n \alpha_j \rho_j^T)\sum_{i=1}^n \rho'_i \otimes \omega_i^T\right]\\
    &=\text{Tr}_{\text{in}}\left[\sum_{i=1}^n\sum_{j=1}^n \rho'_i \otimes \alpha_j \rho_j^T\omega_i^T\right]\\
    &= \sum_{i=1}^n\sum_{j=1}^n\rho_i'\alpha_j\Tr[\rho_j\omega_i]\\
    &= \sum_{i=1}^n \alpha_i \rho_i'.
\end{split}
\end{equation}

Similarly, for a process tensor's action on a generic sequence of operations $\mathbf{A}_{k-1:0}$:
\begin{gather}
\label{eq:LI-action}
\begin{split}
        &\mathcal{T}_{k:0}\left[\mathbf{A}_{k-1:0}\right] = \text{tr}_{\text{in}} \left[\left(\hat{\mathbf{A}}_{k-1:0} \otimes\mathbb{I}_{\text{out}}\right)^\text{T}
        \Upsilon_{k:0}\right]\\
        &=\text{tr}_{\text{in}}\left[\left(\bigotimes_{i=0}^{k-1}\hat{\mathcal{A}}_i^\text{T} \otimes \mathbb{I}\right)\sum_{\vec{\nu}} (\mathbf{\Delta}^{\vec{\nu}}_{k-1:0})^\text{T}\otimes \rho_k^{\vec{\nu}}\right]\\
        &=\text{tr}_{\text{in}}\left[
        \sum_{\vec{\mu}}
        \alpha^{\vec{\mu}}
        \bigotimes_{i=0}^{k-1} 
        \hat{\mathcal{B}}_i^{\mu_i \text{T}}
        \sum_{\vec{\nu}} 
        \bigotimes_{j=0}^{k-1}
        \Delta_j^{\nu_j \text{T}} \otimes \rho_k^{\vec{\nu}}\right]\\
        &=\text{tr}_{\text{in}}\left[
        \sum_{\vec{\mu},\vec{\nu}} 
        \alpha^{\vec{\mu}}
        \bigotimes_{i,j=0}^{k-1} \{\hat{\mathcal{B}}_i^{\mu_i \text{T}} \Delta_j^{\nu_j \text{T}}\}\otimes \rho_k^{\vec{\nu}}\right]\\
         &=\sum_{\vec{\mu},\vec{\nu}} 
         \alpha^{\vec{\mu}}
         \prod_{i,j=0}^{k-1}  \,
        \text{tr}\left[
        \hat{\mathcal{B}}_i^{\mu_i} \Delta_j^{\nu_j}\right]
        \rho_k^{\vec{\nu}}\\
        &=\sum_{\vec{\mu},\vec{\nu}}
        \alpha^{\vec{\mu}}
        \prod_{i=0}^{k-1}  \, \delta_{\vec{\mu}\vec{\nu}} \, \rho_k^{\vec{\nu}} \\
        &= \sum_{\vec{\mu}} \alpha^{\vec{\mu}}\rho^{\vec{\mu}}_k\\
        &=\rho_k(\textbf{A}_{k-1:0}).
\end{split}
\end{gather}
The direct calculation of each expansion coefficient is therefore given by 
\begin{gather}
\begin{split}
    \alpha^{\vec{\mu}} =& \text{tr}\left[\hat{\mathbf{A}}_{k-1:0}\mathbf{\Delta}_{k-1:0}^{\vec{\mu}}\right]\\
    =& \text{tr}\left[
    \bigotimes_{i=0}^{k-1} \hat{\mathcal{A}}_{i}
    \Delta^{(\mu,i)}\right]\\
    =& \prod_{i=0}^{k-1}
    \text{tr}\left[
     \hat{\mathcal{A}}_{i}
    \Delta_i^{\mu_i }
    \right] = \prod_{i=0}^{k-1} \alpha_i^{\mu_i}.
    \end{split}
\end{gather}

\subsection{Maximum likelihood, cost evaluation, and gradient}
Full details and benchmarking of the \texttt{pgdb} algorithm for QPT can be found in Ref.~\cite{QPT-projection}. Here, we provide the pseudocode in this context, which forms the basis for our implementation of MLE-PTT.

\begin{algorithm}[H]
\caption{\texttt{pgdb}}
\begin{algorithmic}[1]
\State $j=0,n=d_S^{2k+1}$
\State \text{Initial estimate}:
$\Upsilon_{k:0}^{(0)} = \mathbb{I}_{n\times n}/n$
\State \text{Set metaparameters: }$\alpha=2n^2/3,\gamma=0.3$
\While{$f(\Upsilon_{k:0}^{(j)}) - f(\Upsilon_{k:0}^{(n+1)}) > 1\times 10^{-6}$}
\State $D^{(j)} =\text{Proj}_{S_n^+\cap \mathcal{V}}\left(\Upsilon_{k:0}^{(j)} - \mu \nabla f(\Upsilon_{k:0}^{(j)})\right) - \Upsilon_{k:0}^{(j)}$
\State $\beta = 1$
\While{$f(\Upsilon_{k:0}^{(j)}) + \beta D^{(k)}) > f(\Upsilon_{k:0}^{(j)}) + \gamma \beta \left\langle D^{(j)}, \nabla f(\Upsilon_{k:0}^{(j)})\right\rangle$}
\State $\beta = 0.5\beta$
\EndWhile
\State $\Upsilon_{k:0}^{(j+1)} = \Upsilon_{k:0}^{(j)} + \beta D^{(j)}$
\State $j = j+1$
\EndWhile
\State\Return $\Upsilon_{k:0}^{(\text{est})} = \Upsilon_{k:0}^{(j+1)}$
\end{algorithmic}
\end{algorithm}
$\text{Proj}_{S_n^+\cap\mathcal{V}}(\cdot)$ here is the projection subroutine described in Section~\ref{sec:MLPTT}. Although we have fixed the gradient step size here to be the same as in~\cite{QPT-projection}, we find this to be slightly problem-dependent in terms of its performance. The reason is that the larger the step, the less physical $\Upsilon_{k:0}^{(j)} - \mu \nabla f(\Upsilon_{k:0}^{(j)})$ tends to be, increasing the run-time of the projection subroutine. In general, we find that decreasing $\mu$ to favour the runtime of the projection is overall favourable to the performance of the algorithm.\par

The process tensor action described in Equation~\eqref{eq:LI-action} is pedagogically useful, however in practice, we compute the action of some process tensor $\mathcal{T}_{k:0}$ on a sequence of control operations $\mathbf{A}_{k-1:0}$ via the projection of its Choi state onto as in Equation \eqref{PT-vector-est}. Because the input operations are always tensor product (omitting the case of correlated instruments), this can be performed fast as a tensor network contraction. In this form, computation of the cost and the gradient is significantly sped up in comparison to multiplying out the full matrices.\par 
Writing the Choi state of a process tensor $\Upsilon_{k:0}$ explicitly with its indices as a rank $2(2k+1)$ tensor, we have $2k+1$ subsystems alternating with outputs from the $j$th step ($\mathfrak{o}_j$) and inputs to the $(j+1)$th step ($\mathfrak{i}_j$), i.e.
\begin{equation}
    \Upsilon_{k:0} \equiv (\Upsilon_{k:0})_{k_{\mathfrak{o}_k},k_{\mathfrak{i}_k},\cdots,k_{\mathfrak{o}_0}}^{b_{\mathfrak{o}_k},b_{\mathfrak{i}_k},\cdots,b_{\mathfrak{o}_0}},
\end{equation}
where $b$ is shorthand for bra, and $k$ is shorthand for ket.
The basis operation at time step $j$ has indices (we write its transpose) $(\mathcal{B}_j^{\mu_j})_{b_{i_{j+1},b_{o_j}}}^{k_{i_{j+1}},k_{o_j}}$, meanwhile the POVM element $\Pi_i$ is written $(\Pi_i)_{k_{o_k}}^{b_{o_k}}$. Consequently, the full tensor of predicted probabilities for all basis elements is given by
\begin{equation}
\label{TN-cost}
    p_{i,\vec{\mu}} = \sum_{\substack{k_{\mathfrak{o}_k},k_{\mathfrak{i}_k},\cdots,k_{\mathfrak{i}_1},k_{\mathfrak{o}_0} \\ b_{\mathfrak{o}_k},b_{\mathfrak{i}_k},\cdots,b_{\mathfrak{i}_1},b_{\mathfrak{o}_0}}} (\Upsilon_{k:0})_{k_{\mathfrak{o}_k},k_{\mathfrak{i}_k},\cdots,k_{\mathfrak{o}_0}}^{b_{\mathfrak{o}_k},b_{\mathfrak{i}_k},\cdots,b_{\mathfrak{o}_0}}(\Pi_i)_{k_{\mathfrak{o}_k}}^{b_{\mathfrak{o}_k}} (\mathcal{B}_{k-1}^{\mu_{k-1}})_{b_{\mathfrak{i}_{k}},b_{\mathfrak{o}_{k-1}}}^{k_{\mathfrak{i}_{k}},k_{\mathfrak{o}_{k-1}}}(\mathcal{B}_{k-2}^{\mu_{k-2}})_{b_{\mathfrak{i}_{k-1}},b_{\mathfrak{o}_{k-2}}}^{k_{\mathfrak{i}_{k-1}},k_{\mathfrak{o}_{k-2}}}\cdots (\mathcal{B}_{k-1}^{\mu_{k-1}})_{b_{\mathfrak{i}_{1}},b_{\mathfrak{o}_{0}}}^{k_{\mathfrak{i}_{1}},k_{\mathfrak{o}_{0}}}
\end{equation}
We use the quantum information Python library \texttt{QUIMB} \cite{quimb} to perform this, and all future tensor contractions straightforwardly. The cost function is then evaluation as in Equation \eqref{ML-cost} in the same way: through an element-wise logarithm of $p_{i,\vec{\mu}}$ followed by contraction with the data tensor $n_{i,\vec{\mu}}$. Since the cost function is linear in $\Upsilon_{k:0}$, computing the gradient $\nabla f/\nabla \Upsilon_{k:0}$ is simply $\nabla p_{i,\vec{\mu}}/\nabla \Upsilon_{k:0} : (n/p)^{i,\vec{\mu}}$, which expands to:
\begin{equation}
    \frac{\nabla f}{\nabla \Upsilon_{k:0}} = \sum_{i,\vec{\mu}} \left[(\mathcal{B}_{k-1}^{\mu_{k-1}})_{b_{\mathfrak{i}_{k},b_{\mathfrak{o}_{k-1}}}}^{k_{\mathfrak{i}_{k}},k_{\mathfrak{o}_{k-1}}}(\mathcal{B}_{k-2}^{\mu_{k-2}})_{b_{\mathfrak{i}_{k-1},b_{\mathfrak{o}_{k-2}}}}^{k_{\mathfrak{i}_{k-1}},k_{\mathfrak{o}_{k-2}}}\cdots (\mathcal{B}_{0}^{\mu_{0}})_{b_{\mathfrak{i}_{1},b_{\mathfrak{o}_{0}}}}^{k_{\mathfrak{i}_{1}},k_{\mathfrak{o}_{0}}}\right] \frac{n_{i,\vec{\mu}}}{p_{i,\vec{\mu}}}
\end{equation}
i.e. Equation \eqref{TN-cost} without the inclusion of $\Upsilon_{k:0}$. \par

\section{Approximate conditional Markov order}
\label{appendix:mo}

In order to estimate a CMO process tomographically, we employ MLE-PTT as described by the circuits in Figure~\ref{fig:cmo_circs}. The action of the reduced process tensors under a finite conditional Markov order model is best posed as a tensor network contraction, so that tasks such as the optimisation in Section~\ref{sec:applications} can be performed quickly. We show this here. Note that the following extravagant working is equivalent to Figure \ref{fig:stitching-cmo}, but we write it out in full generality in order to make the indices explicit and replication more straightforward. First, the conditional reduced states of each of the process tensors must be taken by contracting the relevant control operations (including final measurement) into the process tensors. These are:

\begin{equation}
    \begin{split}
        &\left(\Upsilon_{\ell:0}^{(\mathbf{B}_{\ell-1:0}^{\vec{\mu}})}\right)_{k_{o_\ell}}^{b_{o_\ell}} =  \sum_{\substack{k_{o_0},k_{i_1},k_{o_1},\cdots,k_{i_{\ell-1}} \\ b_{o_0},b_{i_1},b_{o_1},\cdots,b_{i_{\ell-1}}}}\left(\Upsilon_{\ell:0}\right)_{{k_{o_\ell},k_{i_\ell},\cdots,k_{i_{1}}},k_{o_{0}}}^{b_{o_\ell},b_{i_\ell},\cdots,b_{i_{1}},b_{o_{0}}}(\mathcal{B}_{\ell-1}^{\mu_{\ell-1}})_{k_{i_{\ell}},k_{o_{\ell-1}}}^{b_{i_{\ell}},b_{o_{\ell-1}}}\cdots (\mathcal{B}_{0}^{\mu_{0}})_{k_{i_{1}},k_{o_{0}}}^{b_{i_{1}},b_{o_{0}}},\\
        &\left(\Upsilon_{j:j-\ell}^{(\mathbf{B}_{j-2:j-\ell}^{\vec{\mu}})}\right)_{k_{o_j},k_{i_{j}}}^{b_{o_j},b_{i_{j}}} = \\ &\sum_{\substack{k_{o_{j-\ell}},k_{i_{j-\ell+1}},k_{o_{j-\ell+1}},\cdots,k_{i_{j-1}}\\ b_{o_{j-\ell}},b_{i_{j-\ell+1}},b_{o_{j-\ell+1}},\cdots,b_{i_{j-1}}}}\left(\Upsilon_{j:j-\ell}\right)_{{k_{o_j},k_{i_j},\cdots,k_{i_{j-\ell+1}},k_{o_{j-\ell}}}}^{b_{o_j},b_{i_j},\cdots,b_{i_{j-\ell+1}},b_{o_{j-\ell}}}\delta_{k_{o_{j-1}}}^{b_{o_{j-1}}}
        (\mathcal{B}_{j-2}^{\mu_{j-2}})_{k_{i_{\ell-1}},k_{o_{\ell-2}}}^{b_{i_{\ell-1}},b_{o_{\ell-2}}}\cdots (\mathcal{B}_{j-\ell}^{\mu_{j-\ell}})_{k_{i_{j-\ell+1}},k_{o_{j-\ell}}}^{b_{i_{j-\ell+1}},b_{o_{j-\ell}}},\\
        &\left(\Upsilon_{k:k-\ell}^{(\mathbf{B}_{k-2:k-\ell}^{\vec{\mu}},\Pi_i)}\right)_{k_{i_k}}^{b_{i_k}} = \\ &\sum_{\substack{k_{o_{k-\ell}},k_{i_{k-\ell+1}},k_{o_{k-\ell+1}},\cdots,k_{i_{k-1},k_{o_k}}\\ b_{o_{k-\ell}},b_{i_{k-\ell+1}},b_{o_{k-\ell+1}},\cdots,b_{i_{k-1}},b_{o_k}}}\left(\Upsilon_{k:k-\ell}\right)_{{k_{o_k},k_{i_k},\cdots,k_{i_{k-\ell+1}},k_{o_{k-\ell}}}}^{b_{o_j},b_{i_j},\cdots,b_{i_{k-\ell+1}},b_{o_{k-\ell}}}(\Pi_i)_{k_{o_k}}^{b_{o_k}}\delta_{k_{o_{k-1}}}^{b_{o_{k-1}}}
        (\mathcal{B}_{k-2}^{\mu_{k-2}})_{k_{i_{\ell-1}},k_{o_{\ell-2}}}^{b_{i_{\ell-1}},b_{o_{\ell-2}}}\cdots (\mathcal{B}_{k-\ell}^{\mu_{k-\ell}})_{k_{i_{k-\ell+1}},k_{o_{k-\ell}}}^{b_{i_{k-\ell+1}},b_{o_{k-\ell}}}.
    \end{split}
\end{equation}
Then, the tensor of predicted probabilities $p_{i,\vec{\mu}}$ is obtained by stitching each conditional process tensor together with the overlapping control operations. That is:

\begin{equation}
\begin{split}
    \sum_{\substack{k_{o_\ell}, k_{i_{\ell+1}}, \cdots, k_{i_k} \\ b_{o_\ell}, b_{i_{\ell+1}}, \cdots, b_{i_k}}}&
    \left(\Upsilon_{k:k-\ell}^{(\mathbf{B}_{k-2:k-\ell}^{\vec{\mu}},\Pi_i)}\right)_{k_{i_k}}^{b_{i_k}}(\mathcal{B}_{k-1}^{\mu_{k-1}})_{k_{i_k},k_{o_{k-1}}}^{b_{i_k},b_{o_{k-1}}}\\
    &\left( \prod_{j=\ell+1}^{k-1}\left(\Upsilon_{j:j-\ell}^{(\mathbf{B}_{j-2:j-\ell}^{\vec{\mu}})}\right)_{k_{o_j},k_{i_{j}}}^{b_{o_j},b_{i_{j}}}(\mathcal{B}_{j-1}^{\mu_{j-1}})_{k_{i_j},k_{o_{j-1}}}^{b_{i_j},b_{o_{j-1}}}\right)(\mathcal{B}_{\ell}^{\mu_{\ell}})_{k_{i_{\ell+1}},k_{o_{\ell}}}^{b_{i_{\ell+1}},b_{o_{\ell}}}\left(\Upsilon_{\ell:0}^{(\mathbf{B}_{\ell-1:0}^{\vec{\mu}})}\right)_{k_{o_\ell}}^{b_{o_\ell}}.
    \end{split}
\end{equation}

Which is precisely the generalisation of the strategy presented in Figure \ref{fig:stitching-cmo}. When evaluated from left to right, this can be performed efficiently since every contraction is a rank-2 tensor with a rank-4 tensor. Note that in this instance, the index vector $\vec{\mu}$ does not run from $(0,0,\cdots,0)$ to $(d_S^4, d_S^4,\cdots,d_S^4)$ but rather for each block of memory, it contains all $d_S^{4\ell}$ combinations of basis elements, with all other operations fixed at $\mu_0$. There are therefore $(k-\ell+1)\cdot d_S^{4\ell}$ values taken by $\vec{\mu}$.
\section{Identifying minimal unitary basis overlap}
\label{appendix:muub}
The effects of basis overlap in quantum tomography on the reconstruction have been discussed both with respect to conventional QST and QPT, and more recently with respect to the process tensor. 
In particular, the process tensor has shown itself to be highly sensitive to any overlap in its control basis. 
With access to all 16 dimensions of superoperator space, a mutually unbiased basis can be constructed in the form of a symmetric IC-POVM followed by an update~\cite{PhysRevA.70.052321}. 
However, in the limited case of a unitary-only basis, the ideal method is less straightforward. 
A randomly chosen unitary basis has been shown to adversely affect the reconstruction fidelity by as much as 30\%. 
Selecting a basis with mutual overlap here would be ideal, akin to the notion of a SIC-POVM in conventional quantum state tomography. 
However, it has been shown that MUUBs do not exist in dimension 10 (the dimension for single qubit channels). 
Because of this limitation, we numerically search for a basis which minimises its average overlap with the remainder of the set. This procedure is performed as follows:\par 
We parametrise these gates using the standard \texttt{qiskit} unitary parametrisation: 
\begin{gather}
    u(\theta,\phi,\lambda) = \begin{pmatrix}
\cos(\theta/2) & -\text{e}^{i\lambda}\sin(\theta/2) \\
\text{e}^{i\phi}\sin(\theta/2) & \text{e}^{i\lambda+i\phi}\cos(\theta/2) 
\end{pmatrix}.
\label{unitary-param}
\end{gather}
For two unitaries $u$ and $v$, let $\mathcal{U}$ and $\mathcal{V}$ be their superoperator equivalent, according to some representation. The overlap between the two channels is given by the Hilbert-Schmidt inner product: 
\begin{equation}
    \langle A, B\rangle_{\text{HS}} := \Tr[A^\dagger B].
\end{equation}
Importantly, this quantity is independent of representation, allowing us to select a form most desirable for computation. To this effect, we use the row-vectorised convention for states. Here, operations are given by $\mathcal{U} = u\otimes u^\ast$. The inner product between two unitaries parametrised as in~(\ref{unitary-param}) is then:
\begin{gather}
    \begin{split}
        \langle \mathcal{U},\mathcal{V}\rangle_{\text{HS}} &= \Tr[\mathcal{U}^\dagger \mathcal{V}]\\
        &= \Tr[(u\otimes u^\ast)^\dagger \cdot (v\otimes v^\ast)] \\
        &= \Tr[(u^\dagger\cdot v)\otimes (u^T\cdot v^\ast)]\\
        &= \Tr[u^\dagger v]\cdot \Tr[u^\dagger v]^\ast\\
        &= \left|\Tr[u^\dagger v]\right|^2
    \end{split}
    \label{HSIP-unitary}
\end{gather}
If we write $u=u(\theta_1,\phi_1,\lambda_1)$ and $v = v(\theta_2,\phi_2,\lambda_2)$, then (\ref{HSIP-unitary}) can be straightforwardly written (after some simplification) as

\begin{gather}
    \begin{split}
        & \Tr[u^\dagger v] = \cos\frac{\theta_1}{2}\cos\frac{\theta_2}{2} + \text{e}^{i(\phi_2 - \phi_1)}\sin\frac{\theta_1}{2}\sin\frac{\theta_2}{2} + \text{e}^{i(\lambda_2-\lambda_1)}\sin\frac{\theta_1}{2}\sin\frac{\theta_2}{2} + \text{e}^{i(\lambda_2 + \phi_2 - \lambda_1 - \phi_1)}\cos\frac{\theta_1}{2}\cos\frac{\theta_2}{2}\\
        & \Rightarrow \langle \mathcal{U},\mathcal{V}\rangle_{\text{HS}} = 
        4\cos^2\left(\frac{1}{2}(\lambda_1 - \lambda_2 + \phi_1 - \phi_2)\right)\cos^2\left(\theta_1 - \theta_2\right)
    \end{split}
\end{gather}
This simple expression for the inner product of any two single-qubit unitaries allows us to construct an objective function for the straightforward mutual minimisation of overlap between all ten elements of the basis set. Let 
\begin{gather}
    \mathscr{U}(\vec\theta,\vec\phi,\vec\lambda) = \{\mathcal{U}_i\}_{i=1}^{10} \equiv \{(\theta_i,\phi_i,\lambda_i)\}_{i=1}^{10}
\end{gather}
be our parametrised basis set. A basis set with the least mutual overlap can then be found by minimising the sum of the squares of each unitary with the remainder of the set. This minimises both the average overlap and the variance of overlaps with the remainder of the set. That is, by computing:
\begin{gather}
    \begin{split}
        \argmin_{(\vec\theta,\vec\phi,\vec\lambda)} \sum_{i=1}^{10}\sum_{j>i}\left(\langle \mathcal{U}_i,\mathcal{U}_j\rangle_{\text{HS}}\right)^2 = \sum_{i=1}^{10}\sum_{j>i}16\cos^4\left(\frac{1}{2}(\lambda_i - \lambda_j + \phi_i - \phi_j)\right)\cos^4\left(\theta_i - \theta_j\right)
    \end{split}
\end{gather}
One such ideal set can be found in Table~\ref{tab:muub-params}. This is the basis set used for the data obtained in the main text. Its overlaps with respect to the Hilbert-Schmidt inner product are listed in Table~\ref{tab:muub-overlaps}.
\begin{table}[h!]
\begin{tabular}{@{}lp{1.5cm}p{1.5cm}p{1.5cm}p{0cm}@{}}
\hline
                   & $\theta$ & $\phi$ & $\lambda$  &  \\ \hline
$\mathcal{U}_1$    & 1.1148   & 1.5606    & 0.8160  &  \\
$\mathcal{U}_2$    & -2.1993  & -2.0552   & -0.3564 &  \\
$\mathcal{U}_3$    & 0.9616   & -0.8573   & 1.2333  &  \\
$\mathcal{U}_4$    & 2.2655   & -2.7083   & 0.3154  &  \\
$\mathcal{U}_5$    & -0.1013  & -0.5548   & -1.1472 &  \\
$\mathcal{U}_6$    & 1.8434   & 0.8074    & -1.1772 &  \\
$\mathcal{U}_7$    & -2.2036  & 1.9589    & 2.4002  &  \\
$\mathcal{U}_8$    & -1.2038  & -0.2023   & 1.2355  &  \\
$\mathcal{U}_9$    & 2.1791   & 3.2836    & 2.3524  &  \\
$\mathcal{U}_{10}$ & -1.3116  & 2.3082    & 0.2882  &  \\ \hline
\end{tabular}
\caption{A set of thirty parameter values which constitute a set of ten unitary gates with minimal average mutual overlap.}
\label{tab:muub-params}
\end{table}

\begin{table}[]
\centering
\begin{tabular}{@{}lllllllllll@{}}
\toprule
$\text{Tr}[\mathcal{U}_i^{\text{T}} \mathcal{U}_j]$ & $\mathcal{U}_1$ & $\mathcal{U}_2$ & $\mathcal{U}_3$ & $\mathcal{U}_4$ & $\mathcal{U}_5$ & $\mathcal{U}_6$ & $\mathcal{U}_7$ & $\mathcal{U}_8$ & $\mathcal{U}_9$ & $\mathcal{U}_{10}$ \\ \midrule
$\mathcal{U}_1$ & 1 &  &  &  &  &  &  &  &  &  \\
$\mathcal{U}_2$ & 0.19688 & 1 &  &  &  &  &  &  &  &  \\
$\mathcal{U}_3$ & 0.19688 & 0.11111 & 1 &  &  &  &  &  &  &  \\
$\mathcal{U}_4$ & 0.16758 & 0.19688 & 0.19688 & 1 &  &  &  &  &  &  \\
$\mathcal{U}_5$ & 0.16758 & 0.19688 & 0.19688 & 0.16758 & 1 &  &  &  &  &  \\
$\mathcal{U}_6$ & 0.19688 & 0.11111 & 0.11111 & 0.19688 & 0.19688 & 1 &  &  &  &  \\
$\mathcal{U}_7$ & 0.03286 & 0.19688 & 0.19688 & 0.16758 & 0.16758 & 0.19688 & 1 &  &  &  \\
$\mathcal{U}_8$ & 0.16758 & 0.19688 & 0.19688 & 0.16758 & 0.03286 & 0.19688 & 0.16758 & 1 &  &  \\
$\mathcal{U}_9$ & 0.19688 & 0.11111 & 0.11111 & 0.19688 & 0.19688 & 0.11111 & 0.19688 & 0.19688 & 1 &  \\
$\mathcal{U}_{10}$ & 0.16758 & 0.19688 & 0.19688 & 0.03286 & 0.16758 & 0.19688 & 0.16758 & 0.16758 & 0.19688 & 1 \\ \midrule
Average & 0.24907 & 0.25146 & 0.25146 & 0.24907 & 0.24907 & 0.25146 & 0.24907 & 0.24907 & 0.25146 & 0.24907 \\ \bottomrule
\end{tabular}%
\caption{Hilbert-Schmidt overlap between each element of the numerically constructed (approximate) MUUB. We find this to be the most uniformly overlapping unitary basis possible, thus optimal for PTT.}
\label{tab:muub-overlaps}
\end{table}

\section{IBM Quantum Analyses}
\label{appendix:experiments}
The quantum device procedures in this work were carried out on IBM Quantum cloud devices: \emph{ibmq\_boeblingen}, \emph{ibmq\_johannesburg}, \emph{ibmq\_valencia}, \emph{ibmq\_bogota}, \emph{ibmq\_manhattan}, \emph{ibmq\_montreal}, and \emph{ibmq\_guadalupe}. Data from the first three is the same as the data featured in Ref~\cite{White-NM-2020}, with the exception of the MUUB jobs on \emph{ibmq\_valencia}, which was newly taken for this work. The detail of these runs can be found within that reference. All other data taken from the remainder of the devices was newly collected for this work. All devices are fixed frequency superconducting transmon quantum computers. Below, we step through the circuits conducted in Sections~\ref{se:ReFi} and~\ref{sec:applications} on these devices.

\subsection{Reconstruction fidelities}
Reconstruction fidelity experimentally validates the quality of a model by comparing predictions made by the model with data generated by the device, where the data is not used to create the model. The process tensor establishes a mapping from a sequence of control operations to a final state. Therefore, in this context, the comparison is between the final state predicted by a process tensor model subject to a sequence of operations (from outside the basis set), with the actual state reconstructed when the same sequence of operations is run on the device. The results of Figure~\ref{fig:RF_boxplot} are reconstruction fidelity distributions for a number of sequences of three random unitaries.\par 
The procedure for constructing a single qubit three-step process tensor is as follows: 
\begin{enumerate}
    \item Initialise the system,
    \item Wait some time $T_1$,
    \item Apply basis element $\mathcal{U}_{\mu_0}$,
    \item Wait some time $T_2$,
    \item Apply basis element $\mathcal{U}_{\mu_1}$,
    \item Wait some time $T_3$,
    \item Apply basis element $\mathcal{U}_{\mu_2}$,
    \item Wait some time $T_4$,
    \item Measure in $X$, $Y$, and $Z$ bases.
\end{enumerate}
The total number of circuits here is $10\times10\times 10\times 3 = 3000$. Finally, the data is then processed according to the LI/MLE processing methods stipulated in the main text. For LI, this means constructing the density matrix corresponding to each basis sequence. For MLE, this means shaping the data into a $(6,10,10,10)$ array, where the first dimension corresponds to each of the six effects in the POVM $\{\ket{+}\!\bra{+}, \ket{i+}\!\bra{i+},\ket{0}\!\bra{0},\ket{-}\!\bra{-}, \ket{i-}\!\bra{i-},\ket{1}\!\bra{1}\}$. Note that in general, the times $T_i$ can be chosen to be different, and include whatever background dynamics and circuit structure the experimenter is interested in. This is simply a choice of quantum stochastic process being studyied. In the cases of \emph{ibmq\_manhattan} and \emph{ibmq\_bogota}, the circuit structures are as described in Section~\ref{sec:applications}.\par 

Once the process tensor data is collected and an LI/ML model constructed, a number of random unitary circuits are generated, with each gate chosen by the \texttt{scipy.stats.unitary\_group.rvs()} function. These gates are then run on the real devices, following the same circuit structure. QST is then performed at the end of each sequence. The conditional state is computed by contracting the random sequence into the process tensor model. Finally, the state fidelity is computed between the predicted and the actual states. This then forms the data sets shown in Figure~\ref{fig:RF_boxplot}. With all things equal, the average reconstruction fidelity will necessarily increase with an increased number of shots per circuit. It will also decrease if the final states are noisier or more mixed, since this will add to the variance of sampling statistics.\par 

Once the model is validated to the desired level of accuracy, it becomes a useful tool for optimal control of the non-Markovian system. In Section~\ref{ssec:state-improvement} we use the same process tensors from \emph{ibmq\_manhattan} and \emph{ibmq\_bogota} to show a circuit-by-circuit improvement of the fidelities of states generated by IBM Quantum devices. So as to avoid readout error obfuscating any results, or overstating any improvements, we first performed gate set tomography (GST) using the \texttt{pyGSTi} software package~\cite{pygsti} in order to estimate the actual POVM giving $X$, $Y$, and $Z$ projections on the device. This POVM was then used both in reconstructing the states, and in the PTT estimate. 
\subsection{Conditional Markov order circuit improvement}
For each of the conditional Markov order tests, a five step process with $\ell = 1$, $\ell = 2$, and $\ell = 3$ is considered. This amounts to reconstructing, respectively, 5, 4, and 3 memory block process tensors. The structure of the circuits is similar to the three step process tensor, however not all circuit elements are varied. For example, with $\ell=3$, this means reconstructing the three process tensors corresponding to circuit structure $\mathcal{U}_{\mu_0}-\mathcal{U}_{\mu_1}-\mathcal{U}_{\mu_2}-\Pi_i$, $\mathcal{U}_0 - \mathcal{U}_{\mu_0}-\mathcal{U}_{\mu_1}-\mathcal{U}_{\mu_2} -\Pi_i$, and $\mathcal{U}_0 -\mathcal{U}_0 - \mathcal{U}_{\mu_0}-\mathcal{U}_{\mu_1}-\mathcal{U}_{\mu_2}- \Pi_i$. For $\ell = 2$, a subset of the same data can be reused: fixing $\mu_0=0$ and varying $\mu_1$ and $\mu_2$, for example. The only additional information required is that a projective measurement needs to be made in position 2 of the circuit in order to determine the state at the end of the first $\ell=2$ memory block. A similar process follows for determination of $\mathbf{\Upsilon}_{5:0}^1$, with an extra memory block process tensor constructed with a projective measurement at position 1. This totals $3 \times (10\times10\times 10 \times 3) = 9000$ circuits for $\mathbf{\Upsilon}_{5:0}^3$, an extra $10\time10\times 3 = 300$ circuits for $\mathbf{\Upsilon}_{5:0}^2$, and an extra $10\times 3 = 30$ circuits for $\mathbf{\Upsilon}_{5:0}^1$.\par 
Since the state is being propagated along in our finite Markov order stitching procedure, it is important that it is well-characterised without measurement error. To this effect, we use GST again to estimate our POVM. This is more essential than before, since now our PTT construction is contingent on inputting the correct form of the operation. This is also true of the unitary gates we apply, however single qubit error rates are $\mathcal{O}(10^{-4})$, compared with measurement errors of $\mathcal{O}(10^{-2})$, and so a far smaller assumption.
\end{document}